\documentclass[12pt,preprint]{emulateapj}
\usepackage{url}
\usepackage{graphicx}
\usepackage{natbib}
\usepackage{color}
\definecolor{purple}{rgb}{.9,0,.1}

\begin{document}

\title{The many-faceted light curves of young disk-bearing stars \\ in Upper Sco and $\rho$ Oph observed by $K2$ Campaign 2}

\author{
Ann Marie Cody\altaffilmark{1,2},
Lynne A. Hillenbrand\altaffilmark{3}
}

\altaffiltext{1}{NASA Ames Research Center, Moffett Field, CA 94035}
\altaffiltext{2}{Bay Area Environmental Research Institute, 625 2nd St., Suite 209, Petaluma, CA 94952}
\altaffiltext{3}{Department of Astronomy, California Institute of Technology, Pasadena CA 91125}

\begin{abstract}

The $K2$ Mission has photometrically monitored thousands of stars at high precision and cadence in a series of $\sim$80-day campaigns focused on sections of the ecliptic plane.  During its second campaign, $K2$ targeted over 1000 young stellar objects (YSOs) in the $\sim$1--3~Myr $\rho$~Ophiuchus and 5--10~Myr Upper Scorpius regions. From this set, we have carefully vetted photometry from {\em WISE} and {\em Spitzer} to identify those YSOs with infrared excess indicative of primordial circumstellar disks. We present here the resulting comprehensive sample of 288 young disk-bearing stars from B through M spectral types and analysis of their associated $K2$ light curves. Using statistics of periodicity and symmetry, we categorize each light curves into eight different variability classes, notably including ``dippers'' (fading events), ``bursters'' (brightening events), stochastic, and quasi-periodic types. Nearly all (96\%) of disk-bearing YSOs are identified as variable at 30-minute cadence with the sub-1\% precision of {\em K2}. Combining our variability classifications with (circum)stellar properties, we find that the bursters, stochastic sources, and the largest amplitude quasi-periodic stars have larger infrared colors, and hence stronger circumstellar disks. They also tend to have larger H$\alpha$ equivalent widths, indicative of higher accretion rates. The dippers, on the other hand, cluster toward moderate infrared colors and low H$\alpha$. Using resolved disk observations, we further find that the latter favor high inclinations, apart from a few notable exceptions with close to face-on disks. These observations support the idea that YSO time domain properties are dependent on several factors including accretion rate and view angle.

\end{abstract}

  
\section{Introduction}

Pre-main-sequence stars are highly variable due to mechanisms operating in or near the stellar photosphere, in the magnetosphere and
innermost circumstellar disk regions, or in the disk atmosphere over a broader range of stellocentric radii.  Specific processes include: rotation of hot and cold photospheric star spots, fluctuations in the accretion flow from the circumstellar disk onto the star (the rate and/or the geometry), and asymmetry in the density or the geometry of the disk material that traverses our line-of-sight as the star, magnetosphere, and inner disk all co-rotate.  Variability thus probes a diverse set of physical processes. 
While the phenomenology and interpretation of young star variability has received attention since the work of \cite{joy1945}, followed by, e.g., \cite{herbig1954} and \cite{walker1954},
it is only in the last several years that high-quality light curves have become commonplace for young stars.  Multi-week duration and high cadence observations \citep[e.g.][]{cody2010,cody2011,cody2014} complement work spanning longer multi-year duration but at much lower cadence \citep[e.g.][]{findeisen2015, parks2014, rice2012}.

The \cite{cody2014} study of 1-3 Myr old stars in NGC 2264, based on high-quality space-based photometry from $CoRoT$ \cite{baglin2006}, established the light curve morphology infrastructure that we use as the basis for this work. Metrics established therein enable ranking of sources along two axes: one a scale ranging from pure periodicity to complete stochasticity in terms of pattern repetition within the light curve, and the other a scale of flux asymmetry, ranging from predominantly fading or dipping (exhibiting short-term fades) to predominantly rising or bursting (exhibiting short-term flux increases). 

Subsequent work with Kepler in its $K2$ mission \citep{howell2014} has further defined and refined sub-categories of young star variability within the basic framework. Specifically, the $K2$ Field 2 pointing encompassed the Upper Scorpius region of recent star formation, as well as the molecular cloud near $\rho$ Ophiuchus in which star formation is ongoing. 
Periodic objects are studied in detail by \cite{rebull2018} in the context of pre-main sequence rotation evolution.  Sub-categories known as ``scallop shells'' and ``persistent/transient flux dips" are discussed by \cite{stauffer2017} and \cite{david2017}, and are interpreted as structures associated with the magnetospheric region in rapidly rotating, diskless, low mass young stars. ``Bursters", discussed by \cite{cody2017} are interpreted as discrete accretion events lasting hours to about a week, and can be quasi-periodic, while ``dippers'' are interpreted as obscuring dust clumps that are associated with either disk structure or orbiting bodies and semi-periodically cross the line of sight. Herein we present a comprehensive light curve census for the disk-bearing members of Upper Sco and $\rho$ Oph.

Extinction is quite high towards ``$\rho$ Oph" but some cluster members, typically those of higher mass, are bright enough for study with $K2$. The sizable young ``Upper Sco" association \cite[see][for a review]{pm2008} is essentially gas free but there is a small amount of dust extinction ($A_V < 1$). The association samples a wide range in mass -- from mid-B type stars having several to ten solar masses, all the way down to late M-type, very low mass stars and sub-stellar mass objects.  Notable studies include the early kinematic work that culminated in \cite{preibisch2002} as well as contemporaneous x-ray \citep[e.g.][]{kohler2000} and wide-field optical \citep[e.g.][]{ardila2000} studies, through to the most recent additions to the stellar population by e.g. \cite{rizzuto2011,rizzuto2015}. For $\rho$ Oph, \cite{wilking2008} provides a compilation of accepted members.  Based on this literature (plus in a few cases, original unpublished survey work), the young star community generated sets of stars that were submitted for observation with $K2$.

Relative to NGC 2264, which was observed with high precision and cadence photometry by $CoRoT$, the older 3-10 Myr old Upper Scorpius region observed by $K2$ offers a somewhat more evolved disk population to investigate.  The younger 1-3 Myr old $\rho$ Oph region that is adjacent to Upper Sco should be comparable to the NGC 2264 population in terms of disk fraction, though the population is more heavily extincted.  We consider both Upper Sco and $\rho$~Oph together when defining the sample for investigation and categorizing the variability, but do assess the distribution among the different light curve families separately, in comparing to NGC 2264.

The outline of the paper is as follows.  \S2 describes the sample selection, beginning with all objects proposed for observation in the $K2$ Field 2 Campaign and down-selecting to consider only disk-bearing sources, similar to in \cite{cody2014}. \S3 discusses the $K2$ data processing and photometry, and \S4 the selection of variable light curves.  The variability classes are assigned in \S5 based on the ``Q" and ``M" metrics for periodicity and flux asymmetry, respectively.  \S6 presents and the relationship between light curve morphology and circumstellar disk properties.   Finally, \S7 presents comparison of the present results to other recent studies of $K2/C2$ data, and \S8 our summary and discussion.

\section{Sample}

To parallel our efforts in NGC 2264, we aimed to assemble and analyze as complete a set of light curves as possible for young, low-mass disk-bearing stars among the $K2$ Campaign 2 targets.
We initially selected objects submitted under programs GO2020, GO2047, GO2052, GO2056, GO2063, and GO2085, all specifically targeting Upper Scorpius and $\rho$~Ophiuchus members or candidate members. At a later point, we became aware that aperiodic variability is exhibited among some of the targets in additional programs observing cool dwarf stars, suggestive of young star behavior. As a result, we added objects from programs GO2104, GO2051, GO2054, GO2069, GO2029, GO2106, GO2089, GO2092, GO2049, GO2045, GO2107, GO2075, and GO2114 to our sample, but only if their proper motions were suggestive of membership in the Upper Sco region. This resulted in a set of 2072 potential young stars, upon which we made further cuts based on WISE photometry. One eliminated likely field star interlopers (as described in \S2.1 with 587 stars removed) and another selected only the warm inner-disk-bearing systems (see \S2.2 with 1137 stars removed). The set of light curves considered in this work (Table 1) includes 340 young stars with presumably primordial circumstellar disks,
of which 288 are considered to have suitably good quality light curves for further analysis.  

We spatially divided this sample into $\rho$~Oph and Upper Sco sets by considering the coordinates of each star. There is an overdensity of K2 sources in a $1.2\arcdeg\times 1.2\arcdeg$ square surrounding the position RA=146.79, Dec=-24.60. The square shape is likely an artifact of target selection for various {\em K2} programs. Nevertheless, we have used it as our boundary to roughly separate the young stars near the $\rho$~Oph core from the older ones in the less nebulous surrounding regions (see further discussion in \cite{cody2017}. We find that this method yields a similar sample than that generated with an $A_V>2$ extinction contour.

\subsection{Selection of young stars in Upper Sco and $\rho$ Oph}

Known members of Upper Sco and $\rho$ Oph regions have already been vetted  by, e.g., \cite{ardila2000,slesnick2006, luhman2012,lodieu2013,rizzuto2015}. However, the Upper Sco
population is known to be incompletely cataloged, and there remain hundreds of photometric and proper motion selected candidate members to be confirmed or rejected. Many of these were included in the K2 Campaign 2 proposals. For $\rho$ Oph, \cite{wilking2008} provide a compilation of accepted members which is thought to be relatively complete given the more compact distribution on the sky and the decades of deep infrared survey work in this region. 

\begin{figure}
\epsscale{1.2}
\plotone{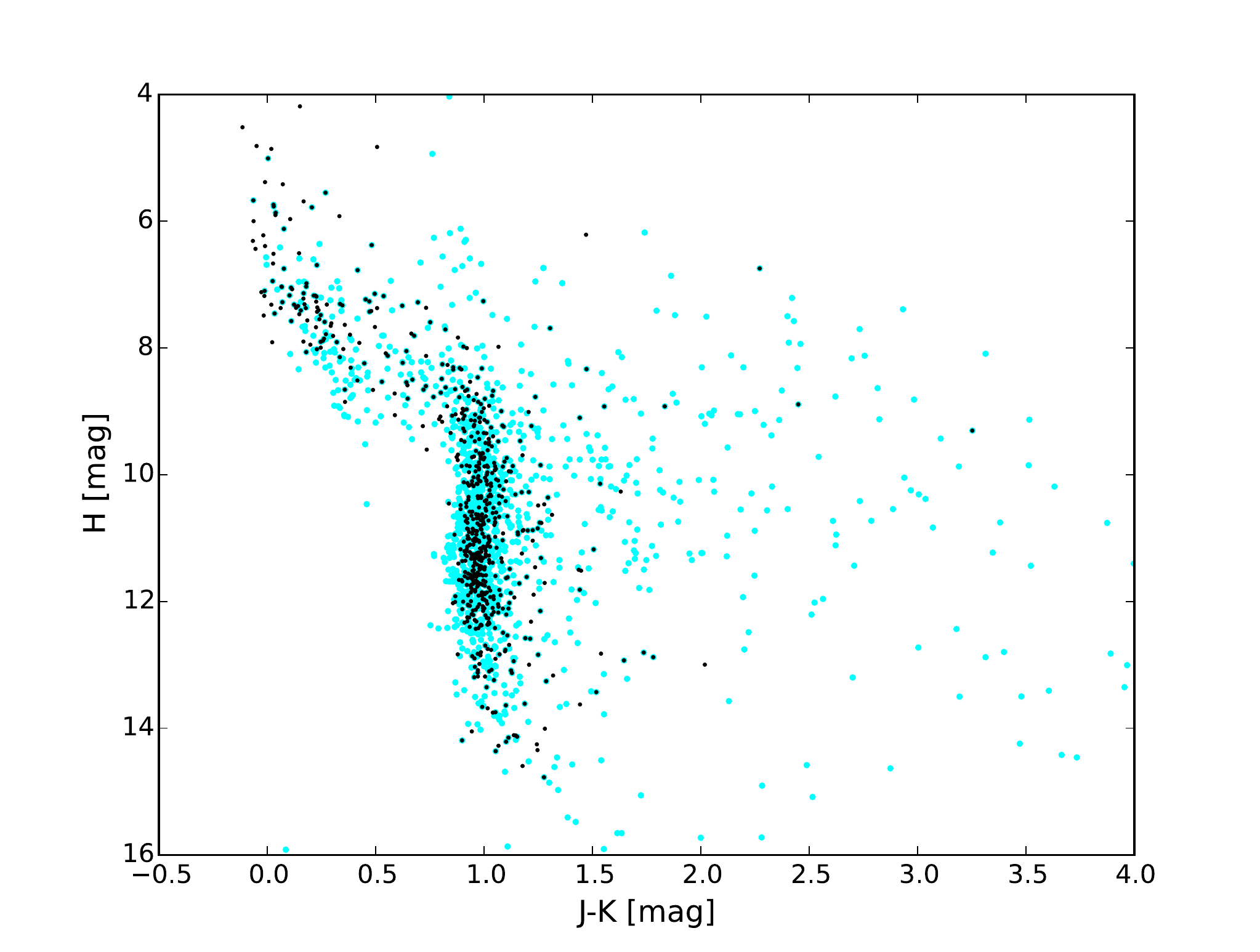}
\caption{
A $2MASS$ near-infrared color magnitude diagram of K2 Campaign 2 young star candidates (cyan) 
compared to known Upper Sco members from LM12 (black).
The distinct sequence of pre-main sequence stars retains some field star contaminants
due to the overlap in colors between young pre-main sequence and older post-main sequence stars.}
\label{cmd_nir}
\end{figure}

\begin{figure}
\epsscale{1.2}
\plotone{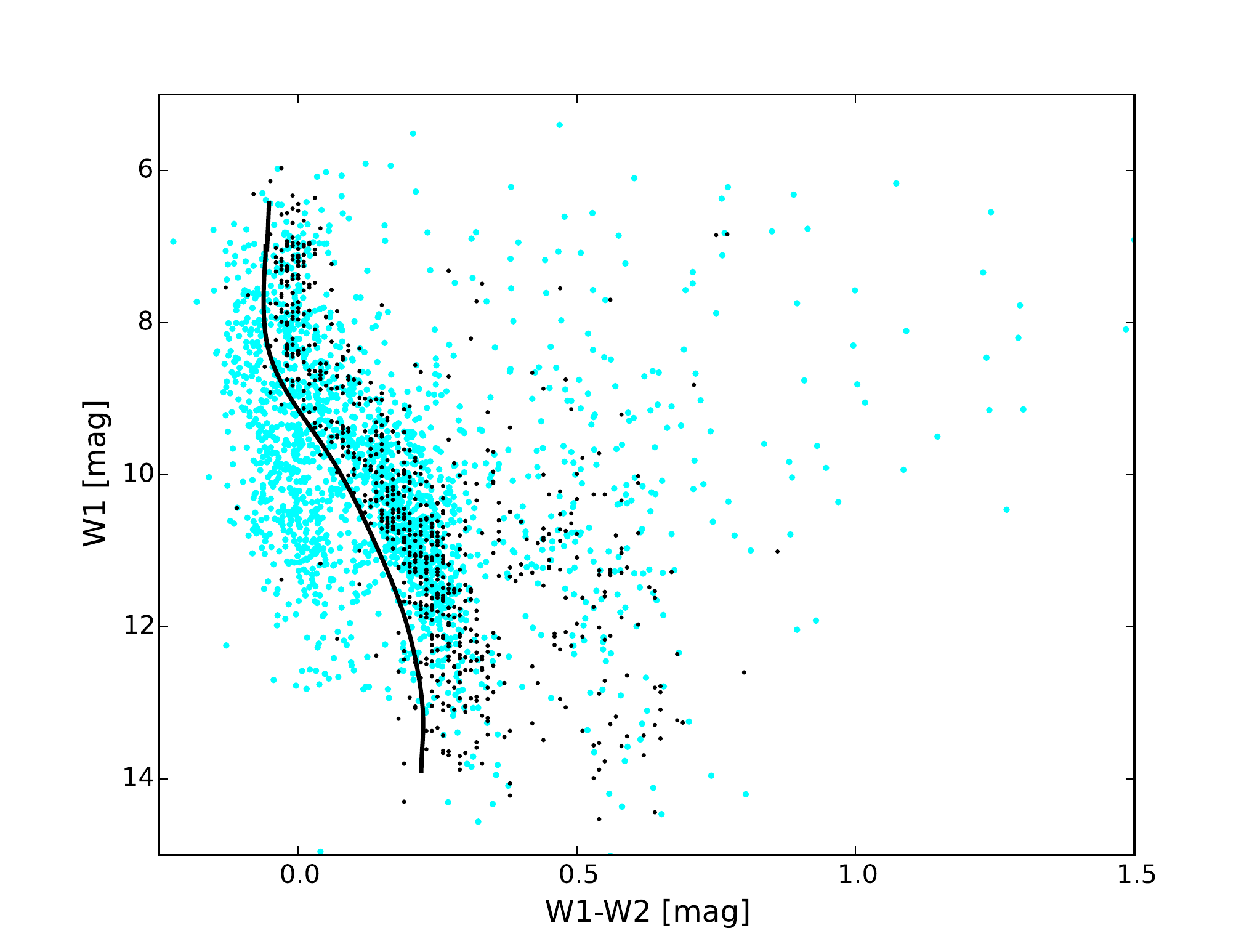}
\caption{Similar to Figure~\ref{cmd_nir}, the mid-infrared $WISE$ infrared color magnitude diagram shows all $K2$ targets in cyan, overlaid with the set of LM12 Upper Sco members in black.  A large set of the targeted $K2$ stars are too blue to be in the Upper Sco region or $\rho$ Oph cluster. The black points blueward of $W1-W2$=0.3 display a distinct sequence, which we have fitted, and subsequently shifted to model the blue boundary. This boundary then serves as the cut-off between likely and unlikely young stars-- the latter of which we remove from our sample.}
\label{cmd_mir}
\end{figure}

To assess which of the (mainly Upper Sco) candidates have photometry consistent with the expected 1--10 Myr old sequence of cluster members, we plotted near-infrared and mid-infrared color-magnitude diagrams for all objects with prior reports of being young and/or having proper motions consistent with membership in the Sco-Cen region, and compared them with the $K2$ samples (Figures \ref{cmd_nir} and \ref{cmd_mir}).  Infrared data from the 2MASS and WISE missions is available for nearly all of the stars in our sample. Over 65\% of these candidates do {\em not} appear in Luhman \& Mamajek's (2012; LM12) work. We therefore queried the IRSA database \footnote{https://irsa.ipac.caltech.edu} to match objects to infrared and near-infrared counterparts, using a 6\arcsec\ matching radius to account for the large WISE pixels. This photometry is provided with flags, and we eliminated measurements for which the {\em WISE} band 1 or 2 signal-to-noise ratio (SNR) was less than 10, or for which the band 3 SNR was less than 7. In addition, we discarded photometry reported as affected by diffraction spike contamination, persistence, halo, or optical ghosts. The remaining datapoints were used to clean the $K2$ sample of field stars, as described below.

A near-infrared $H$ versus $J-K$ color-magnitude diagram (Figure \ref{cmd_nir}) displays a clean and nearly vertical cluster sequence at low masses, with stars previously 
vetted by LM12 forming a narrower sequence among the broader set of $K2$ observed stars.  There are few stars blueward of $J-K=0.8$, apart from the bright end ($H < 9$). 
There are many red outliers beyond $J-K=1.3$ and these are likely disk-bearing objects (see \S4). Although contaminants are present, the near-infrared color-magnitude 
diagram does not appear to have many interlopers.

A mid-infrared $W1$ versus $W1-W2$ color-magnitude diagram (Figure \ref{cmd_mir}), on the other hand, reveals that there is indeed a large population of field objects in the 
$K2$ candidate Upper Sco and $\rho$ Oph sample.  A clear cluster sequence once again emerges, but now there are many $K2$ lightcurve sources lying blueward of that sequence. 
Unlike the near-infrared colors -- which become bluer by $\sim$0.2-0.8 mag during pre-main sequence evolution and redder in the post-main sequence -- the mid-infrared colors 
exhibit little evolution, in fact no more than 0.05 mag at any given mass over the entirety of 1-1000 Myr of evolution.  Furthermore, the range in color with mass is 
relatively small, $<$0.3 mag. This means that at any given color, sources having the same or similar brightness are at the same distance, essentially regardless of age.  
Sources at different distances are shifted vertically by the distance modulus, again essentially independent of age.  The resulting effect is a cleaner break between 
age and distance degeneracies in the mid-infrared relative to near-infrared color-magnitude diagrams, where the spread in color at a given magnitude masks the distance 
modulus effects.

We fit a polynomial to the sequence of disk-less stars from LM12 in the WISE $W1$ versus $W1-W2$ color-magnitude diagram, and then shifted it until 96\% of these points like on the red side of the curve. All $K2$ objects blueward of this curve are hereafter removed from the young star sample. This rejection reduces the sample from an initial 2072 to 1485.  Remaining field star contaminants are likely older objects with a distance close to that of Upper Sco members.

\subsection{Selection of stars with inner disks}

\begin{figure}
\epsscale{1.2}
\plotone{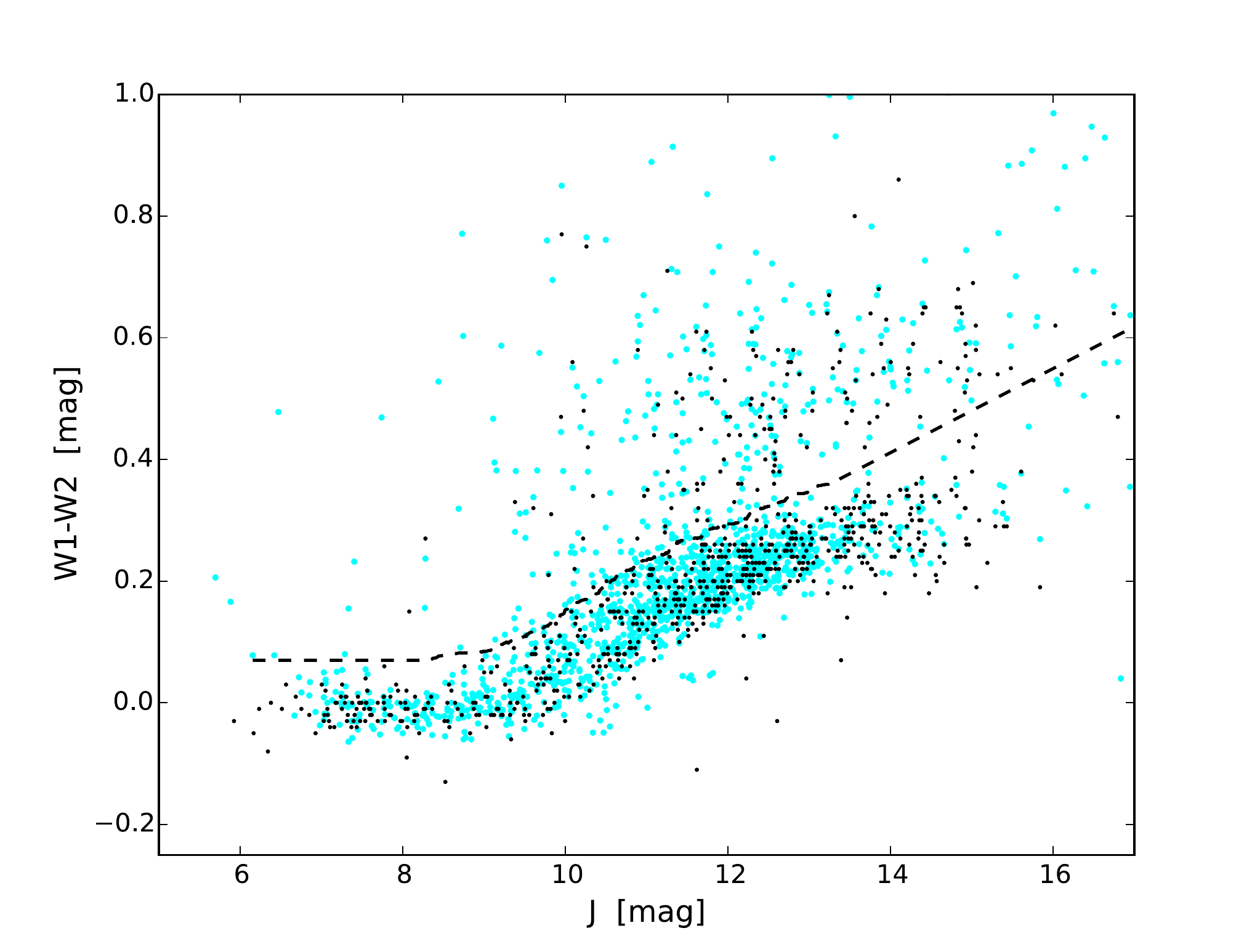}
\plotone{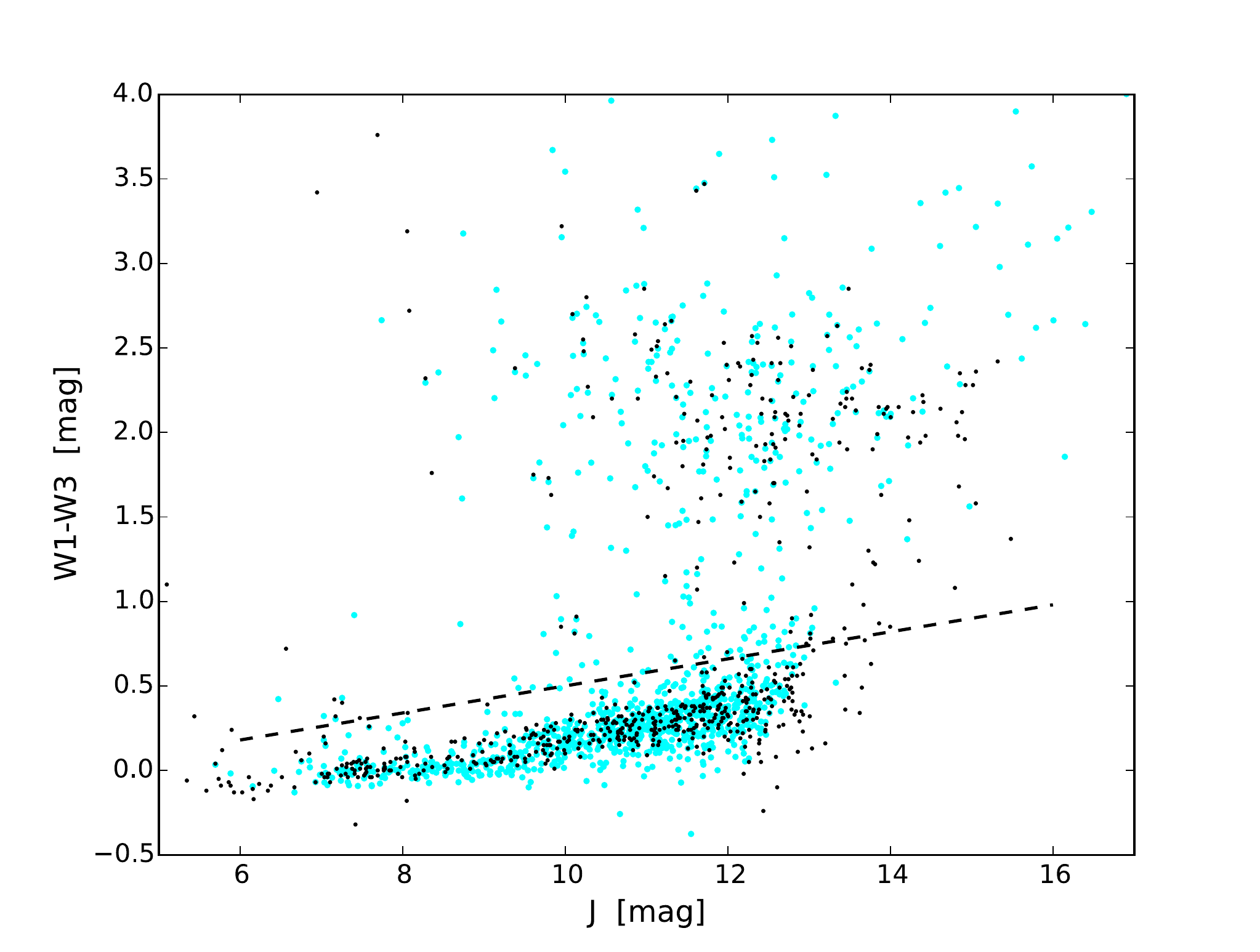}
\plotone{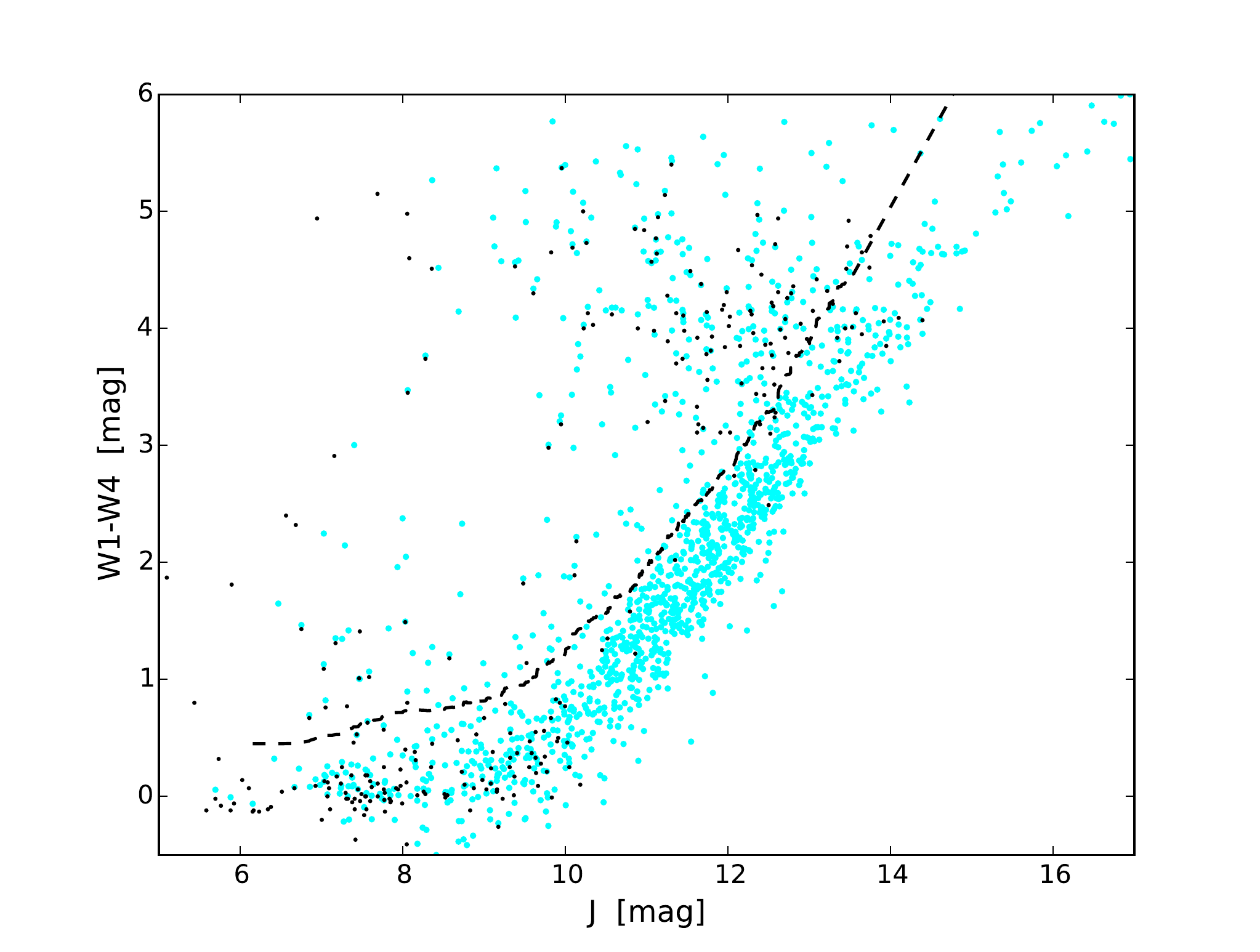}
\caption{We use $WISE$ colors, plotted against 2MASS $J$-band magnitude, to discern which of our young star sample has infrared excesses 
indicative of circumstellar disks. Shown here are both the K2 sample (cyan), and the LM12 sample (black).  For each set of $WISE$ colors in 
the K2 sample, we determine the median trend as a function of $J$. We then shift that median upward by 1$\sigma$ to denote the boundary 
between disk-bearing and diskless stars.}
\label{wise_excess}
\end{figure}

With our cleaned young star sample, we then identified objects with evidence of inner, largely primordial, circumstellar disks. We searched for infrared excesses in infrared color-magnitude diagrams utilizing $WISE$ colors versus $J$-band magnitudes, which exhibit a clearly bimodal sequence of stars with and without excesses, as shown in Figure~\ref{wise_excess}. To define the border between the disk-bearing stars and naked photospheres, we first computed a running median of $W1-W2$ values as a function of $J$ magnitude, then shifted this median curve upward such that it in lay above 84\% of the points (i.e., median plus one standard deviation). We took 
this boundary as the criterion for selecting stars with $W1-W2$ excesses, identifying 326 likely disk bearing stars in this way. Not all disks will appear as infrared excess sources in the $W1-W2$ color (4.5 $\mu$m excess), however, so we repeated this exercise for $W1-W3$ (8 $\mu$m excess) and $W1-W4$ (24 $\mu$m excess), where the longer wavelength photometry is available. This netted 77 and 65 additional disk-bearing stars, respectively, namely those lacking strong shorter wavelength excesses due to inner cleared regions in their disks.

Our disk selection method includes fairly liberal color-magnitude cut-offs, and we find 158 stars for which a potential infrared excess is seen only in one band. Of these, 11 appear in either LM12 or \cite{dawson2013}. Only three of this set are classified by those authors as 
disk-bearing; the other eight are all listed as class III. It is difficult to envision a scenario under which a disk could appear as an W2-only excess; debris disks may appear at $W3$ and $W4$, but we are interested only in primordial disks that retain significant gas. Therefore we removed these objects from the sample. We inspected the $WISE$ images for targets with excesses only in $W3$ or $W4$, and of those that were more than 1-$\sigma$ above the color-magnitude cut-off, the majority had suspicious nebulous background. Only three had clear $W3$-band excesses, as seen in the image and SED, and two of those objects were too faint to properly centroid and produce a light curve. We retained the remaining target (EPIC 203440253) in the disk sample. None of the $W4$-only sources were included in our young disk sample, as they are likely debris disks. 

Through further literature searches, we found that a few asymptotic giant branch (AGB) stars made it into the sample due to their very red colors. We eliminated eight such objects (EPIC ids 203812608, 203871153, 203882149, 203902911, 203962241, 203963116, 203970821, and 204232199) from the disk list based on reported spectroscopic data indicating AGB status or otherwise an M spectral and III luminosity class. 

The disk-bearing sample was augmented with some sources that were missing usable $WISE$ photometry (e.g., due to source crowding or bright infrared background) but are previously known disk-bearing sources from {\em Herschel}, {\em Spitzer} Space Telescope, or submillimeter data that are more sensitive than $WISE$ to cool dust. We performed a literature search on all objects with missing $WISE$ data to identify potential such cases.  As a result, the following stars were added to the disk hosting set: EPIC 203806628 \cite{rebollido2015}, EPIC 203833873 \cite{isella2009}, EPIC 204514546 \cite{honda2015}, EPIC 205684783 \cite{andrews2007}, and EPIC 210282534 \cite{najita2015}. We augment this list with disk-bearing sources found by \cite{gutermuth2009} in $\rho$ Oph using {\em Spitzer}: EPIC 203860592, EPIC 203863066, EPIC 203924502, and EPIC 203934728. In addition, there were four sources (EPIC 204281210, EPIC 204399980, EPIC 205249328, and EPIC 205327575) for which we had declined to use the {\em WISE} data, but LM12 declared them to be disk bearing. 

We created spectral energy distributions (SEDs) with all available photometry for each source, to determine which, if any, of the $WISE$-flagged objects should be retained 
in the disk-bearing sample. In addition to the four sources named above from LM12, one star with little literature attention showed an SED with an infrared slope consistent 
with a young star excess: EPIC 203725791. Upon further inspection of its light curve, we opted to add it to the sample.

We have compared our disk selection results with those of LM12, who also identified infrared excess sources using all four $WISE$ bands. There are 474 stars in the K2 
Campaign 2 young star dataset (of which they list 140 as having a disk). After rejecting sources with excesses in a single band, in no cases do we find evidence of a 
disk where LM12 do not. Moreover, as noted above there are just four cases for which LM12 report an infrared excess at one of the $WISE$ bands, whereas we do not detect it. 

We retain our no-disk classifications for these. Thus overall, 99\% of our classifications are in agreement with LM12's results.

\subsection{Selection of stars with good $K2$ light curves}

After the above adjustments, from the initial set of $K2$ Campaign 2 targets from which we down-selected to 1485 likely young cluster members, we are left with 340 disk bearing stars; 217 are in Upper Sco, and 123 are in $\rho$~Oph.  The resulting composite disk fraction (23\%) is consistent with the value of 27\% quoted by \cite{erickson2011} for stars between the embedded $\rho$~Oph core and the more dispersed Upper Sco region.  We list the names of each object in Table~1 along with available spectral types.  A ``blend" flag is given if the $K2$ photometry contains more than one source.  And a ``sample'' flag indicates whether they source is retained in the final light curve sample after the further considerations below.

A further, practical limitation to our analysis below comes from considering the quality of the $K2$ photometry. Our photometric extraction methods (see Section~3) to 
produce light curves perform fairly well on bright stars, but they often fail to centroid on low signal-to-noise objects with Kepler magnitudes less than $\sim$16. Many of 
these are near the sub-stellar limit, and a fraction of them are in our disk-bearing sample. However, we are unable to study the variability properties of those objects 
missing reliable $K2$ photometry. Thus, we were compelled to disregard another 45 faint stars from the disk set. There were a further seven extremely bright stars that were 
too saturated to produce a systematics-free light curve. We also removed these objects from the disk set.

After removal of the faint members, we were left with 288 disk-bearing stars for analysis of variability properties.

\subsection{Present $K2$ Upper Sco sample relative to previous $CoRoT$ sample in NGC 2264}

Considering the 176 young classical T Tauri stars already probed by {\em CoRoT} in NGC~2264 \citep{cody2014}, with our sample of 288 identified Upper Sco and $\rho$ Oph members with $K2$ data, we increase the set of disk-bearing YSOs with high quality light curves to over 450.  Relative to the $CoRoT$ NGC 2264 sample, the $K2$ Upper Sco and $\rho$ Oph sample has greater breadth in stellar mass.  There are 15 stars in the BAF spectral type range (all of these were too bright for {\em CoRoT} in NGC 2264) and low-mass targets all the way down to M8 (whereas the NGC 2264 set reached only to spectral type of M5). It is also important to keep in mind possible age spreads among the {\em K2} stars, as compared to previous samples. This is not only because $\rho$~Ophiuchus is significantly younger than Upper Sco, but also because there may be a range of ages in Sco itself \citep{fang2017}. We explore potential age and mass dependences of variability in \S6.

\LongTables
\begin{deluxetable*}{cccccccc}
\tabletypesize{\scriptsize}
\tablecolumns{7}
\tablewidth{0pt}
\tablecaption{Young inner disk-bearing stars in $K2$ Campaign 2}
\tablehead{
\colhead{EPIC id} & \colhead{2MASS id} & \colhead{SpT} &  \colhead{Reference} & \colhead{Blend} & \colhead{Region} & \colhead{Sample}\\ 
\colhead{} & \colhead{} & \colhead{} & \colhead{} & \colhead{} & \colhead{} & \colhead{flag} \\
}
\startdata
202610930 & J16232307-2901331 & - & - & - & Sco & Y \\
202876718 & J16181616-2802300 & A0V & \cite{debruijne1999} & - & Sco & Y \\
203082998 & J16244448-2719036 & - & - & - & Sco & Y \\
203083616 & J16145253-2718557 & - & - & - & Sco & Y \\
203318214 & J16112601-2631558 & M2.0 & \cite{rizzuto2015} & - & Sco & Y \\
203337814 & J16164756-2628178 & - & - & - & Sco & Y \\
203343161 & J16245587-2627181 & M5.5 & \cite{ansdell2016} & - & Sco & Y \\
203377650 & J16165225-2620387 & - & - & - & Sco & Y \\
203382255 & J16144265-2619421 & - & - & - & Sco & Y \\
203385048 & J16181618-2619080 & M4.5 & \cite{luhman2012} & - & Sco & Y \\
203410665 & J16253849-2613540 & K7.0 & \cite{ansdell2016} & - & Sco & Y \\
203417549 & J16213469-2612269 & K5e & \cite{skiffBMK} & - & Sco & Y \\
203429083 & J15570350-2610081 & - & - & - & Sco & Y \\
203440253 & J16252883-2607538 & M2.5 & \cite{rizzuto2015} & - & Sco & Y \\
203465909 & J16274905-2602437 & - & - & - & Sco & Y \\
203542463 & J16180868-2547126 & - & - & - & Sco & Y \\
203559274 & J16175432-2543435 & - & - & - & Sco & Y \\
203604427 & J16290873-2534240 & - & - & - & Sco & Y \\
203637940 & J16262774-2527247 & M0 & \cite{walter1994} & - & Sco & Y \\
203649927 & J16240289-2524539 & B8II & \cite{debruijne1999} & - & Sco & Y \\
203664569 & J16163345-2521505 & M0.5 & \cite{luhman2012} & - & Sco & Y \\
203690414 & J16011398-2516281 & M4.0 & \cite{rizzuto2015} & - & Sco & Y \\
203703016 & J16145244-2513523 & M3.5 & \cite{rizzuto2015} & Y & Sco & Y \\
203710077 & J15554883-2512240 & G3 & \cite{luhman2012} & - & Sco & Y \\
203712588 & J16251521-2511540 & - & - & - & Sco & Y \\
203716389 & J16251727-2511054 & - & - & - & Oph & Y \\
203725791 & J16012902-2509069 & M3.5 & \cite{rizzuto2015} & Y & Sco & Y \\
203726323 & J16134880-2509006 & M5.0 & \cite{lodieu2006} & - & Sco & F \\
203749770 & J16271273-2504017 & M1.0 & \cite{rizzuto2015} & - & Oph & Y \\
203750883 & J16133650-2503473 & M3.5 & \cite{luhman2012} & - & Sco & Y \\
203770366 & J16150524-2459351 & M5.25 & \cite{luhman2012} & - & Sco & Y \\
203770559 & J16250208-2459323 & M4.5 & \cite{erickson2011} & Y & Oph & Y \\
203770673 & J16145928-2459308 & M4.25 & \cite{luhman2012} & - & Sco & Y \\
203774126 & J16295459-2458459 & A2V & \cite{debruijne1999} & - & Sco & Y \\
203785905 & J16281385-2456113 & M0 & \cite{erickson2011} & - & Oph & Y \\
203786695 & J16245974-2456008 & M3.5 & \cite{erickson2011} & Y & Oph & Y \\
203789325 & J16174768-2455251 & - & - & - & Sco & Y \\
203789507 & J15570490-2455227 & - & - & - & Sco & Y \\
203791768 & J16271836-2454537 & M3.75 & \cite{erickson2011} & - & Oph & Y \\
203794605 & J16302339-2454161 & em & \cite{skiffBMK} & - & Sco & Y \\
203795359 & J16282992-2454062 & - & - & - & Oph & F \\
203797163 & J16280011-2453427 & M5 & \cite{erickson2011} & - & Oph & Y \\
203801323 & J16255893-2452483 & M4 & \cite{erickson2011} & Y & Oph & Y \\
203806628 & J16271513-2451388 & M2 & \cite{reboussin2015} & - & Oph & Y \\
203810851 & J15575444-2450424 & - & - & - & Sco & Y \\
203822485 & J16272297-2448071 & M4.25 & \cite{erickson2011} & - & Oph & Y \\
203822946 & J16251891-2448006 & M3 & \cite{erickson2011} & - & Oph & Y \\
203824153 & J16285407-2447442 & M1.5 & \cite{erickson2011} & - & Oph & Y \\
203826403 & J16264441-2447138 & M4 & \cite{erickson2011} & - & Oph & Y \\
203833873 & J16265843-2445318 & K7 & \cite{wahhaj2010} & Y & Oph & Y \\
203837701 & J16262189-2444397 & M6.5 & \cite{manara2015} & - & Oph & Y \\
203842632 & J16271382-2443316 & M3.5 & \cite{manara2015} & - & Oph & Y \\
203843009 & J16075567-2443267 & M5.5 & \cite{slesnick2008} & - & Sco & Y \\
203843841 & J16273982-2443150 & - & - & - & Oph & F \\
203843911 & J16262367-2443138 & K5 & \cite{reboussin2015} & - & Oph & Y \\
203848625 & J16202863-2442087 & - & - & - & Sco & Y \\
203848661 & J16255754-2442082 & K2 & \cite{erickson2011} & - & Oph & Y \\
203849739 & J16262753-2441535 & - & - & - & Oph & Y \\
203850058 & J16270659-2441488 & M5 & \cite{manara2015} & - & Oph & Y \\
203850425 & J16272146-2441430 & - & - & - & Oph & F \\
203850605 & J16271951-2441403 & M0IVe & \cite{pecaut2016} & - & Oph & Y \\
203851860 & J16294427-2441218 & M5 & \cite{walter1994} & Y & Sco & Y \\
203852282 & J16273311-2441152 & K8 & \cite{luhman1999} & - & Oph & Y \\
203856041 & J16273894-2440206 & M2.5 & \cite{manara2015} & - & Oph & F \\
203856109 & J16095198-2440197 & - & - & - & Sco & Y \\
203856244 & J16264125-2440179 & M5 & \cite{erickson2011} & - & Oph & Y \\
203860070 & J16272648-2439230 & M1 & \cite{reboussin2015} & - & Oph & F \\
203860546 & J16272943-2439161 & - & - & - & Oph & F \\
203860592 & J16273942-2439155 & K5 & \cite{reboussin2015} & - & Oph & Y \\
203862309 & J16274270-2438506 & M2 & \cite{erickson2011} & - & Oph & Y \\
203863066 & J16273863-2438391 & M8 & \cite{manara2015} & - & Oph & Y \\
203864032 & J16264897-2438252 & M3.5 & \cite{manara2015} & - & Oph & Y \\
203867167 & J16254767-2437394 & M3.5 & \cite{erickson2011} & - & Oph & Y \\
203867975 & J16270233-2437272 & - & - & - & Oph & F \\
203868595 & J16270943-2437187 & - & - & - & Oph & F \\
203870022 & J16273832-2436585 & M0 & \cite{reboussin2015} & - & Oph & Y \\
203870058 & J16281650-2436579 & M4 & \cite{reboussin2015} & - & Oph & Y \\
203874287 & J16265904-2435568 & M4 & \cite{luhman1999} & - & Oph & F \\
203876897 & J16150807-2435184 & - & - & - & Sco & Y \\
203877533 & J16243969-2435091 & M4.5 & \cite{erickson2011} & - & Oph & Y \\
203878861 & J16271213-2434491 & M2 & \cite{manara2015} & - & Oph & F \\
203878912 & J16264419-2434483 & - & - & - & Oph & F \\
203881373 & J16260931-2434121 & A0 & \cite{luhman1999} & - & Oph & Y \\
203881640 & J16270910-2434081 & K8 & \cite{luhman1999} & - & Oph & Y \\
203884731 & J16273267-2433239 & - & - & - & Oph & F \\
203887087 & J16281379-2432494 & M4 & \cite{erickson2011} & - & Oph & Y \\
203888154 & J16273285-2432348 & - & - & - & Oph & F \\
203889938 & J16072625-2432079 & M3.5 & \cite{luhman2012} & - & Sco & Y \\
203891430 & J16275180-2431455 & - & - & - & Oph & F \\
203891751 & J16274629-2431411 & M7.5 & \cite{manara2015} & - & Oph & Y \\
203892903 & J16224539-2431237 & - & - & - & Sco & Y \\
203893434 & J16272738-2431165 & M0 & \cite{luhman1999} & - & Oph & Y \\
203893891 & J16285694-2431096 & M5.5 & \cite{erickson2011} & - & Oph & Y \\
203894375 & J16264214-2431029 & - & - & - & Oph & F \\
203895738 & J16273812-2430429 & M4 & \cite{luhman1999} & - & Oph & Y \\
203895983 & J16041893-2430392 & M2.5 & \cite{rizzuto2015} & - & Sco & Y \\
203896277 & J16273718-2430350 & A0 & \cite{erickson2011} & - & Oph & Y \\
203899786 & J16252434-2429442 & M4.5 & \cite{erickson2011} & - & Oph & Y \\
203901938 & J16271003-2429133 & - & - & - & Oph & F \\
203902450 & J16271848-2429059 & M1.5 & \cite{luhman1999} & - & Oph & F \\
203903767 & J16262295-2428461 & - & - & - & Oph & F \\
203904212 & J16262083-2428395 & - & - & - & Oph & F \\
203904213 & J16275525-2428395 & - & - & - & Oph & Y \\
203904426 & J16270597-2428363 & M6.5 & \cite{luhman1999} & - & Oph & F \\
203904870 & J16270410-2428299 & - & - & - & Oph & F \\
203905576 & J16261886-2428196 & M0 & \cite{reboussin2015} & - & Oph & Y \\
203905625 & J16284527-2428190 & M3.75 & \cite{erickson2011} & - & Oph & Y \\
203905980 & J16284703-2428138 & M4.5 & \cite{erickson2011} & - & Oph & Y \\
203908052 & J16273018-2427433 & K8 & \cite{luhman1999} & - & Oph & F \\
203909356 & J16260704-2427241 & - & - & - & Oph & F \\
203909577 & J16272844-2427210 & K6.5 & \cite{luhman1999} & - & Oph & F \\
203909943 & J16270457-2427156 & - & - & - & Oph & F \\
203912136 & J16110360-2426429 & M8 & \cite{luhman2012} & - & Sco & Y \\
203912674 & J16253958-2426349 & M2 & \cite{erickson2011} & - & Oph & Y \\
203913635 & J16265444-2426207 & M0 & \cite{luhman1999} & - & Oph & F \\
203913804 & J16275558-2426179 & M2 & \cite{erickson2011} & - & Oph & Y \\
203914316 & J16261882-2426105 & M7 & \cite{manara2015} & - & Oph & F \\
203914960 & J16262152-2426009 & M7 & \cite{manara2015} & - & Oph & Y \\
203915424 & J16272658-2425543 & M8 & \cite{erickson2011} & - & Oph & Y \\
203916376 & J16274987-2425402 & A7 & \cite{erickson2011} & - & Oph & Y \\
203917608 & J16274978-2425219 & - & - & - & Oph & Y \\
203917711 & J16260137-2425203 & - & - & - & Oph & F \\
203919315 & J16273084-2424560 & M3.25 & \cite{erickson2011} & - & Oph & Y \\
203920354 & J16262357-2424394 & K7 & \cite{manara2015} & - & Oph & Y \\
203922515 & J16262226-2424070 & M8 & \cite{manara2015} & - & Oph & F \\
203923185 & J16252622-2423566 & G7 & \cite{erickson2011} & - & Oph & Y \\
203924150 & J16271168-2423419 & - & - & - & Oph & F \\
203924502 & J16260302-2423360 & K1 & \cite{luhman1999}\tablenotemark{a} & - & Oph & Y \\
203925443 & J16281475-2423225 & K0: & \cite{erickson2011} & - & Oph & Y \\
203926424 & J16264502-2423077 & M0 & \cite{reboussin2015} & - & Oph & Y \\
203926667 & J16262138-2423040 & - & - & - & Oph & F \\
203926890 & J16263778-2423007 & K8 & \cite{luhman1999} & - & Oph & F \\
203927902 & J16283266-2422449 & G7 & \cite{erickson2011} & - & Oph & Y \\
203928175 & J16282333-2422405 & K5 & \cite{erickson2011} & - & Oph & Y \\
203928921 & J16265497-2422296 & K8 & \cite{luhman1999} & - & Oph & F \\
203929332 & J16261684-2422231 & K6 & \cite{erickson2011} & - & Oph & Y \\
203930599 & J16274028-2422040 & K5 & \cite{erickson2011} & - & Oph & Y \\
203931628 & J16221989-2421482 & A1III/IV & \cite{debruijne1999} & - & Sco & Y \\
203932787 & J16265839-2421299 & - & - & - & Oph & F \\
203933268 & J16255965-2421223 & M5.5 & \cite{erickson2011} & - & Oph & Y \\
203934728 & J16262335-2420597 & G6 & \cite{erickson2011} & - & Oph & Y \\
203935066 & J16261033-2420548 & M0 & \cite{reboussin2015} & - & Oph & F \\
203935537 & J16255615-2420481 & K5 & \cite{reboussin2015} & - & Oph & Y \\
203936815 & J16264285-2420299 & M1 & \cite{erickson2011} & Y & Oph & Y \\
203937317 & J16261706-2420216 & K7.5 & \cite{ansdell2016} & - & Oph & Y \\
203938167 & J16151239-2420091 & - & - & - & Sco & Y \\
203938591 & J16264923-2420029 & K6 & \cite{luhman1999} & - & Oph & Y \\
203941210 & J16272622-2419229 & - & - & - & Oph & F \\
203941868 & J16271027-2419127 & G3 & \cite{reboussin2015} & - & Oph & Y \\
203943710 & J16250062-2418442 & - & - & - & Oph & Y \\
203944338 & J16265863-2418346 & M4.5 & \cite{manara2015} & - & Oph & F \\
203945512 & J16271372-2418168 & - & - & - & Oph & Y \\
203946909 & J16273742-2417548 & M7.5 & \cite{manara2015} & - & Oph & Y \\
203947119 & J16260457-2417514 & - & - & - & Oph & F \\
203947305 & J16244104-2417488 & M2 & \cite{erickson2011} & - & Oph & Y \\
203950167 & J16230923-2417047 & - & - & - & Sco & Y \\
203953466 & J16262407-2416134 & K6 & \cite{reboussin2015} & - & Oph & Y \\
203954898 & J16263682-2415518 & M0 & \cite{erickson2011} & - & Oph & Y \\
203955457 & J16253673-2415424 & K4 & \cite{erickson2011} & - & Oph & Y \\
203956650 & J16283256-2415242 & M3.5 & \cite{erickson2011} & - & Oph & Y \\
203962599 & J16265677-2413515 & K7 & \cite{erickson2011} & - & Oph & Y \\
203969672 & J16270907-2412007 & M2.5 & \cite{erickson2011} & - & Oph & Y \\
203969721 & J16264643-2412000 & G3.5 & \cite{erickson2011} & - & Oph & Y \\
203971352 & J16281271-2411355 & M6 & \cite{manara2015} & - & Oph & Y \\
203972079 & J16245729-2411240 & M3.5 & \cite{erickson2011} & - & Oph & Y \\
203981774 & J16262097-2408518 & M2 & \cite{erickson2011} & - & Oph & Y \\
203982074 & J16260289-2408474 & F3 & \cite{erickson2011} & - & Oph & Y \\
203987773 & J16261877-2407190 & M3.25 & \cite{erickson2011} & - & Oph & Y \\
203995761 & J16281673-2405142 & K6 & \cite{erickson2011} & - & Oph & Y \\
204078097 & J16095852-2345186 & M6 & \cite{luhman2012} & - & Sco & Y \\
204094503 & J16084836-2341209 & M5 & \cite{luhman2012} & - & Sco & Y \\
204103213 & J16142144-2339146 & M7 & \cite{lodieu2011} & Y & Sco & F \\
204107757 & J15560104-2338081 & M5.5 & \cite{luhman2012} & - & Sco & Y \\
204108293 & J15591135-2338002 & M6.5 & \cite{luhman2012} & - & Sco & Y \\
204120066 & J16083048-2335109 & M8.25 & \cite{luhman2012} & Y & Sco & F \\
204130613 & J16145026-2332397 & - & - & - & Sco & Y \\
204137184 & J16020517-2331070 & M4.5 & \cite{ansdell2016} & - & Sco & Y \\
204141928 & J16002323-2329595 & M6.5+M6.5 & \cite{luhman2012} & Y & Sco & F \\
204142243 & J16222497-2329553 & em & \cite{skiffBMK} & - & Sco & Y \\
204147776 & J15581270-2328364 & G2IV & \cite{pecaut2016} & - & Sco & Y \\
204160652 & J16224680-2325331 & M1 & \cite{wahhaj2010} & Y & Sco & Y \\
204161056 & J16254289-2325260 & em & \cite{skiffBMK} & - & Sco & Y \\
204176565 & J16221852-2321480 & K/M & \cite{skiffBMK} & - & Sco & Y \\
204181799 & J16135434-2320342 & M4.5 & \cite{lodieu2011} & Y & Sco & Y \\
204182919 & J16023587-2320170 & M3.5 & \cite{rizzuto2015} & - & Sco & Y \\
204187094 & J16111907-2319202 & M5 & \cite{lodieu2011} & - & Sco & Y \\
204187469 & J16251052-2319145 & K6-7 & \cite{guenther2007} & - & Sco & Y \\
204193996 & J15575396-2317416 & - & - & - & Sco & Y \\
204206295 & J16264741-2314521 & - & - & - & Sco & Y \\
204211116 & J16214199-2313432 & - & - & - & Sco & Y \\
204226548 & J15582981-2310077 & M3 & \cite{carpenter2014} & - & Sco & Y \\
204231861 & J16145131-2308515 & - & - & - & Sco & Y \\
204233955 & J16072955-2308221 & - & - & - & Sco & Y \\
204239132 & J16225177-2307070 & A1III/IV & \cite{debruijne1999} & - & Sco & Y \\
204245509 & J16141107-2305362 & K2 & \cite{carpenter2014} & - & Sco & Y \\
204248645 & J16024575-2304509 & M5 & \cite{luhman2012} & - & Sco & Y \\
204250417 & J16151361-2304261 & M6.5 & \cite{slesnick2008} & - & Sco & Y \\
204256494 & J16243654-2303000 & - & - & - & Sco & Y \\
204262368 & J16012652-2301343 & - & - & - & Sco & Y \\
204268916 & J16243520-2300022 & - & - & - & Sco & Y \\
204274536 & J16233283-2258468 & - & - & - & Sco & Y \\
204274743 & J15572986-2258438 & M4 & \cite{carpenter2014} & - & Sco & Y \\
204277211 & J16014086-2258103 & M4 & \cite{luhman2012} & Y & Sco & Y \\
204278916 & J16020757-2257467 & M2.5 & \cite{luhman2012} & - & Sco & Y \\
204281210 & J15583692-2257153 & G5IVe & \cite{pecaut2016} & - & Sco & Y \\
204290918 & J16211848-2254578 & - & - & - & Sco & Y \\
204317053 & J16024142-2248419 & M5.5 & \cite{slesnick2008} & - & Sco & Y \\
204329690 & J16220194-2245410 & - & - & - & Sco & Y \\
204342099 & J16153456-2242421 & M1 & \cite{preibisch1998} & Y & Sco & Y \\
204344180 & J16143287-2242133 & M6.5 & \cite{lodieu2011} & - & Sco & Y \\
204347422 & J16195140-2241266 & - & - & - & Sco & Y \\
204347824 & J16243182-2241207 & - & - & Y & Sco & Y \\
204360645 & J16032277-2238206 & - & - & - & Sco & Y \\
204360807 & J16215741-2238180 & - & - & - & Sco & Y \\
204365840 & J16320136-2237081 & M5.5 & \cite{luhman2012} & Y & Sco & Y \\
204372172 & J16205022-2235387 & A9V & \cite{debruijne1999} & - & Sco & Y \\
204388640 & J16020429-2231468 & - & - & - & Sco & Y \\
204395393 & J16001844-2230114 & M4.5 & \cite{luhman2012} & - & Sco & Y \\
204397408 & J16081081-2229428 & M5.75 & \cite{lodieu2011} & - & Sco & Y \\
204397879 & J16093229-2229360 & - & - & - & Sco & Y \\
204398857 & J16093164-2229224 & M2.0 & \cite{rizzuto2015} & - & Sco & Y \\
204399980 & J16131158-2229066 & A8III/IV & \cite{debruijne1999} & - & Sco & Y \\
204401119 & J16110737-2228501 & M5.75 & \cite{lodieu2011} & - & Sco & Y \\
204408707 & J16202291-2227041 & - & - & - & Sco & Y \\
204409463 & J16125528-2226542 & M5.5 & \cite{lodieu2011} & - & Sco & Y \\
204413641 & J15562477-2225552 & M4 & \cite{carpenter2014} & - & Sco & Y \\
204419255 & J16095804-2224348 & - & - & - & Sco & B \\
204428864 & J16081566-2222199 & M3.25 & \cite{lodieu2011} & - & Sco & Y \\
204434363 & J16075039-2221021 & M5.75 & \cite{lodieu2011} & - & Sco & Y \\
204435866 & J16192393-2220412 & - & - & - & Sco & Y \\
204440603 & J16142312-2219338 & M5.75 & \cite{lodieu2011} & - & Sco & Y \\
204447221 & J16094098-2217594 & M0 & \cite{preibisch1998} & - & Sco & Y \\
204449274 & J16222160-2217307 & M5 & \cite{slesnick2008} & - & Sco & Y \\
204449389 & J16082733-2217292 & - & - & - & Sco & Y \\
204467371 & J16154914-2213117 & - & - & - & Sco & Y \\
204467584 & J16111705-2213085 & M5 & \cite{lodieu2011} & - & Sco & Y \\
204469637 & J16200616-2212385 & M3.5 & \cite{rizzuto2015} & - & Sco & Y \\
204472612 & J16083455-2211559 & - & - & - & Sco & Y \\
204487447 & J16103069-2208229 & - & - & - & Sco & Y \\
204489514 & J16030161-2207523 & M4.75 & \cite{luhman2012} & - & Sco & Y \\
204495624 & J16104259-2206212 & - & - & - & Sco & Y \\
204496657 & J15570641-2206060 & M4 & \cite{carpenter2014} & - & Sco & Y \\
204501712 & J16105691-2204515 & - & - & - & Sco & Y \\
204508462 & J16194711-2203112 & M5.0 & \cite{rizzuto2015} & - & Sco & Y \\
204512343 & J15572109-2202130 & - & - & - & Sco & Y \\
204514546 & J15564002-2201400 & A8 & \cite{1988mcts.book.....H} & - & Sco & Y \\
204517888 & J16023227-2200486 & M5: & \cite{luhman2012} & - & Sco & F \\
204530046 & J16105011-2157481 & - & - & - & Sco & Y \\
204538777 & J16032625-2155378 & M5 & \cite{luhman2012} & - & Sco & Y \\
204565982 & J16270942-2148457 & M4.5 & \cite{slesnick2008} & - & Sco & Y \\
204578601 & J16193976-2145349 & M6 & \cite{luhman2012} & - & Sco & Y \\
204581550 & J16123414-2144500 & - & - & - & Sco & Y \\
204584778 & J16152516-2144013 & - & - & - & Sco & Y \\
204602441 & J16092136-2139342 & M5 & \cite{luhman2012} & - & Sco & Y \\
204607034 & J16024152-2138245 & M4.75 & \cite{luhman2012} & - & Sco & Y \\
204611292 & J16082870-2137198 & M5: & \cite{luhman2012} & - & Sco & Y \\
204615647 & J16132190-2136136 & M1.5 & \cite{rizzuto2015} & - & Sco & Y \\
204630363 & J16100501-2132318 & M0.0 & \cite{rizzuto2015} & - & Sco & Y \\
204637622 & J16042097-2130415 & M3.5 & \cite{luhman2012} & - & Sco & Y \\
204638512 & J16042165-2130284 & K2 & \cite{luhman2012} & - & Sco & Y \\
204649301 & J16100608-2127440 & M8 & \cite{luhman2012} & Y & Sco & F \\
204651122 & J16122289-2127158 & - & - & - & Sco & Y \\
204662993 & J16192923-2124132 & F2/3 V & \cite{debruijne1999} & - & Sco & Y \\
204757338 & J16072747-2059442 & M4.75 & \cite{luhman2012} & - & Sco & Y \\
204769599 & J16002669-2056316 & M5 & \cite{luhman2012} & - & Sco & Y \\
204776782 & J16152083-2054372 & - & - & - & Sco & Y \\
204807722 & J15570146-2046184 & - & - & - & Sco & Y \\
204810161 & J16221481-2045398 & - & - & - & Sco & Y \\
204811478 & J15555600-2045187 & M5 & \cite{luhman2012} & - & Sco & Y \\
204817605 & J16120505-2043404 & M1.0 & \cite{rizzuto2015} & - & Sco & Y \\
204830786 & J16075796-2040087 & em & \cite{skiffBMK} & - & Sco & Y \\
204832936 & J15564244-2039339 & M3.5 & \cite{rizzuto2015} & - & Sco & Y \\
204856535 & J16070014-2033092 & M2.75 & \cite{luhman2012} & - & Sco & Y \\
204860656 & J16104391-2032025 & - & - & Y & Sco & Y \\
204864076 & J16035767-2031055 & K5 & \cite{luhman2012} & - & Sco & Y \\
204870258 & J15594426-2029232 & M5.0 & \cite{rizzuto2015} & - & Sco & Y \\
204871202 & J16090071-2029086 & M5 & \cite{luhman2012} & - & Sco & Y \\
204871862 & J16070169-2028579 & M5.25 & \cite{luhman2012} & - & Sco & Y \\
204874314 & J16353913-2028195 & - & - & - & Sco & Y \\
204894208 & J16002945-2022536 & - & - & - & Sco & Y \\
204906020 & J16070211-2019387 & M5 & \cite{carpenter2014} & Y & Sco & Y \\
204908189 & J16111330-2019029 & M3 & \cite{luhman2012} & - & Sco & Y \\
204932990 & J16115091-2012098 & M3.5 & \cite{luhman2012} & - & Sco & Y \\
204933717 & J16072240-2011581 & M5.5 & \cite{slesnick2008} & - & Sco & Y \\
204939243 & J16153220-2010236 & M1.5 & \cite{rizzuto2015} & - & Sco & Y \\
204940701 & J16122737-2009596 & M4.5 & \cite{luhman2012} & Y & Sco & Y \\
204951022 & J16203026-2007037 & - & - & - & Sco & Y \\
204951731 & J16203056-2006518 & B9.5Va & \cite{skiffBMK} & - & Sco & B \\
204964091 & J16200549-2003228 & B9II/III & \cite{debruijne1999} & - & Sco & Y \\
204966512 & J16200397-2002413 & A0V & \cite{debruijne1999} & - & Sco & B \\
204982702 & J16095206-1958065 & - & - & - & Sco & Y \\
205000676 & J16220961-1953005 & - & - & - & Sco & Y \\
205002311 & J16474733-1952319 & F3V & \cite{pecaut2012} & - & Sco & B \\
205008727 & J16193570-1950426 & - & - & - & Sco & Y \\
205024407 & J15583620-1946135 & M4.0 & \cite{rizzuto2015} & - & Sco & Y \\
205037578 & J16041740-1942287 & M3.5 & \cite{luhman2012} & Y & Sco & Y \\
205038557 & J16035793-1942108 & M2 & \cite{preibisch2002} & - & Sco & Y \\
205051240 & J16140792-1938292 & - & - & - & Sco & Y \\
205061092 & J16145178-1935402 & - & - & - & Sco & Y \\
205063210 & J16073915-1935041 & - & - & - & Sco & Y \\
205064383 & J16122183-1934445 & B9V & \cite{debruijne1999} & - & Sco & Y \\
205068630 & J16111095-1933320 & M5 & \cite{preibisch2002} & Y & Sco & Y \\
205080089 & J16124410-1930102 & B9V & \cite{debruijne1999} & - & Sco & B \\
205080616 & J16082324-1930009 & K9 & \cite{carpenter2014} & - & Sco & Y \\
205086621 & J16114534-1928132 & M5Ve & \cite{skiffBMK} & - & Sco & Y \\
205088645 & J16111237-1927374 & M5 & \cite{preibisch2002} & - & Sco & Y \\
205089268 & J16092089-1927259 & A0V & \cite{debruijne1999} & - & Sco & B \\
205091879 & J16115763-1926389 & - & - & - & Sco & Y \\
205092303 & J16092054-1926318 & M5.5 & \cite{luhman2012} & - & Sco & Y \\
205092842 & J16120239-1926218 & - & - & - & Sco & F \\
205110000 & J16154416-1921171 & K5:Ve & \cite{torres2006} & - & Sco & Y \\
205115701 & J16100541-1919362 & M5.75 & \cite{luhman2012} & - & Sco & Y \\
205145188 & J16102819-1910444 & M4 & \cite{preibisch2002} & - & Sco & Y \\
205151387 & J16090075-1908526 & M1.0 & \cite{ansdell2016} & - & Sco & Y \\
205152244 & J16090002-1908368 & M5 & \cite{carpenter2014} & - & Sco & Y \\
205154017 & J16064385-1908056 & M0.0 & \cite{rizzuto2015} & - & Sco & Y \\
205156547 & J16121242-1907191 & - & - & - & Sco & Y \\
205158239 & J16142029-1906481 & K5 & \cite{carpenter2014} & - & Sco & Y \\
205160565 & J16142091-1906051 & - & - & - & Sco & Y \\
205164892 & J16102857-1904469 & M3 & \cite{preibisch1998} & - & Sco & Y \\
205164999 & J16130235-1904450 & M6 & \cite{luhman2012} & - & Sco & Y \\
205165965 & J16130996-1904269 & - & - & - & Sco & Y \\
205179845 & J16143367-1900133 & M2 & \cite{preibisch2002} & - & Sco & Y \\
205182200 & J16123916-1859284 & K2.5IV & \cite{pecaut2016} & - & Sco & Y \\
205198363 & J16153341-1854249 & - & - & - & Sco & Y \\
205208701 & J16064266-1851140 & M8 & \cite{luhman2012} & - & Sco & Y \\
205218826 & J16093653-1848009 & M3 & \cite{preibisch2002} & - & Sco & Y \\
205238942 & J16064794-1841437 & M0.0 & \cite{rizzuto2015} & - & Sco & Y \\
205241182 & J16104636-1840598 & M4.5 & \cite{slesnick2008} & - & Sco & Y \\
205249328 & J16113134-1838259 & K5 & \cite{reboussin2015} & - & Sco & Y \\
205327575 & J16382865-1813136 & B9.5IV & \cite{debruijne1999} & - & Sco & B \\
205345560 & J16062383-1807183 & - & - & - & Sco & Y \\
205364526 & J16124893-1800525 & M3 & \cite{preibisch2002} & - & Sco & Y \\
205366676 & J16095933-1800090 & M4 & \cite{carpenter2014} & - & Sco & Y \\
205375290 & J16111534-1757214 & M1 & \cite{preibisch1998} & - & Sco & Y \\
205383125 & J16095361-1754474 & M3 & \cite{preibisch2002} & - & Sco & Y \\
205519771 & J16071403-1702425 & M3.5 & \cite{rizzuto2015} & - & Sco & Y \\
205684783 & J16340916-1548168 & G5 & \cite{reboussin2015} & - & Sco & Y \\
210282528 & J16333496-1832540 & M3.5 & \cite{rizzuto2015} & - & Sco & Y \\
210282534 & J16265850-2445368 & K2 & \cite{reboussin2015} & Y & Oph & Y \\
\enddata
\tablecomments{\label{tab:obstable} Stars with inner disks observed in K2 Campaign 2, in order of EPIC id. In column 5, a 'Y' appears for blends-- cases in which ground-based photometry indicates another star or stars contaminating the K2 aperture. In column 7, 'Y' denotes objects that are included in our ultimate sample of disk-bearing stars. Objects with 'F' were too faint for K2 photometry, whereas objects with 'B' were too bright; none of these were retained in the ultimate sample of 288. }
\tablenotetext{a}{For EPIC 203924502 we disregard the spectral type of B2V given by Wahhaj et al.\ 2010 for this source, as it is inconsistent with all previously reported spectral types, as well as our own examination of independent spectra.  Comments in the paper suggest that a nearby HII region contaminates Spitzer photometry, which implies possible contamination in the optical spectrum as well if sky subtraction was not handled properly.}
\end{deluxetable*}

\section{K2 Photometry and Light Curves}

With our 288 disk-bearing stars with good quality $K2$ photometry, we now proceed to examine the light curve variability properties. We created our own light curves starting with the target pixel files. As is well known,
unstable pointing of the Kepler telescope during the $K2$ phase has a significant (but surmountable) effect on the mission's light curve quality, and we tested several mitigation methods during light curve extraction. 

All tests were conducted on a set of non-variable Campaign 2 stars encompassing the full range of Kepler magnitudes. 
For each of these targets, we produced aperture photometry with radii of 1, 2, 3, and 4 pixels. In one experiment, we allowed
the apertures to move along with the object centroid (as determined by a flux-weighted moment), and in another experiment, we
fixed the aperture location on the detector and allowed the centroid to wobble within it. The latter method produces light curves
with stronger systematic effects, but these are often just as easily removed with a detrending algorithm. For bright saturated stars,
we additionally produced photometry based on a summation of the background-subtracted flux across the entire pixel stamp. The resulting light curve
was satisfactory (i.e., free of strong systematics) in all but seven cases that we eliminated from the sample, as discussed in Section 2.3 and listed with `B' flags in Table~1.

Before comparing the precision of the raw the light curves, we cleaned them of pointing related systematics using the approach
and code described in \cite{aigrain2016}. In brief, this uses Gaussian process regression to separately model the position and
time-dependent systematics in $K2$ light curves. In general, we find that the photometric performance depends on both object
magnitude and detector position. 

We have estimated our own $K_p$ magnitudes directly from spacecraft-measured stellar flux, as opposed to adopting the values from the Ecliptic Plane Input Catalog (EPIC) which are interpolated and sometimes extrapolated from magnitudes measured at other wavelengths. The latter can be systematically too bright for young stars with disks, since many of these only have near-infrared photometry and the disk contributes emission at these wavelengths. We recalibrated stellar magnitudes for the entire $\rho$~Oph/Upper Sco sample by deriving the median relation between $K_p$ and (log) measured flux from the target pixel files of quiet field stars.

The best RMS photometric precisions achieved are approximately 0.13 millimagnitudes for a star with $K_p$=10.0 and 1.8~mmag 
for a star with $K_p$=16.0. These values are measured over the {\em entire} light curve {\em after} detrending for pointing systematics
and time-dependent drift in the light curve values. We have also measured the so-called CDPP 
(``combined differential photometric precision''), which is an assessment of the detectability of signals on six-hour timescales, and a standard
metric of the {\em Kepler} Mission (Christensen et al.). CDPP values ranged from 39 for a 10th magnitude star to 417 for a 16th magnitude star. 
Some targets had better precision with moving apertures, while others had better precision for fixed apertures. For each target,
we selected the light curve with the best performance by selecting the one with the lowest CDPP value after detrending both position-
and time-dependent effects. After making this choice, we reverted to the version that was position detrended only, so as to preserve
intrinsic stellar variability.

We display the full set of disk bearing light curves in Figure~\ref{alllcs}.  

\section{Variability selection} 

Most young accreting stars are variable, though at $K2$ photometric precision, most stars
of any age are measurably variable.  In order to detect generic aperiodic as well as
periodic variability, we require a photometrically quiet control sample.
Since we did not process the entire $K2$ Campaign 2 set, we used the set of stars rejected as 
young cluster members (as described in \S3) as a potential benchmark for typical
field star variability.

We initially attempted to assess variability by examining the standard deviations of cluster stars in comparison to the ``field'' sample, as a function of magnitude. This is a standard method for picking out outliers with excess variability.
Unfortunately, some of the unlikely cluster members turned out to be variable as well at $K2$ precision, and furthermore, there was no clear boundary between variables and non-variables as a function of magnitude. Even the true noise floor 
was difficult to discern, perhaps because of systematic effects remaining in the data.

\begin{figure*}
\epsscale{1.0}
\plotone{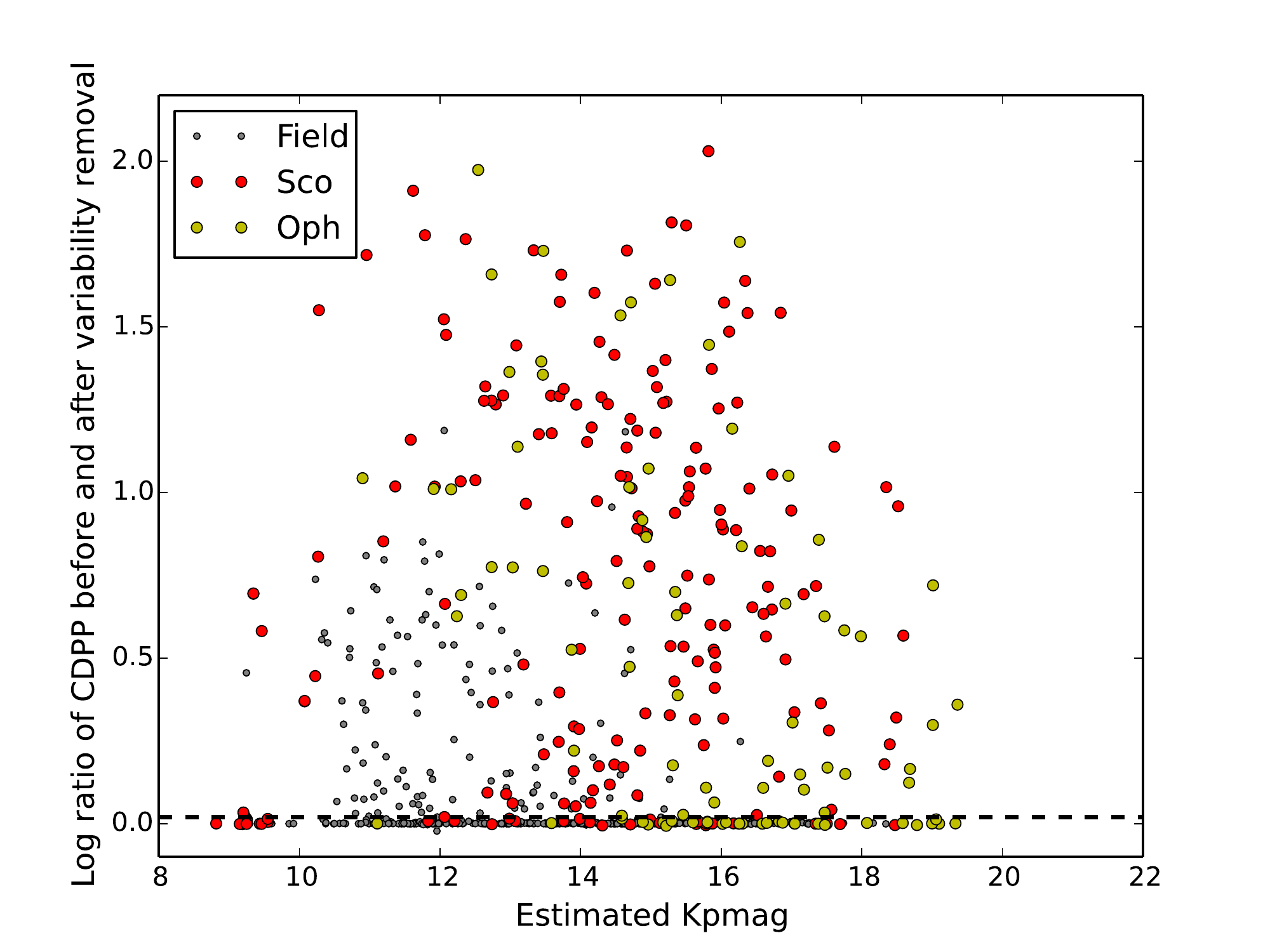}
\caption{We plot the ratio of the CDPP (a measure of statistical spread in flux on 6-hour timescales) before and after variability detrending, against estimated Kepler magnitudes for both our 288 member disk-bearing young star sample, along with a control sample of field stars that were rejected as Upper Sco/$\rho$~Ophiuchus members (see \S2.1). The samples depicted here bifurcate into a group with similar CDPP before and after detrending (i.e., zero in the ordinate), and a group with much lower CDPP after detrending (positive values along the ordinate). Variable stars were selected based on a cut-off value of 0.02, plotted as the horizontal dashed line.}
\label{varselect}
\end{figure*}

\begin{figure*}
\epsscale{1.0}
\plotone{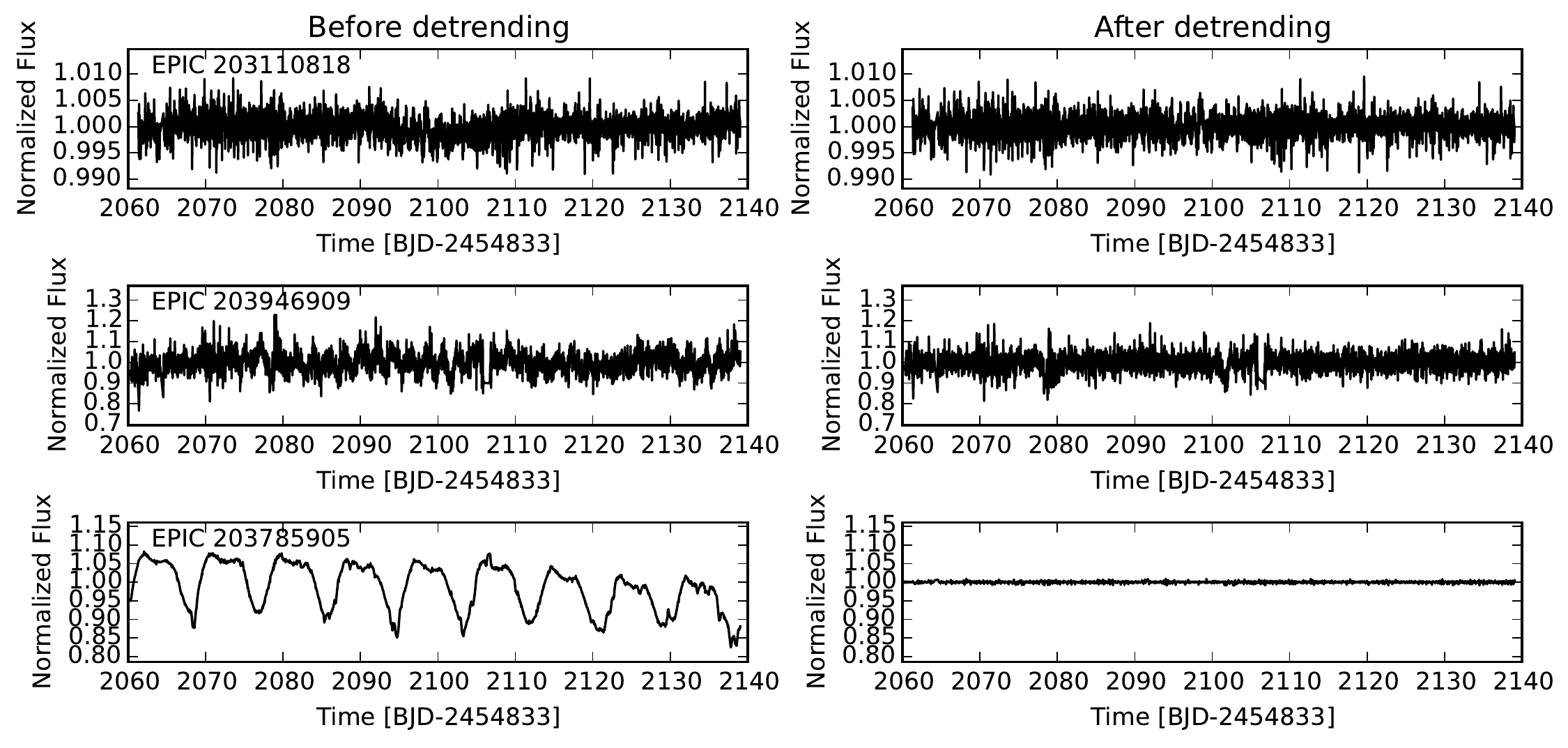}
\caption{Illustrative lightcurves at various CDPP ratios, for three example stars having similar brightness, $K_p$=14.5-15.5. The left side illustrates the light curve before variability removal, and the right after this detrending, for log CDPP ratios of 0.0 (top), 0.02 (middle), and 0.2 (bottom).}
\label{cdpplcs}
\end{figure*}

We thus identified a better metric for discerning true variables, both aperiodic and periodic. The detrending code provided by \cite{aigrain2016} provides not only a method to remove pointing systematics, but also an additional option to flatten out all variability via Gaussian process modeling. For most of our light curve assessments, we did not use this option, as it destroys the intrinsic variability by design. However, we found that a comparison of the noise levels before and after flattening provided a  measure 
of how variable a given light curve is. To quantify the noise levels, we used an estimate of the CDPP on 6-hour timescales, as described in \cite{aigrain2016}. Our final variability metric was the logarithm of the ratio of the CDPP before flattening to the CDPP after flattening. We plot this statistic against source brightness, expressed as $K_p$ magnitude, in Figure~\ref{varselect}.

Many disk-bearing stars clearly stand out in Figure~\ref{varselect} as having high amplitude variations, with CDPP ratios in excess of 100. To quantitatively gauge 
variability, we created a histogram of log CDPP ratios for the field star sample. Most of these are non-variable, and hence have log CDPP ratios near zero. But there is a 
population of presumably variable objects at higher ratios. We note a sharp cut-off between the variable and non-variable populations around $\log$~CDPP$\sim$0.015. To be 
conservative, we select as variable any object with a log ratio larger than 0.02; this cut-off is drawn as a horizontal dashed line in Figure~\ref{varselect}. Examples 
of light curves at different log CDPP ratio values are shown before and after detrending in Figure~\ref{cdpplcs}.

The only drawback to using the CDPP ratio to select variables is that variability appearing only on long timescales goes undetected. This is because the algorithm we have employed to measure CDPP \citep{aigrain2016} initially removes trends on timescales of more than a few days. We therefore added to the variability sample targets for which the overall photometric amplitude was more than a factor of 10 times the standard deviation of the trend-removed light curve.

After these assessments, We find 268 variables (77 in $\rho$~Oph and 191 in Upper Sco) among the disk-bearing sample. 
The field stars (gray) with high CDPP ratio are objects that display variations on the $<$1\% level that are inconsistent with white noise. Some of these may be real variables, and others may have light curves contaminated by systematics. 

There is a handful of additional young cluster members that fall below the CDPP ratio cut-off, but are (quasi-)periodically variable, as revealed in a periodogram and/or 
autocorrelation analysis. These were likely missed because of flukes in the detrending regimen. We identified such objects during the variability classification 
process, as described below and in cite{cody2014}. 
We identify a total of 9 periodic or quasi-periodic young disk-bearing stars not 
selected by the CDPP ratio method.  This includes 3 in $\rho$ Oph and 6 in Upper Sco 
that were below the generic variability cut-off but are included in our final variable tally.

The overall variability fraction of inner disk-bearing stars is thus 96\% . Broken down into the two clusters, we find $94_{-4}^{+2}$\% (80/85) of $\rho$~Oph 
disk-bearing targets and $97_{-2}^{+1}$\% (197/203) of Upper Sco disk-bearing targets are variable at the precision of $K2$ photometry. 
While the variability fractions are not particularly distinguishable, there are small differences between the two regions in terms of the {\it type} of variability exhibited (see Table 3 presented below).

We list the variability status of all disk-bearing stars (illustrated in Figure~\ref{alllcs}) in Table~2, including the variability type, amplitude, timescale, and metrics regarding the degree of periodicity and flux asymmetry about the mean value. The derivation of these values is discussed below. 

\LongTables
\begin{deluxetable*}{ccccccc}
\tabletypesize{\scriptsize}
\tablecolumns{7}
\tablewidth{0pt}
\tablecaption{Variability properties of young disk-bearing stars in $K2$ Campaign 2}
\tablehead{
\colhead{EPIC} & \colhead{2MASS} & \colhead{Variability} &  \colhead{Amplitude} & \colhead{Timescale} & \colhead{$Q$} & \colhead{$M$}\\ 
\colhead{id} & \colhead{id} & \colhead{type} &  \colhead{(Norm. Flux)} & \colhead{(d)} & \colhead{} & \colhead{}\\
}
\startdata
202610930 & J16232307-2901331 & B & 0.18 & 11.91 & 1.0 & -0.35 \\
202876718 & J16181616-2802300 & QPS & 0.002 & 1.95 & 0.58 & 0.09 \\
203082998 & J16244448-2719036 & N & 0.06 & 13.16 & 0.85 & -0.01 \\
203083616 & J16145253-2718557 & QPD & 0.33 & 1.35 & 0.75 & 0.53 \\
203318214 & J16112601-2631558 & P & 0.13 & 4.17 & 0.03 & 0.19 \\
203337814 & J16164756-2628178 & P & 0.06 & 0.78 & 0.06 & -0.21 \\
203343161 & J16245587-2627181 & APD & 0.09 & 2.23 & 0.77 & 0.62 \\
203377650 & J16165225-2620387 & QPS & 0.08 & 61.46 & 0.83 & -0.14 \\
203382255 & J16144265-2619421 & B & 0.10 & 78.66 & 1.0 & -1.14 \\
203385048 & J16181618-2619080 & MP & 0.03 & 1.08 & 0.16 & 0.13 \\
203410665 & J16253849-2613540 & APD & 0.24 & 7.36 & 0.91 & 0.88 \\
203417549 & J16213469-2612269 & S & 0.45 & 18.7 & 0.83 & 0.37 \\
203429083 & J15570350-2610081 & APD & 0.17 & 25.36 & 0.82 & 0.64 \\
203440253 & J16252883-2607538 & QPS & 0.05 & 6.76 & 0.32 & 0.29 \\
203465909 & J16274905-2602437 & QPS & 0.79 & 3.79 & 0.37 & 0.2 \\
203542463 & J16180868-2547126 & QPD & 0.09 & 1.56 & 0.72 & 0.44 \\
203559274 & J16175432-2543435 & QPD & 0.10 & 1.84 & 0.5 & 0.56 \\
203604427 & J16290873-2534240 & U & 0.04 & 95.25 & 1.0 & -0.36 \\
203637940 & J16262774-2527247 & QPS & 0.49 & 3.29 & 0.78 & -0.13 \\
203649927 & J16240289-2524539 & QPS & 0.003 & 2.63 & 0.48 & -0.39 \\
203664569 & J16163345-2521505 & APD & 0.08 & 47.57 & 1.0 & 1.04 \\
203690414 & J16011398-2516281 & MP & 0.02 & 1.81 & 0.55 & 0.11 \\
203703016 & J16145244-2513523 & QPS & 0.09 & 4.9 & 0.24 & 0.01 \\
203710077 & J15554883-2512240 & MP & 0.03 & 3.73 & 0.38 & -0.11 \\
203712588 & J16251521-2511540 & QPD & 0.13 & 3.29 & 0.53 & 0.52 \\
203716389 & J16251727-2511054 & B & 0.46 & 31.08 & 0.83 & -0.42 \\
203725791 & J16012902-2509069 & B & 0.26 & 10.95 & 0.87 & -0.28 \\
203749770 & J16271273-2504017 & APD & 0.18 & 11.16 & 0.91 & 0.72 \\
203750883 & J16133650-2503473 & APD & 0.29 & 4.94 & 0.9 & 0.37 \\
203770366 & J16150524-2459351 & QPD & 0.08 & 1.81 & 0.42 & 0.47 \\
203770559 & J16250208-2459323 & QPS & 0.19 & 12.5 & 0.34 & 0.32 \\
203770673 & J16145928-2459308 & P & 0.07 & 2.29 & 0.07 & 0.1 \\
203774126 & J16295459-2458459 & QPS & 0.002 & 1.95 & 0.29 & -0.39 \\
203785905 & J16281385-2456113 & QPD & 0.18 & 8.93 & 0.17 & 0.49 \\
203786695 & J16245974-2456008 & B & 0.11 & 35.65 & 1.0 & -0.59 \\
203789325 & J16174768-2455251 & QPS & 0.09 & 1.28 & 0.43 & 0.06 \\
203789507 & J15570490-2455227 & B & 0.14 & 29.89 & 0.97 & -0.41 \\
203791768 & J16271836-2454537 & APD & 0.72 & 4.0 & 0.84 & 0.37 \\
203794605 & J16302339-2454161 & B & 0.47 & 4.46 & 0.55 & -0.25 \\
203797163 & J16280011-2453427 & QPS & 0.10 & 14.71 & 0.3 & -0.06 \\
203801323 & J16255893-2452483 & MP & 0.07 & 1.26 & 0.55 & 0.12 \\
203806628 & J16271513-2451388 & APD & 0.20 & 7.31 & 0.97 & 1.1 \\
203810851 & J15575444-2450424 & QPD & 0.56 & 4.24 & 0.56 & 0.08 \\
203822485 & J16272297-2448071 & B & 0.31 & 20.33 & 0.84 & -0.29 \\
203822946 & J16251891-2448006 & MP & 0.17 & 0.68 & 0.56 & -0.12 \\
203824153 & J16285407-2447442 & QPD & 0.28 & 11.9 & 0.59 & 0.32 \\
203826403 & J16264441-2447138 & P & 0.09 & 3.97 & 0.07 & 0.03 \\
203833873 & J16265843-2445318 & S & 0.38 & 14.55 & 0.83 & -0.18 \\
203837701 & J16262189-2444397 & QPS & 0.17 & 2.63 & 0.3 & -0.06 \\
203842632 & J16271382-2443316 & S & 0.22 & 50.36 & 0.87 & -0.13 \\
203843009 & J16075567-2443267 & QPD & 0.14 & 1.35 & 0.53 & 0.69 \\
203843911 & J16262367-2443138 & QPD & 0.35 & 8.93 & 0.31 & 0.1 \\
203848625 & J16202863-2442087 & QPS & 0.05 & 8.06 & 0.17 & -0.13 \\
203848661 & J16255754-2442082 & L & 0.07 & 78.76 & 1.0 & 0.03 \\
203849739 & J16262753-2441535 & QPS & 0.23 & 39.64 & 0.84 & 0.14 \\
203850058 & J16270659-2441488 & QPD & 0.22 & 2.87 & 0.66 & 0.03 \\
203850605 & J16271951-2441403 & MP & 0.18 & 3.91 & 0.38 & -0.14 \\
203851860 & J16294427-2441218 & MP & 0.04 & 3.52 & 0.42 & 0.07 \\
203852282 & J16273311-2441152 & B & 0.12 & 79.48 & 1.0 & -0.3 \\
203856109 & J16095198-2440197 & B & 0.10 & 55.84 & 0.92 & -1.0 \\
203856244 & J16264125-2440179 & QPD & 0.08 & 4.03 & 0.7 & 0.29 \\
203860592 & J16273942-2439155 & QPD & 0.20 & 6.25 & 0.08 & 0.4 \\
203862309 & J16274270-2438506 & QPD & 0.16 & 4.31 & 0.67 & 0.38 \\
203863066 & J16273863-2438391 & QPS & 0.14 & 4.1 & 0.75 & -0.14 \\
203864032 & J16264897-2438252 & B & 0.56 & 13.22 & 0.93 & -0.25 \\
203867167 & J16254767-2437394 & P & 0.09 & 3.16 & 0.13 & -0.12 \\
203870022 & J16273832-2436585 & QPS & 0.18 & 12.5 & 0.57 & -0.24 \\
203870058 & J16281650-2436579 & S & 0.24 & 10.24 & 0.96 & -0.19 \\
203876897 & J16150807-2435184 & N & 0.004 & 7.4 & 1.0 & -0.03 \\
203877533 & J16243969-2435091 & P & 0.13 & 1.62 & 0.08 & -0.09 \\
203881373 & J16260931-2434121 & MP & 0.02 & 0.28 & 0.66 & 0.13 \\
203881640 & J16270910-2434081 & QPS & 0.52 & 2.4 & 0.6 & 0.05 \\
203887087 & J16281379-2432494 & QPD & 0.21 & 6.76 & 0.38 & 0.96 \\
203889938 & J16072625-2432079 & S & 0.05 & 8.19 & 0.94 & -0.13 \\
203891751 & J16274629-2431411 & U & 0.21 & 77.8 & 1.0 & -0.14 \\
203892903 & J16224539-2431237 & APD & 0.31 & 24.38 & 1.0 & 0.6 \\
203893434 & J16272738-2431165 & APD & 0.14 & 8.34 & 1.0 & 0.26 \\
203893891 & J16285694-2431096 & QPS & 0.19 & 3.97 & 0.56 & -0.43 \\
203895738 & J16273812-2430429 & QPS & 0.12 & 1.36 & 0.71 & -0.16 \\
203895983 & J16041893-2430392 & QPS & 0.11 & 2.45 & 0.74 & 0.08 \\
203896277 & J16273718-2430350 & L & 0.08 & 21.68 & 1.0 & 0.19 \\
203899786 & J16252434-2429442 & B & 0.48 & 5.95 & 0.61 & -0.83 \\
203904213 & J16275525-2428395 & N & 0.08 & 23.54 & 0.95 & -0.04 \\
203905576 & J16261886-2428196 & B & 2.11 & 23.82 & 1.0 & -0.66 \\
203905625 & J16284527-2428190 & B & 0.18 & 24.07 & 1.0 & -0.31 \\
203905980 & J16284703-2428138 & QPS & 1.13 & 1.89 & 0.36 & -0.29 \\
203912136 & J16110360-2426429 & QPS & 0.16 & 1.41 & 0.77 & -0.0 \\
203912674 & J16253958-2426349 & S & 0.44 & 28.05 & 0.82 & -0.11 \\
203913804 & J16275558-2426179 & S & 0.26 & 13.53 & 1.0 & -0.37 \\
203914960 & J16262152-2426009 & U & 0.08 & 37.86 & 1.0 & 0.04 \\
203915424 & J16272658-2425543 & QPS & 0.36 & 2.87 & 0.42 & -0.06 \\
203916376 & J16274987-2425402 & L & 0.03 & 6.93 & 1.0 & 0.1 \\
203917608 & J16274978-2425219 & QPS & 0.38 & 2.58 & 0.71 & 0.09 \\
203919315 & J16273084-2424560 & APD & 0.2 & 64.67 & 0.94 & 0.86 \\
203920354 & J16262357-2424394 & N & 0.69 & 12.77 & 1.0 & -0.04 \\
203923185 & J16252622-2423566 & N & 0.09 & 40.29 & 1.0 & 0.02 \\
203924502 & J16260302-2423360 & P & 0.07 & 3.47 & 0.18 & 0.27 \\
203925443 & J16281475-2423225 & L & 0.36 & 37.51 & 1.0 & 0.09 \\
203926424 & J16264502-2423077 & S & 0.46 & 9.3 & 0.88 & -0.12 \\
203927902 & J16283266-2422449 & APD & 0.17 & 0.69 & 0.5 & 0.58 \\
203928175 & J16282333-2422405 & B & 2.41 & 4.39 & 0.54 & -0.66 \\
203929332 & J16261684-2422231 & QPS & 0.12 & 3.57 & 0.2 & -0.07 \\
203930599 & J16274028-2422040 & S & 0.11 & 21.02 & 0.88 & -0.11 \\
203931628 & J16221989-2421482 & N & 0.002 & 41.25 & 0.97 & 0.1 \\
203933268 & J16255965-2421223 & QPD & 0.08 & 2.38 & 0.49 & 0.35 \\
203934728 & J16262335-2420597 & QPS & 0.03 & 5.81 & 0.74 & -0.15 \\
203935537 & J16255615-2420481 & B & 0.13 & 19.78 & 1.0 & -0.31 \\
203936815 & J16264285-2420299 & QPD & 0.55 & 8.93 & 0.66 & 0.79 \\
203937317 & J16261706-2420216 & APD & 0.07 & 7.8 & 0.85 & 0.81 \\
203938167 & J16151239-2420091 & QPS & 0.07 & 1.61 & 0.31 & -0.06 \\
203938591 & J16264923-2420029 & QP & 0.11 & 2.55 & 0.8 & -0.05 \\
203941868 & J16271027-2419127 & MP & 0.02 & 2.05 & 0.59 & -0.06 \\
203943710 & J16250062-2418442 & N & 0.04 & 69.22 & 1.0 & -0.09 \\
203945512 & J16271372-2418168 & N & 0.14 & 31.85 & 1.0 & -0.22 \\
203946909 & J16273742-2417548 & S & 0.16 & 28.69 & 0.84 & 0.0 \\
203947305 & J16244104-2417488 & L & 0.06 & 45.89 & 1.0 & -0.18 \\
203950167 & J16230923-2417047 & APD & 0.13 & 8.09 & 0.94 & 0.8 \\
203953466 & J16262407-2416134 & B & 0.17 & 34.45 & 1.0 & -0.83 \\
203954898 & J16263682-2415518 & B & 2.17 & 20.83 & 0.61 & -1.35 \\
203955457 & J16253673-2415424 & L & 0.07 & 34.78 & 1.0 & 0.19 \\
203956650 & J16283256-2415242 & MP & 0.09 & 0.68 & 0.53 & 0.24 \\
203962599 & J16265677-2413515 & APD & 0.41 & 15.94 & 0.88 & 0.51 \\
203969672 & J16270907-2412007 & QPD & 0.44 & 4.9 & 0.74 & 0.68 \\
203969721 & J16264643-2412000 & QPS & 0.04 & 3.38 & 0.45 & 0.07 \\
203971352 & J16281271-2411355 & S & 0.31 & 3.68 & 0.84 & 0.14 \\
203972079 & J16245729-2411240 & QPS & 0.04 & 6.41 & 0.23 & 0.12 \\
203981774 & J16262097-2408518 & P & 0.11 & 0.82 & 0.09 & -0.07 \\
203982074 & J16260289-2408474 & L & 0.08 & 78.76 & 1.0 & 0.07 \\
203987773 & J16261877-2407190 & QPS & 0.04 & 6.41 & 0.57 & 0.09 \\
203995761 & J16281673-2405142 & QPD & 0.41 & 5.1 & 0.53 & 0.67 \\
204078097 & J16095852-2345186 & QPS & 0.06 & 1.41 & 0.69 & -0.31 \\
204094503 & J16084836-2341209 & QPS & 0.05 & 1.66 & 0.6 & -0.21 \\
204107757 & J15560104-2338081 & APD & 0.13 & 3.23 & 0.88 & 0.55 \\
204108293 & J15591135-2338002 & QPS & 0.11 & 1.21 & 0.61 & 0.06 \\
204130613 & J16145026-2332397 & B & 0.98 & 10.67 & 0.85 & -0.35 \\
204137184 & J16020517-2331070 & QPD & 0.21 & 2.63 & 0.58 & 1.24 \\
204142243 & J16222497-2329553 & QPD & 0.42 & 6.94 & 0.58 & 0.3 \\
204147776 & J15581270-2328364 & QPS & 0.05 & 1.72 & 0.42 & 0.11 \\
204160652 & J16224680-2325331 & QPS & 0.13 & 2.81 & 0.46 & -0.2 \\
204161056 & J16254289-2325260 & B & 0.12 & 28.18 & 1.0 & -0.55 \\
204176565 & J16221852-2321480 & APD & 0.54 & 7.93 & 0.91 & 0.63 \\
204181799 & J16135434-2320342 & QPS & 0.4 & 2.21 & 0.72 & -0.19 \\
204182919 & J16023587-2320170 & P & 0.14 & 6.76 & 0.04 & -0.04 \\
204187094 & J16111907-2319202 & S & 1.62 & 17.2 & 1.0 & -0.14 \\
204187469 & J16251052-2319145 & B & 0.07 & 5.95 & 0.75 & -0.22 \\
204193996 & J15575396-2317416 & QPS & 0.07 & 15.62 & 0.19 & -0.31 \\
204206295 & J16264741-2314521 & QPD & 0.23 & 14.71 & 0.45 & 0.35 \\
204211116 & J16214199-2313432 & QPD & 0.16 & 16.57 & 0.88 & 0.52 \\
204226548 & J15582981-2310077 & B & 0.27 & 20.17 & 1.0 & -0.53 \\
204231861 & J16145131-2308515 & B & 0.05 & 10.3 & 0.95 & -0.37 \\
204233955 & J16072955-2308221 & B & 0.59 & 39.29 & 0.85 & -0.82 \\
204239132 & J16225177-2307070 & QPS & 0.01 & 4.24 & 0.47 & -0.03 \\
204245509 & J16141107-2305362 & APD & 0.11 & 30.61 & 0.95 & 0.75 \\
204248645 & J16024575-2304509 & APD & 0.03 & 30.44 & 0.9 & 0.4 \\
204250417 & J16151361-2304261 & QPS & 0.15 & 1.34 & 0.7 & -0.19 \\
204256494 & J16243654-2303000 & QPD & 0.14 & 3.29 & 0.54 & 0.5 \\
204262368 & J16012652-2301343 & QPS & 0.02 & 1.87 & 0.32 & -0.09 \\
204268916 & J16243520-2300022 & QPS & 0.08 & 1.75 & 0.42 & -0.48 \\
204274536 & J16233283-2258468 & QPD & 0.49 & 5.1 & 0.24 & 0.57 \\
204274743 & J15572986-2258438 & QPD & 0.18 & 1.77 & 0.57 & 0.46 \\
204277211 & J16014086-2258103 & QPS & 0.67 & 2.55 & 0.52 & -0.14 \\
204278916 & J16020757-2257467 & APD & 0.10 & 39.68 & 0.86 & 0.78 \\
204281210 & J15583692-2257153 & APD & 0.04 & 18.51 & 0.94 & 0.4 \\
204290918 & J16211848-2254578 & QPD & 0.18 & 2.52 & 0.74 & 0.4 \\
204317053 & J16024142-2248419 & QPS & 0.16 & 1.59 & 0.74 & -0.22 \\
204329690 & J16220194-2245410 & QPD & 0.08 & 2.15 & 0.61 & 0.3 \\
204342099 & J16153456-2242421 & B & 0.36 & 11.2 & 0.91 & -0.72 \\
204344180 & J16143287-2242133 & QPD & 0.17 & 1.82 & 0.73 & 0.29 \\
204347422 & J16195140-2241266 & B & 0.24 & 6.94 & 0.75 & -1.1 \\
204347824 & J16243182-2241207 & QPS & 0.08 & 1.58 & 0.66 & 0.2 \\
204360645 & J16032277-2238206 & QPS & 0.03 & 1.89 & 0.62 & -0.07 \\
204360807 & J16215741-2238180 & B & 0.24 & 26.4 & 0.87 & -0.49 \\
204365840 & J16320136-2237081 & QPS & 0.07 & 1.94 & 0.62 & 0.07 \\
204372172 & J16205022-2235387 & MP & 0.002 & 1.97 & 0.68 & -0.14 \\
204388640 & J16020429-2231468 & S & 0.91 & 5.76 & 0.95 & 0.12 \\
204395393 & J16001844-2230114 & B & 0.15 & 27.48 & 0.89 & -0.28 \\
204397408 & J16081081-2229428 & B & 0.07 & 1.64 & 0.59 & -0.68 \\
204397879 & J16093229-2229360 & QPD & 0.10 & 3.68 & 0.39 & 0.32 \\
204398857 & J16093164-2229224 & QPS & 0.07 & 13.16 & 0.21 & 0.04 \\
204399980 & J16131158-2229066 & APD & 0.17 & 14.87 & 0.98 & 1.1 \\
204401119 & J16110737-2228501 & QPS & 0.10 & 1.79 & 0.78 & -0.12 \\
204408707 & J16202291-2227041 & QPD & 0.13 & 3.29 & 0.68 & 0.36 \\
204409463 & J16125528-2226542 & QPS & 0.05 & 1.76 & 0.47 & 0.11 \\
204413641 & J15562477-2225552 & QPS & 0.02 & 1.97 & 0.46 & -0.03 \\
204428864 & J16081566-2222199 & S & 0.02 & 5.19 & 0.89 & 0.0 \\
204434363 & J16075039-2221021 & QPS & 0.05 & 1.59 & 0.41 & -0.26 \\
204435866 & J16192393-2220412 & QPS & 0.07 & 1.4 & 0.65 & -0.2 \\
204440603 & J16142312-2219338 & B & 0.22 & 8.3 & 1.0 & -0.93 \\
204447221 & J16094098-2217594 & QPS & 0.07 & 9.62 & 0.2 & -0.33 \\
204449274 & J16222160-2217307 & QPD & 0.10 & 70.57 & 0.83 & -0.22 \\
204449389 & J16082733-2217292 & QPD & 0.04 & 1.87 & 0.55 & 0.32 \\
204467371 & J16154914-2213117 & P & 0.10 & 1.81 & 0.12 & 0.11 \\
204467584 & J16111705-2213085 & QPD & 0.09 & 2.07 & 0.61 & 0.62 \\
204469637 & J16200616-2212385 & APD & 0.14 & 4.54 & 0.9 & 0.5 \\
204472612 & J16083455-2211559 & APD & 0.06 & 11.91 & 0.94 & 0.26 \\
204487447 & J16103069-2208229 & APD & 0.18 & 8.95 & 0.93 & 0.37 \\
204489514 & J16030161-2207523 & APD & 0.16 & 3.35 & 0.83 & 0.74 \\
204495624 & J16104259-2206212 & QPS & 0.05 & 1.97 & 0.2 & 0.02 \\
204496657 & J15570641-2206060 & QPD & 0.36 & 1.8 & 0.75 & 0.36 \\
204501712 & J16105691-2204515 & P & 0.05 & 2.03 & 0.12 & -0.02 \\
204508462 & J16194711-2203112 & QPS & 0.05 & 12.18 & 0.84 & -0.16 \\
204512343 & J15572109-2202130 & APD & 0.66 & 7.52 & 0.88 & 0.98 \\
204514546 & J15564002-2201400 & APD & 0.41 & 7.33 & 0.89 & 0.52 \\
204530046 & J16105011-2157481 & APD & 0.09 & 55.66 & 0.86 & 0.89 \\
204538777 & J16032625-2155378 & QPD & 0.11 & 61.21 & 0.82 & 0.26 \\
204565982 & J16270942-2148457 & QPS & 0.46 & 1.55 & 0.78 & -0.06 \\
204578601 & J16193976-2145349 & MP & 0.05 & 1.71 & 0.68 & -0.0 \\
204581550 & J16123414-2144500 & QPD & 0.03 & 1.91 & 0.78 & 0.24 \\
204584778 & J16152516-2144013 & QPS & 0.06 & 1.74 & 0.46 & -0.14 \\
204602441 & J16092136-2139342 & B & 0.10 & 15.08 & 0.96 & -0.39 \\
204607034 & J16024152-2138245 & B & 0.13 & 24.62 & 0.9 & -0.41 \\
204611292 & J16082870-2137198 & P & 0.05 & 1.77 & 0.18 & 0.03 \\
204615647 & J16132190-2136136 & APD & 0.05 & 7.81 & 0.77 & 0.3 \\
204630363 & J16100501-2132318 & QPS & 0.09 & 6.58 & 0.47 & -0.07 \\
204637622 & J16042097-2130415 & MP & 0.12 & 1.05 & 0.46 & 0.03 \\
204638512 & J16042165-2130284 & APD & 0.45 & 15.08 & 0.9 & 0.89 \\
204651122 & J16122289-2127158 & QPS & 0.09 & 4.92 & 0.8 & -0.24 \\
204662993 & J16192923-2124132 & N & 0.003 & 41.35 & 1.01 & -0.08 \\
204757338 & J16072747-2059442 & APD & 0.04 & 3.6 & 0.86 & 0.2 \\
204769599 & J16002669-2056316 & QPS & 0.05 & 1.64 & 0.42 & -0.03 \\
204776782 & J16152083-2054372 & QPS & 0.07 & 1.92 & 0.53 & 0.02 \\
204807722 & J15570146-2046184 & QPD & 0.37 & 2.17 & 0.65 & 0.64 \\
204810161 & J16221481-2045398 & QPS & 0.02 & 1.66 & 0.74 & -0.02 \\
204811478 & J15555600-2045187 & QPD & 0.16 & 1.7 & 0.73 & 0.58 \\
204817605 & J16120505-2043404 & P & 0.18 & 9.26 & 0.02 & 0.02 \\
204830786 & J16075796-2040087 & B & 1.14 & 34.57 & 1.0 & -0.67 \\
204832936 & J15564244-2039339 & QPS & 0.04 & 4.24 & 0.31 & 0.22 \\
204856535 & J16070014-2033092 & APD & 0.14 & 58.15 & 1.0 & 0.42 \\
204860656 & J16104391-2032025 & MP & 0.04 & 1.91 & 0.35 & 0.02 \\
204864076 & J16035767-2031055 & APD & 0.19 & 8.36 & 0.84 & 0.21 \\
204870258 & J15594426-2029232 & N & 0.01 & 74.86 & 1.0 & -0.05 \\
204871202 & J16090071-2029086 & APD & 0.11 & 12.57 & 0.9 & 0.15 \\
204871862 & J16070169-2028579 & APD & 0.16 & 5.93 & 0.91 & 0.54 \\
204874314 & J16353913-2028195 & L & 0.90 & 72.94 & 1.0 & 0.0 \\
204894208 & J16002945-2022536 & QPD & 0.08 & 1.7 & 0.79 & 0.47 \\
204906020 & J16070211-2019387 & B & 0.16 & 8.38 & 0.93 & -0.47 \\
204908189 & J16111330-2019029 & B & 0.35 & 19.23 & 0.76 & -0.59 \\
204932990 & J16115091-2012098 & QPD & 0.09 & 2.32 & 0.75 & 0.52 \\
204933717 & J16072240-2011581 & QPS & 0.05 & 1.55 & 0.76 & -0.17 \\
204939243 & J16153220-2010236 & QPD & 0.46 & 8.3 & 0.85 & 0.45 \\
204940701 & J16122737-2009596 & QPS & 0.01 & 8.7 & 0.86 & -0.26 \\
204951022 & J16203026-2007037 & S & 0.001 & 37.08 & 0.84 & -0.29 \\
204964091 & J16200549-2003228 & P & 0.03 & 3.91 & 0.19 & 0.32 \\
204982702 & J16095206-1958065 & QPD & 0.96 & 2.4 & 0.61 & 0.34 \\
205000676 & J16220961-1953005 & QPS & 0.05 & 3.2 & 0.24 & 0.06 \\
205008727 & J16193570-1950426 & B & 0.39 & 39.11 & 1.0 & -0.68 \\
205024407 & J15583620-1946135 & P & 0.04 & 3.01 & 0.15 & -0.08 \\
205037578 & J16041740-1942287 & APD & 0.03 & 22.9 & 0.87 & 0.26 \\
205038557 & J16035793-1942108 & P & 0.06 & 3.85 & 0.09 & 0.02 \\
205051240 & J16140792-1938292 & QPD & 0.06 & 5.32 & 0.33 & 0.32 \\
205061092 & J16145178-1935402 & B & 0.14 & 72.25 & 1.0 & -0.51 \\
205063210 & J16073915-1935041 & QPS & 0.08 & 2.43 & 0.3 & -0.04 \\
205064383 & J16122183-1934445 & N & 0.001 & 34.69 & 0.96 & -0.14 \\
205068630 & J16111095-1933320 & APD & 0.07 & 17.37 & 0.93 & 0.45 \\
205080616 & J16082324-1930009 & APD & 0.25 & 5.32 & 0.68 & 0.21 \\
205086621 & J16114534-1928132 & QPD & 0.15 & 1.75 & 0.73 & 0.71 \\
205088645 & J16111237-1927374 & B & 0.19 & 16.96 & 1.0 & -0.57 \\
205091879 & J16115763-1926389 & MP & 0.10 & 0.36 & 0.58 & -0.12 \\
205092303 & J16092054-1926318 & QPD & 0.11 & 1.72 & 0.68 & 0.3 \\
205110000 & J16154416-1921171 & S & 1.13 & 18.02 & 0.9 & -0.1 \\
205115701 & J16100541-1919362 & APD & 0.38 & 28.56 & 0.88 & 0.13 \\
205145188 & J16102819-1910444 & QPS & 0.04 & 1.76 & 0.76 & 0.14 \\
205151387 & J16090075-1908526 & APD & 0.24 & 10.0 & 0.59 & 0.87 \\
205152244 & J16090002-1908368 & P & 0.05 & 1.81 & 0.21 & -0.1 \\
205154017 & J16064385-1908056 & P & 0.20 & 6.94 & 0.07 & -0.63 \\
205156547 & J16121242-1907191 & B & 0.05 & 7.03 & 1.0 & -1.01 \\
205158239 & J16142029-1906481 & S & 1.47 & 13.18 & 0.87 & -0.11 \\
205160565 & J16142091-1906051 & S & 0.74 & 24.4 & 1.0 & 0.18 \\
205164892 & J16102857-1904469 & P & 0.15 & 6.58 & 0.14 & -0.06 \\
205164999 & J16130235-1904450 & APD & 0.20 & 17.55 & 0.98 & 0.61 \\
205165965 & J16130996-1904269 & S & 0.13 & 1.8 & 0.73 & -0.49 \\
205179845 & J16143367-1900133 & QPS & 0.05 & 2.6 & 0.74 & 0.02 \\
205182200 & J16123916-1859284 & S & 0.16 & 68.85 & 0.82 & 0.05 \\
205198363 & J16153341-1854249 & QPS & 0.12 & 1.91 & 0.79 & -0.02 \\
205208701 & J16064266-1851140 & QPS & 0.13 & 0.91 & 0.71 & -0.03 \\
205218826 & J16093653-1848009 & APD & 0.16 & 7.89 & 0.87 & 0.47 \\
205238942 & J16064794-1841437 & QPD & 0.23 & 9.26 & 0.57 & 0.23 \\
205241182 & J16104636-1840598 & QPD & 0.28 & 7.07 & 0.82 & 0.86 \\
205249328 & J16113134-1838259 & S & 0.85 & 51.77 & 1.0 & -0.05 \\
205345560 & J16062383-1807183 & APD & 0.04 & 21.04 & 1.0 & 0.33 \\
205364526 & J16124893-1800525 & QPS & 0.02 & 2.91 & 0.22 & -0.31 \\
205366676 & J16095933-1800090 & QPS & 0.04 & 3.68 & 0.2 & 0.06 \\
205375290 & J16111534-1757214 & QPS & 0.05 & 6.1 & 0.21 & -0.06 \\
205383125 & J16095361-1754474 & QPD & 0.45 & 2.03 & 0.69 & 0.47 \\
205519771 & J16071403-1702425 & APD & 0.04 & 19.27 & 0.94 & 0.74 \\
205684783 & J16340916-1548168 & QPS & 1.18 & 3.29 & 0.7 & 0.17 \\
210282528 & J16333496-1832540 & QPS & 0.16 & 2.21 & 0.28 & -0.02 \\
210282534 & J16265850-2445368 & S & 0.4 & 14.55 & 0.83 & -0.2 \\
\enddata
\tablecomments{\label{tab:vartable} Variability properties for stars with inner disks observed in K2 Campaign 2. Variability types are determined by eye and supported by statistical measures. The types consist of the following: "P" is for strictly periodic behavior, "MP" is reserved for stars with multiple distinct periods, "QPD" is for quasi-periodic dippers, "QPS" means quasi-periodic symmetric (i.e., quasi-periodic stars that neither burst nor dip), "APD" are aperiodic dippers, "B" is for bursters, "S" is for stochastic stars, "L" is the label for long-timescale behavior that doesn't fall into the other categories, ``U'' is reserved for objects we were unable to classify, and "N" denotes non-variable objects.}
\end{deluxetable*} 

\section{Variability classes}
 
With the full set of disk-bearing variables in hand, we can ask what sort of time domain behavior comprises this sample. In \cite{cody2014}, we devised a set of statistical 
metrics that could separate YSO light curve shapes into different categories. We define them fully in that work and briefly summarize them here. The first of these 
metrics is the flux asymmetry, ``$M$'', which is a measure of the tendency of a light curve to display fading events (positive $M$) or brightening events (negative $M$), or 
a more symmetric, non-skewed light curve that is perhaps a mixture of the two ($M\sim 0$). The value of $M$ is determined by calculating a ``mean'' flux value (by 
averaging the bottom and top 95th percentile points), subtracting off the {\em median} flux value, and then dividing by the estimated white noise level.

\begin{figure*}
\epsscale{1.0}
\plotone{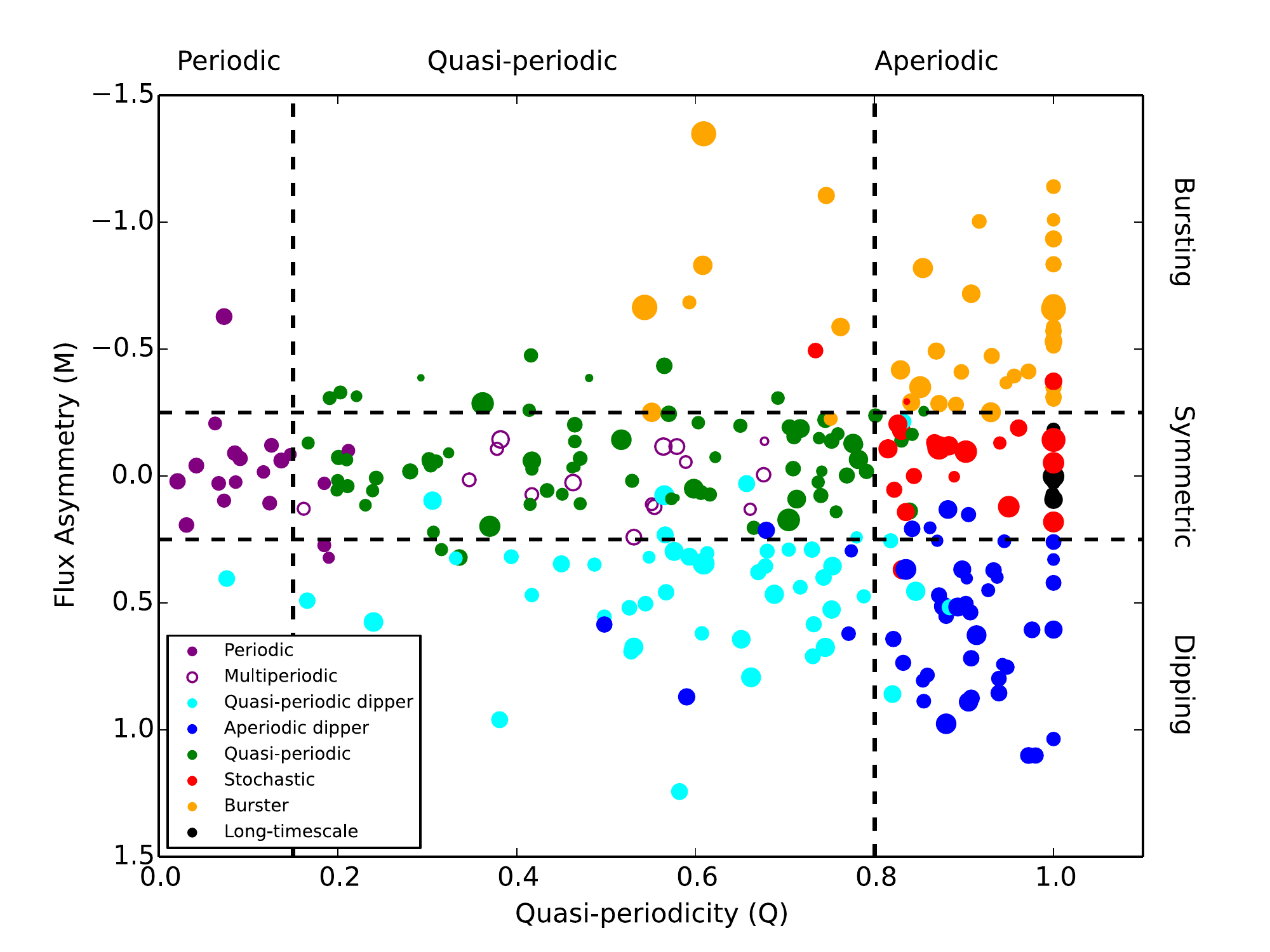}
\caption{$Q$ and $M$ statistics for our sample of disk-bearing stars in Upper Scorpius and $\rho$~Ophiuchus. Non-variable objects are excluded. Colors in this and subsequent 
plots denote different types of variables, as identified by eye; see text.  Point areas in this and subsequent plots are scaled according to variability amplitude to the 
one-third power.}
\label{qmplot}
\end{figure*}

The second metric is the quasi-periodicity, ``Q," which measures on a scale of 0 to 1 how periodic (0) or stochastic (1) the light curve is. It is determined by identifying peaks in the autocorrelation, refining their periods with the Fourier transform periodogram, phasing the light curve to the associated period(s), and then measuring the residual noise after the phased light curve pattern is subtracted out. $Q$ represents the ratio of the residual variance to the original light curve variance. Light curves with low residual noise after removal of the phased pattern are then highly periodic ($Q\sim 0$), whereas those that have larger residuals are quasi-periodic or aperiodic. We computed these $M$ and $Q$ statistics for all disk-bearing light curves. A difference from the \cite{cody2014} implementation  is that, instead of using magnitude values, we use normalized flux. The resulting $Q-M$ diagram is displayed in Figure~\ref{qmplot}, with the variability types assigned by our trained eye also indicated. The values span the space from $Q$=0 to 1 and $M$=-1.4 to 1.4. 

To classify observed variability into different categories based on the $Q$ and $M$ metrics, we have defined boundaries that demarcate different $Q$ and $M$ ranges. Along the flux asymmetry axis, $M<-0.25$ indicates ``bursters'' (preference for brightening events) and $M>0.25$ indicates ``dippers" (preference for fading events). These are same $M$ boundaries promoted in \citet{cody2014}. The quasi-periodicity boundaries are $Q<0.11$ for purely periodic behavior, 
and $Q>0.85$ for purely stochastic behavior. The upper bound is different than that promoted in \cite{cody2014} which was $Q>0.61$ for purely stochastic behavior. It is unclear why this value is so different, although it may be related to different quantification of photometric uncertainty in the $K2$ light curves, as compared to {\em CoRoT}. The dependence of $Q$ on data precision and time sampling will be the subject of future work. We plot the modified boundaries on top of all $Q$ and $M$ values for disk-bearing stars in Figure~\ref{qmplot}. 

\begin{figure*}
\epsscale{1.0}
\plotone{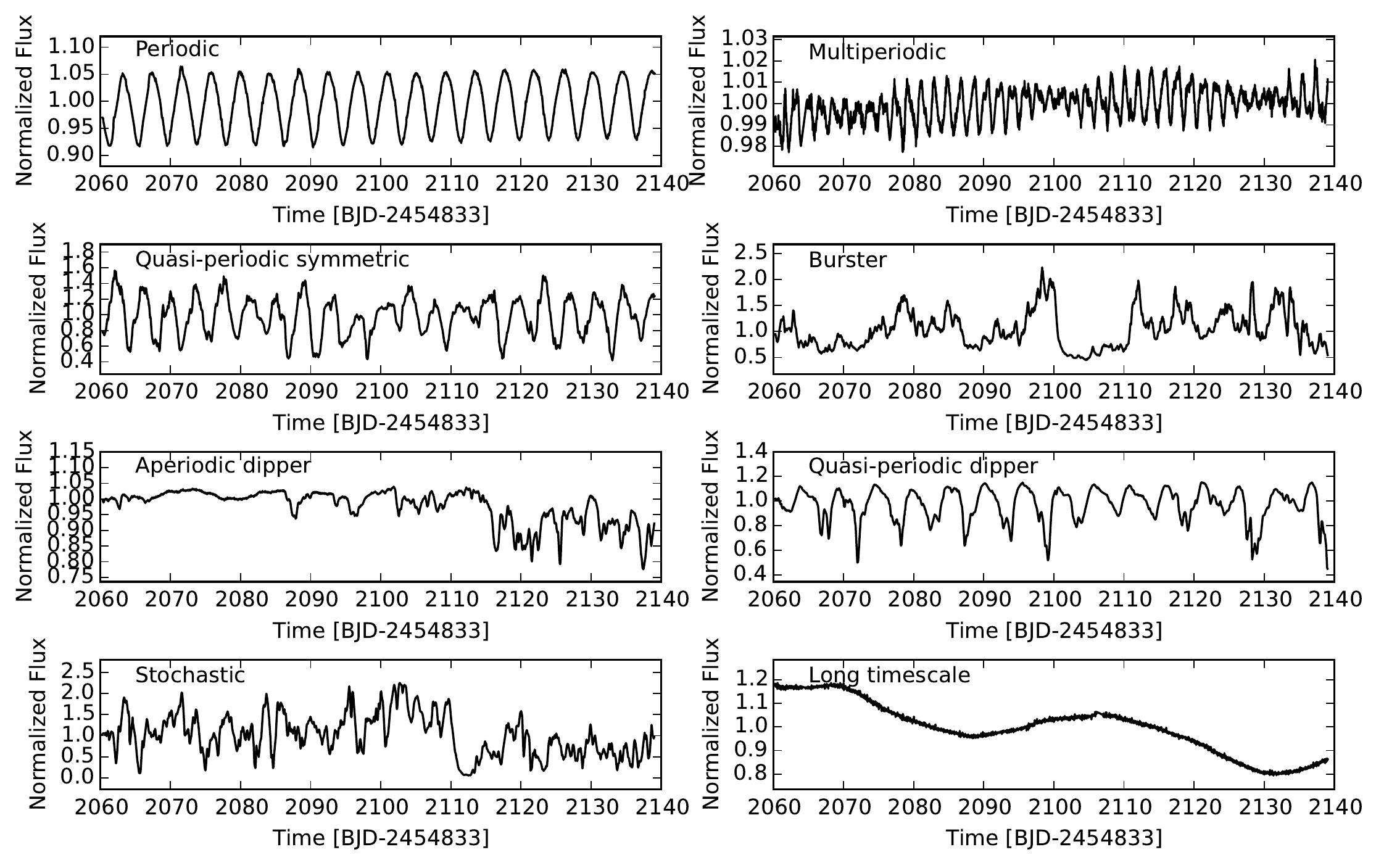}
\caption{Examples of different light curve morphologies seen among disk-bearing stars in {\em K2}'s Campaign 2 observations of $\rho$~Ophiuchus and Upper Scorpius.  These 
examples also appear in Figure~\ref{alllcs} where their EPIC identifiers and their $Q$ and $M$ values are given.}
\label{examples}
\end{figure*}

We thus divide the disk-bearing variables of Upper Sco and $\rho$ Oph into eight different categories (plus two additional groups for non-variable or unclassifiable 
sources), as defined in \cite{cody2014} and shown in Figure~\ref{examples}.  The categories (with denotation elsewhere in this paper given in parentheses) and their 
$Q$ and $M$ ranges are: 
\begin{itemize} 
\item burster (B) \\ 
$M<-0.25$ 
\item purely periodic symmetric (P) \\ 
$Q<0.15$ and $-0.25<M<0.25$ 
\item quasi-periodic symmetric (QPS) \\ 
$0.15<Q<0.85$ and $-0.25<M<0.25$ 
\item purely stochastic (S) \\ 
$Q>0.85$ and $-0.25<M<0.25$ 
\item quasi-periodic dipper (QPD) \\
$0.15<Q<0.85$ and $M>0.25$ 
\item aperiodic dipper (APD) \\
$Q>0.85$ and $M>0.25$ 
\item long timescale (L) 
\item unclassifiable (U) 
\item non-variable (N).  
\end{itemize} 

Relative to \cite{cody2014}, there are no eclipsing binaries identified in the disk-bearing sample studied here.
An additional category is dedicated to ``long-term" variables that showed a trend on $>30$ day timescales. 
Two stars were labeled ``unclassifiable" since their CDPPs indicated variability, but a single brightness bump superimposed on noise or a gradual trend made it impossible to label them as long-term variables or to assign any of the other categories.

As seen from examination of Figure~\ref{alllcs}, the $Q$ and $M$ statistics reflect the visual variability categorization quite well. Problematic borderline cases can occur, 
however, I would for example in which a light curve falls into one class by eye but has a systematic brightness trend that pushes it into another class based on $M$. 
Likewise, there are cases for which an object is periodically variable for part of the time series, but aperiodic for the rest; a viewer may classify this behavior as 
quasi-periodic or aperiodic, but the $Q$ statistic may report otherwise. We therefore continue to rely on the human eye for the final determination of each star's 
variability class, but use the statistics for guidance. All light curves and their classifications are provided in the Appendix, along with a discussion of start that 
displayed multiple types of variability over the duration of observations.

The percentage of disk-bearing stars in each lightcurve group are listed in Table~3.  Uncertainties are generally asymmetric and determined by assuming a binomial 
probability distribution for the number stars observed each group. Notably, there is a somewhat higher number of bursters than reported by \cite{cody2017} due to our 
increased reliance on statistical measurements to identify them in the current work. The sample contains young stars in different environments and potentially at a variety 
of ages; thus we also report the variability category percentages separately for $\rho$~Oph and Upper Sco members. The tallies for the equivalent categories as measured in 
NGC~2264 are also provided, and comparisons discussed in \S7.2.

\begin{deluxetable*}{ccccc}
\tabletypesize{\scriptsize}
\tablecolumns{5}
\tablewidth{0pt}
\tablecaption{Variability types among young disk-bearing stars}
\tablehead{
\colhead{Morphology class} & \colhead{Oph} & \colhead{Sco} & \colhead{Sco/Oph} & \colhead{NGC 2264} \\ 
\colhead{} & \colhead{} & \colhead{} & \colhead{composite} & \colhead{} \\ 
\colhead{} & \colhead{\%} & \colhead{\%} & \colhead{\%} & \colhead{\%} \\
}
\startdata
\multicolumn{5}{c}{Categories based on periodicity and stochasticity}\\
\hline
All Bursters & 14$^{+5}_{-3}$ & 13$^{+3}_{-2}$ & 14$^{+2}_{-2}$ & 13$^{+3}_{-2}$ \\

Aperiodic symmetric (Stochastic) & 12$^{+4}_{-3}$ & 6$^{+2}_{-1}$ & 8$^{+2}_{-2}$ & 13$^{+3}_{-2}$ \\

Quasi-periodic symmetric & 20$^{+5}_{-4}$ & 29$^{+3}_{-3}$ & 26$^{+3}_{-2}$ & 17$\pm$3 \\
               
Aperiodic dippers & 9$^{+5}_{-2}$ & 18$^{+3}_{-2}$ & 16$^{+2}_{-2}$ & 11$^{+3}_{-2}$ \\

Quasi-periodic dippers & 14$^{+5}_{-3}$ & 18$^{+3}_{-2}$ & 17$^{+2}_{-2}$ & 10.5$^{+3}_{-2}$\\  

Periodic symmetric & 6$^{+4}_{-2}$ & 7$^{+2}_{-2}$ & 7$^{+1}_{-2}$ & 3$^{+2}_{-1}$ \\

\\
\multicolumn{5}{c}{Other Categories}\\
\hline
Multiperiodic & 7$^{+4}_{-2}$ & 4$^{+2}_{-1}$ & 5$^{+2}_{-1}$ & 1$^{+2}_{-1}$ \\
Long timescale & 8$^{+4}_{-2}$ & 0$^{+2}_{-0}$ & 3$^{+1}_{-1}$ & 1$^{+2}_{-1}$ \\
Unclassifiable & 2$^{+3}_{-0}$ & 0$^{+2}_{-0}$ & 1$^{+1}_{-1}$ & 11$^{+3}_{-2}$ \\
Non-variable & 6$^{+4}_{-2}$ & 3$^{+2}_{-1}$ & 4$^{+1}_{-1}$ & 19$\pm$3 \\
\enddata
\tablecomments{\label{tab:varfractions} Fraction of young stars in each light curve morphology group, as defined by eye but generally supported by the statistical measures $Q$ and $M$ (\S5). There are eight variability categories, plus two more for non-variable or unclassifiable sources.}
\end{deluxetable*}

The variability timescale given in Table~2 is defined as the period from periodogram analysis, if $Q<0.8$ (i.e., the light curve is [quasi-]periodic). For stars with $Q>0.8$, a timescale is derived in the same manner as described in \citet[][;see \S6.5]{cody2014}.
Amplitudes are measured by determining the normalized flux difference between the 95th and 5th percentile points in the light curve; this is similar to a peak-to-peak amplitude, but is less sensitive to outliers and other errant points. 

\begin{figure}
\epsscale{1.2}
\plotone{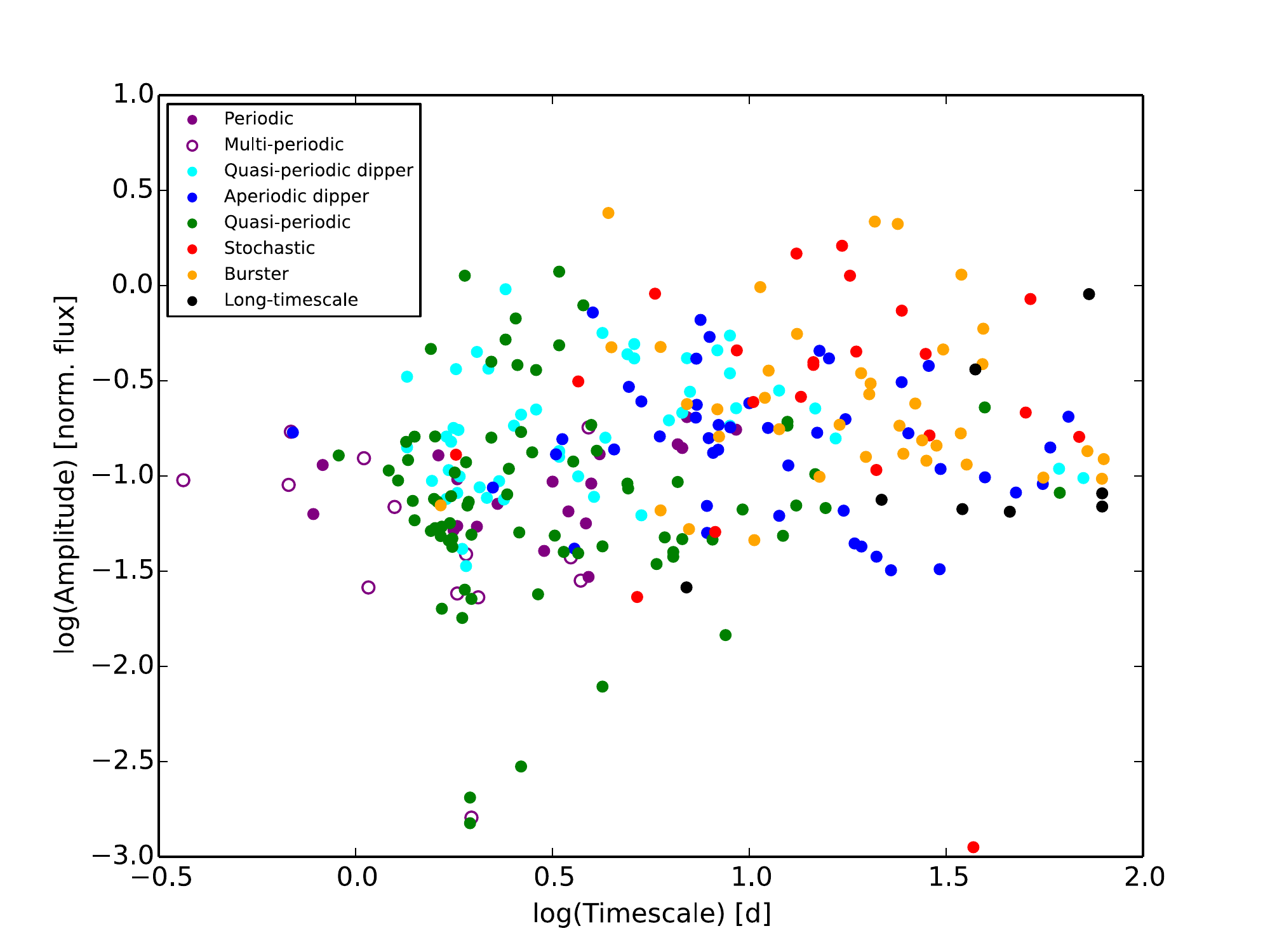}
\caption{Log of the amplitude in normalized flux units versus measured timescale for all disk-bearing light curves. No point size scaling with amplitude is applied in 
this Figure, since one of the axes is amplitude. Although there is significant overlap within the phase-space, some clumping of variability types is apparent.}
\label{amp_vs_time}
\end{figure}

Figure~\ref{amp_vs_time} illustrates the light curve amplitudes as a function of the variability timescale.
The bursting (B) and stochastic (S) sources, taken in aggregate, have the longest timescales (along with the identified long timescale (L) but generally smooth sources).
Overlapping the burster and stochastic sources in timescale are the aperiodic dippers (APDs),
which have longer timescales than the quasi-periodic dippers (QPDs).
Quasi-periodic symmetric (QPS) sources occupy the same range in timescale as the quasi-periodic dippers (QPDs), though have a broader range of amplitudes.
Periodic/multi-periodic (P/MP) sources have the shortest timescales.
The amplitude ranges of the various light curve categories are fairly similar, with periodic and multi-periodic sources 
spanning the narrowest amplitude range. Burster and stochastic
sources perhaps extend to slightly higher amplitudes
than other categories, and quasi-periodic symmetric
sources extend to lower amplitudes than other categories.

\section{Connection between variability and stellar/circumstellar properties}

In this section, we explore relationships between variability type and stellar / circumstellar properties.  Our sample is selected to have disks.  Thus, we are more likely 
to detect the disk-related variability, and less likely to detect processes associated with normal stellar rotation and activity, that generally underlay the disk effects, 
but occur at lower amplitude.

\begin{figure}
\epsscale{1.2}
\plotone{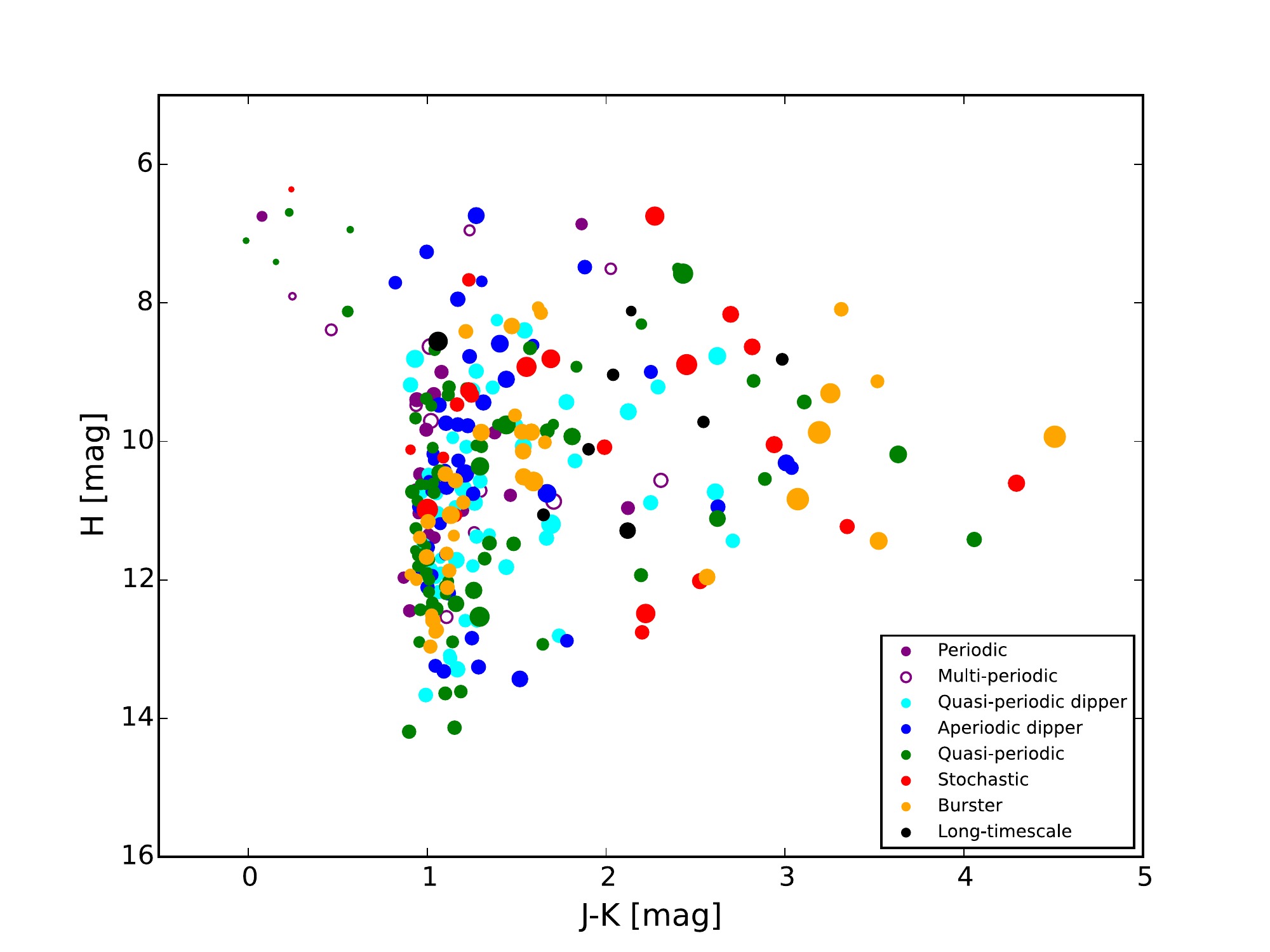}
\caption{
The near-infrared color-magnitude diagram of Figure 1, now colored for the variables of different types. Non-variable objects are excluded.  Location of any individual source in this diagram is influenced by its: mass, age, inner disk properties, and the line-of-sight extinction.}
\label{cmd_nir_var}
\end{figure}

\begin{figure}
\epsscale{1.2}
\plotone{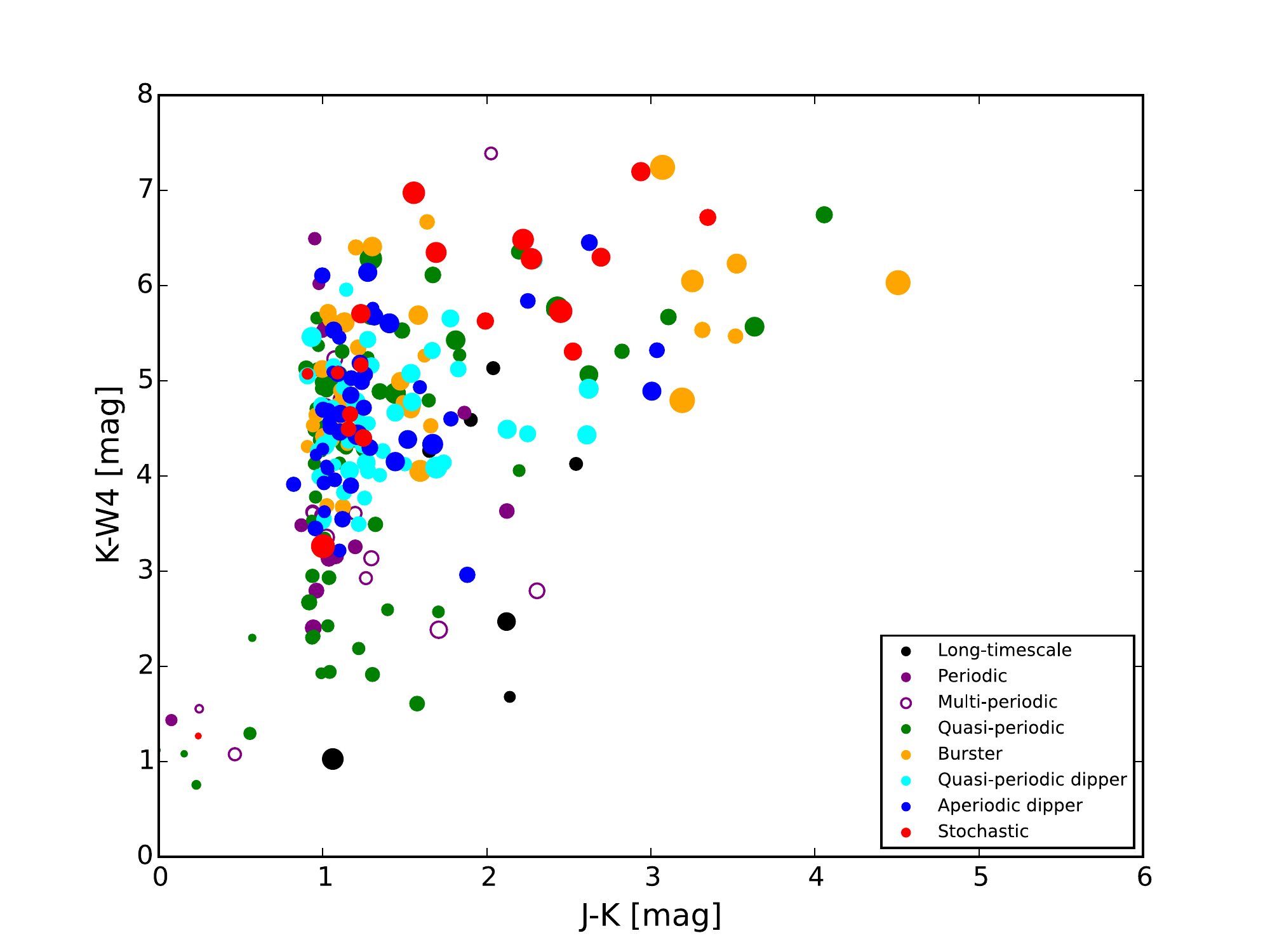}
\plotone{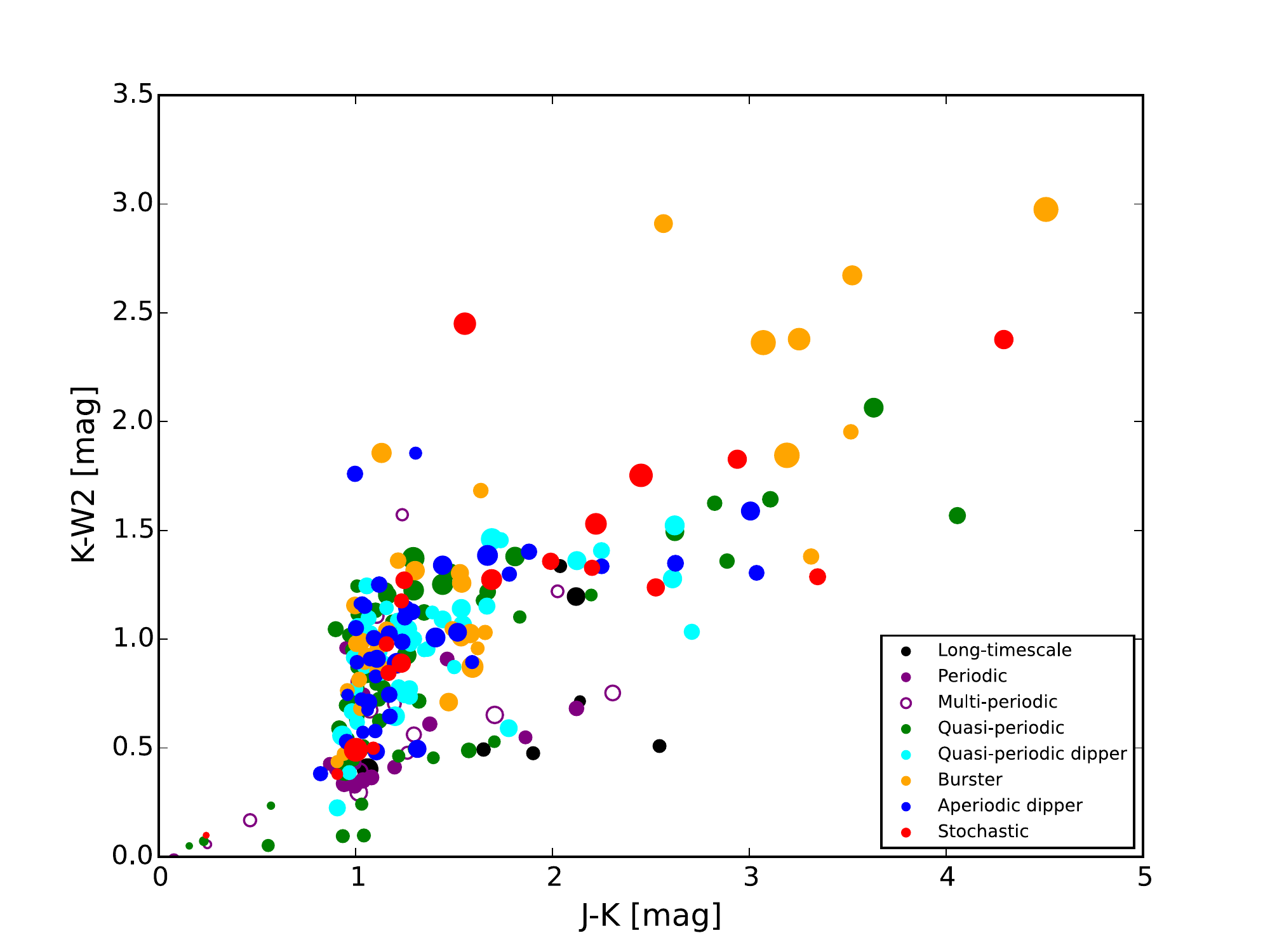}
\caption{For variables of different types, top panel shows
K-W4 (sensitive to disk emission) vs J-K (measuring photospheric color plus inner disk color excess plus reddening effects), 
while bottom shows the same plot with K-W2 along the ordinate. Non-variable objects are excluded.}
\label{kw4_jk_var}
\end{figure}

\begin{figure}
\epsscale{1.2}
\plotone{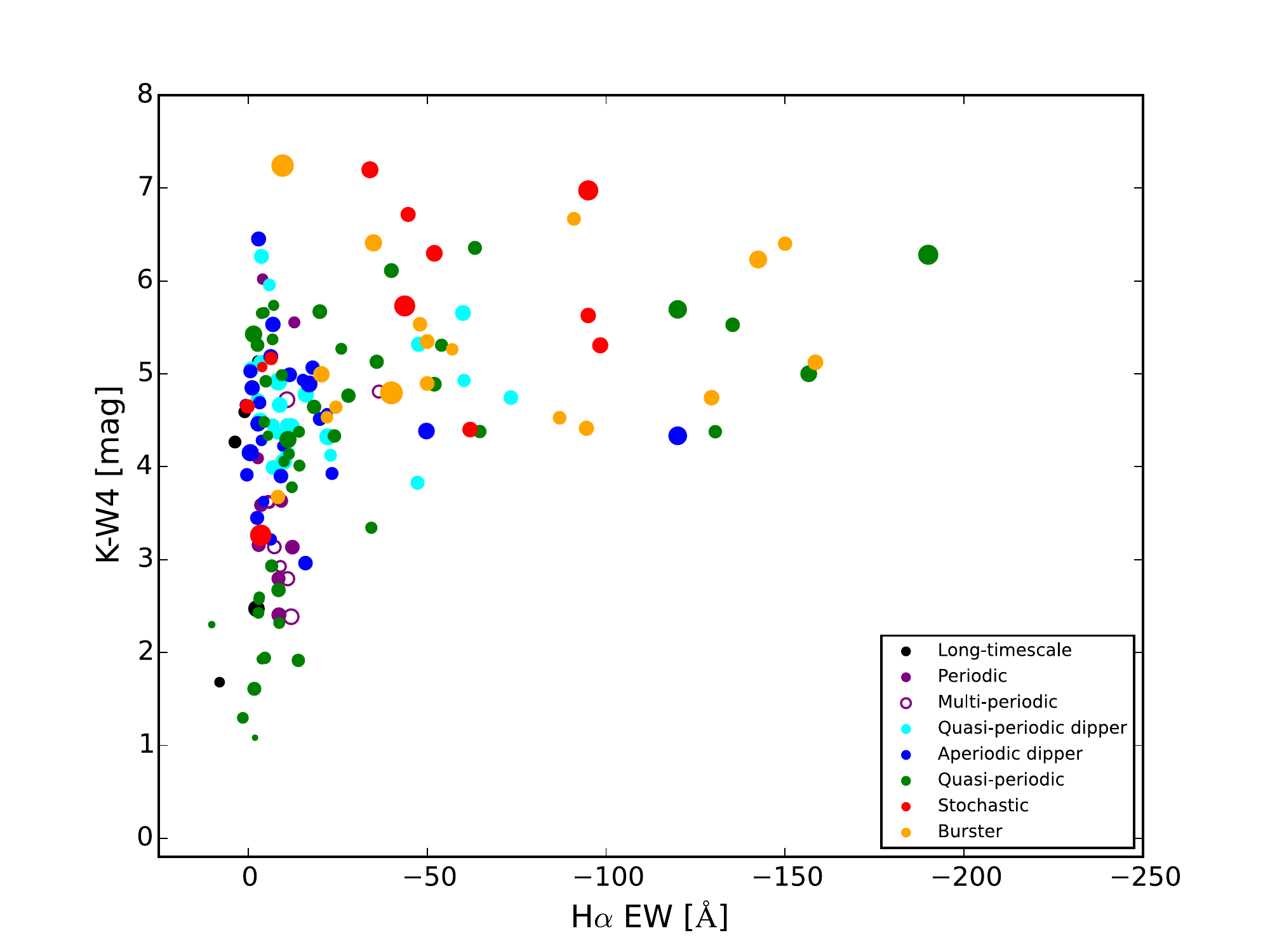}
\caption{
$K$-W4 color versus H$\alpha$ equivalent width 
for disk variables of different types.}
\label{halpha_KW4_var}
\end{figure}

\subsection{Variability and Circumstellar Disks}

Figure~\ref{qmplot}, introduced earlier, illustrates that larger amplitude variables are predominantly found  
among objects with both $Q>0.35$ -- in the quasi-periodic and aperiodic categories --
and $M< 0.25$ -- in the symmetric and bursting categories. 
Figure~\ref{amp_vs_time} just above also shows this segregation of bursting and quasi-periodic symmetric lightcurve types towards higher amplitudes.

As we now demonstrate, these larger-amplitude sources also tend to have stronger $H\alpha$ emission and redder infrared colors - especially for the bursters.  
These findings directly link the presence of accreting gas and the strength of the inner disk,
with photometric variability categories that are attributed to accretion.

Figure \ref{cmd_nir_var} reproduces Figure~\ref{cmd_nir} but now colored by the variability type. 
At bluer $J-K$ color, moving from brighter to fainter $H$ magnitudes is equivalent
to moving from higher to lower mass stars.   There is no apparent systematic relationship
between variability type and stellar mass (see also \S6.2).  Assuming low foreground extinction,
redder $J-K$ color indicates objects with stronger inner disk contributions.
Larger amplitude variability appears to occur for redder objects, and the sources are predominantly bursters, 
stochastic, and quasi-periodic dippers.

Figure \ref{kw4_jk_var} highlights the variability types versus dust excess. For stars with disks, $K-W4$ color probes the disk region associated with few hundred Kelvin 
dust (the terrestrial planet zone for solar-type stars), $K-W2$ color hotter dust (roughly the inner 1 AU for solar type stars), and $J-K$ the region containing the 
hottest, up to 1400 K near the inner disk edge (in the range 0.03-0.1 AU for solar type stars).  For the M-type stars that dominate our sample, the wavelengths probe smaller 
physical radii than the numbers given above for G-type stars. Quasi-periodic dippers are relatively clustered in $K-W4$ color, with redder colors than the periodic sources, 
as well as bluer colors than the bursters and stochastic objects. The same segregation is not as apparent in $K-W2$ colors, suggesting that accretion-related burster and 
stochastic variability is driven by strong mid-infrared excess arising around 1 AU (for solar type stars), rather than being purely inner-disk phenomena.

Figure \ref{halpha_KW4_var} highlights the variability versus accretion strength.
Strong H$\alpha$ emission is seen only among the redder $K-W4$ colors,
solidifying the connection between mid-infrared disks,
accretion-related emission lines, and photometric variability.
Long-timescale variables, periodic sources, and quasi-periodic dippers
are weak accretors.  Quasi-periodic sources have the broadest
distribution across the empirical parameter space.

\begin{figure}
\epsscale{1.2}
\plotone{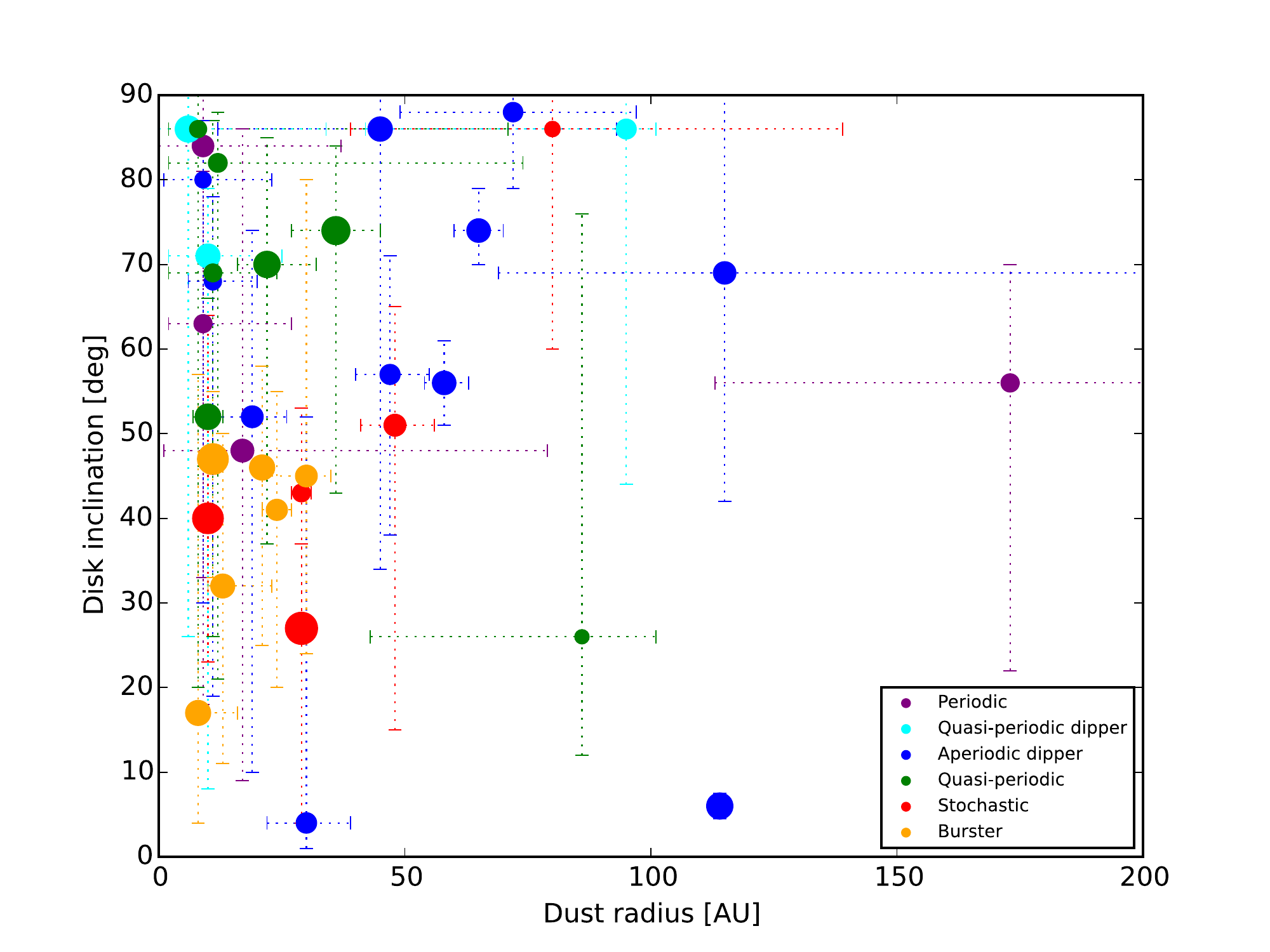}
\plotone{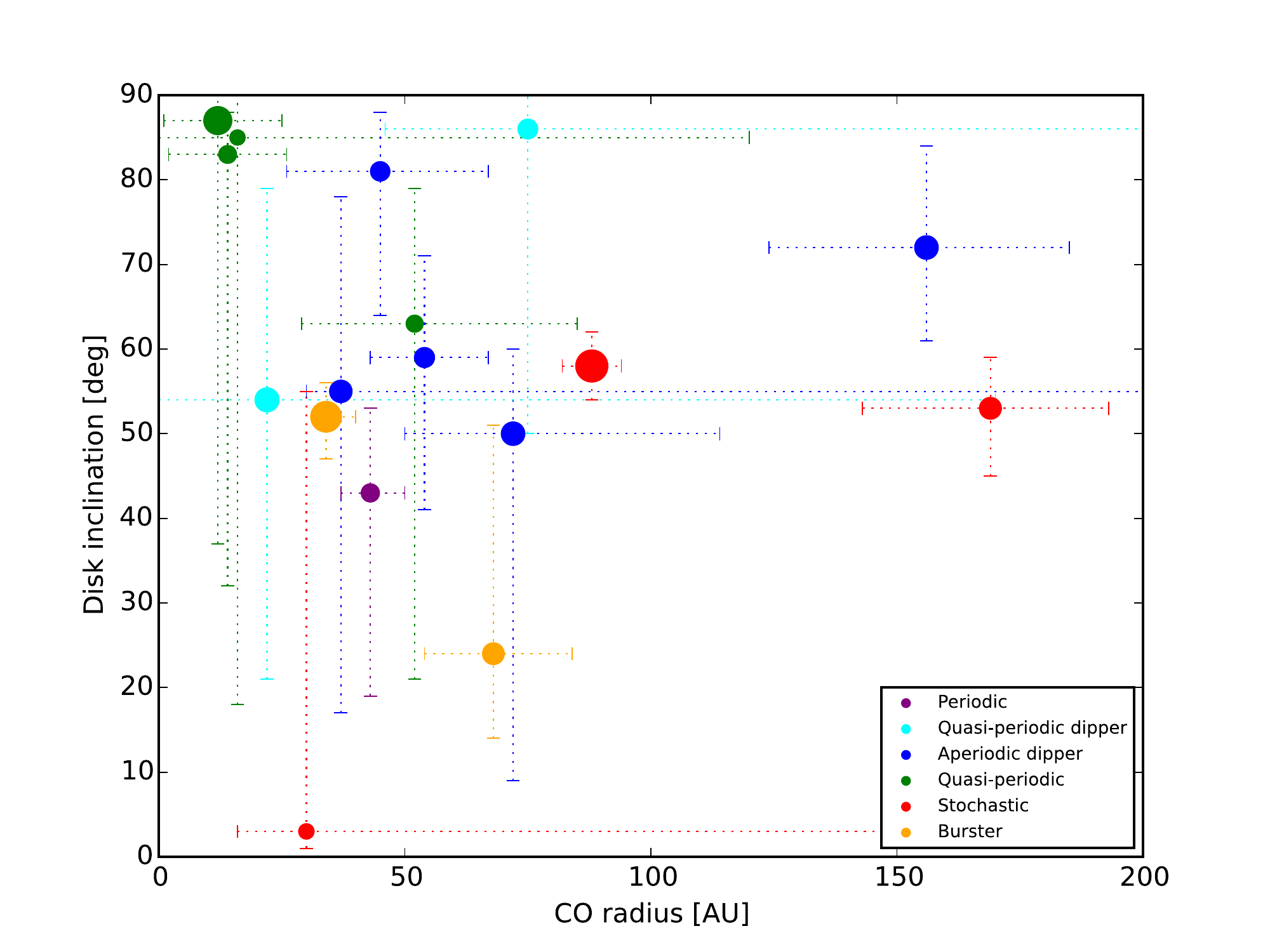}
\caption{Dust continuum (top panel) and CO gas (bottom panel) disk radius, versus disk inclination, based on $ALMA$ observations where they are available for our sample. Point sizes are scaled proportional to light curve amplitude. 
Although the measurements are noisy, there is clear segregation of some variability types in inclination.}
\label{inclination_size}
\end{figure}

Beyond the presence and strength of disks, a critical parameter that determines the observables from star/disk systems is the orientation of the disks. In Figure 
\ref{inclination_size}, we present the variability types in the context of the disk sizes and inclinations to our line-of-sight, as derived by \cite{barenfeld2017} from 
spatially resolved dust and gas images from $ALMA$. Although the formal errors in the inclinations are quite large\footnote{Where disk parameters are derived from both dust 
and gas, the inclinations agree to within about $\pm 15^\circ$ for most, but not all, sources.  Radii can be discrepant at the factor-of-two level, in both directions, again 
with larger outliers present.}, some intriguing trends emerge with respect to the variability categories that we have defined -- without regard to the disk characteristics 
beyond a sample selected based on mid-infrared excess. Dippers of both aperiodic and quasi-periodic flavors tend to have more highly inclined disks ($i > 50$ deg), though 
several such variables are found at $i < 10$ deg (EPIC 204245509 and EPIC 204638512). The quasi-periodic symmetric sources that are neither dipping nor bursting, also tend 
to have larger inclinations. Burster sources, by contrast, have ($i < 50$ deg), and are clustered at small dust radii.  Stochastic sources also tend to have smaller 
inclinations, but have larger disks by about a factor of two.  For both the burster and the stochastic categories, the gas radii are larger than the dust radii. 

Considering the locations of the different variability categories in the above diagrams,
our observations include:

\begin{itemize}
\item The bursters (B) and the stochastic (S) variables have the reddest disk colors, even redder than the dippers (APD).  This is consistent with overall stronger disks for the variables that are dominated
by their accretion behavior.  Dippers may have relatively weaker disks, or different inner-disk geometry.

\item The quasi-periodic dippers (QPD) and aperiodic dippers (APD) have somewhat smaller infrared colors, with only a narrow range of $K-W4$ and $K-W3$ disk color that is 
spanned relative to other lightcurve categories. The colors may be telling us about disk flaring, which could be low (thus confining the color range to only that 
spanned by inclination effects rather than the broader color range allowed by considering vertical disk geometry effects) compared to, e.g., bursters and stochastics with 
typically larger $K-W4$ and $K-W3$ colors (which require some amount of flaring to produce, especially given the tendency towards lower inclinations illustrated in 
Figure~\ref{inclination_size}).

\item The quasi-periodic symmetric (QPS) stars exhibit among the largest $H\alpha$ equivalent widths.  The widths are higher than those of many sources in the burster category, for which \cite{cody2017} established a correlation between burst activity and $H\alpha$ strength.  

\item There may be multiple different sub-categories within the quasi-periodic symmetric (QPS) category, with the higher amplitude sources dominated by large, hot 
accretion spots on top of underlying accretion variations. The lower amplitude sources, on the other hand, may be dominated by cool spot modulation, with inter-cycle 
variations due to a low-level accretion contribution. Venuti et al.\ (2015) found that quasi-periodic objects with larger amplitudes had $u$ and $r$-band variations more 
consistent with hots spots than with cool spots (see their Figure~9). We note that the QPS sources also tend to cluster at the shorter timescales (Figure~\ref{amp_vs_time}), 
similar to both the QPD and P/MP categories. In addition, both the higher and the lower amplitude QPS sources cluster at the higher disk inclinations 
(Figure~\ref{inclination_size}). These findings are consistent with the hypothesis of spot-related variability for QPS sources, as a low inclination view of the star would 
result in little light modulation, due to the spots always being visible.

\item The periodic (P) symmetric sources all have relatively low infrared color,
bluer than the disk-related categories.
This is consistent with the periodic sources lacking strong accretion
that would introduce additional variability effects that swamp the periodic signal.

\item The multi-periodics (MP) seem redder in $J-K$ but not other infrared colors.
This could imply binarity (as opposed to disk effects) as the dominant cause of the variability.

\item Disk inclination is an important parameter in the observed variability type.  Dipper-type sources (both QPD and APD) tend towards higher inclinations, $i > 
50^\circ$, as originally envisioned by, e.g. \cite{bouvier1999}, \cite{cody2010}, \cite{morales2011}, and \cite{stauffer2015}, who invoke dust in the co-rotating inner-disk 
or stellar magnetosphere to explain the flux dips. The QPS sources, as mentioned above, also occupy the high-inclination half of the distribution. Burster-type and 
stochastic sources are not seen among higher inclination sources, but have $i < 50-60^\circ$, consistent with our viewing of the accretion zone more directly, unobscured by 
the disk.

\item  There are a few objects that fall in a very different part of the diagrams compared to objects of similar variability class and amplitude. We highlight the example of EPIC~204187094/2MASS~J16111907-2319202, a brown dwarf in Upper Scorpius. Its light curve shows a high level of erratic variability, suggestive of strong accretion and a significant disk. But surprisingly, its reported H$\alpha$ width and infrared colors are low (see the red dot at H$\alpha < 10$ in Figure~11). We speculate that both variability type and accretion properties change intermittently, and the discrepancy here is due to the non-simultaneous nature of spectroscopic versus photometric data taking.

\end{itemize}

\begin{figure}
\epsscale{1.2}
\plotone{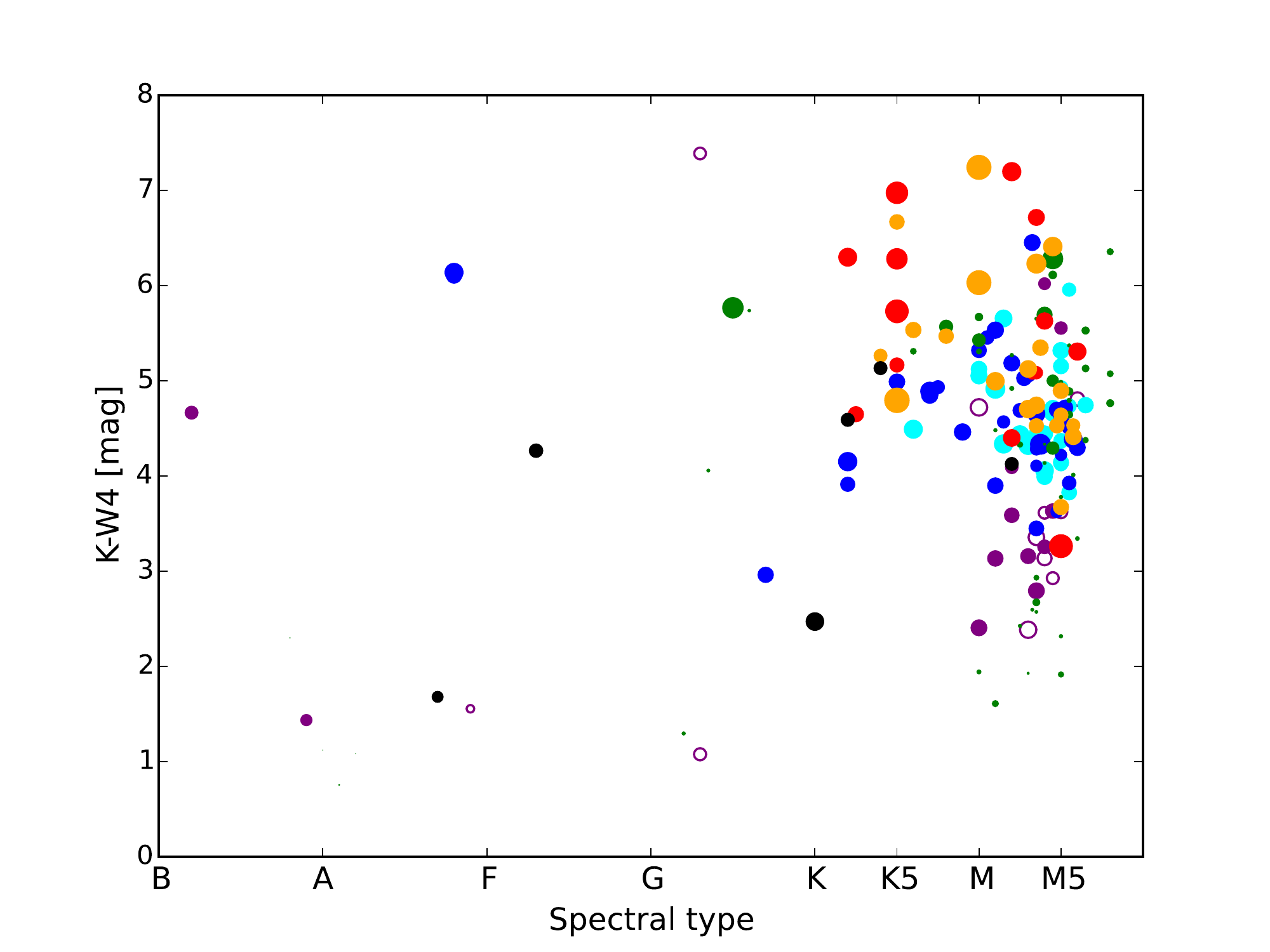}
\plotone{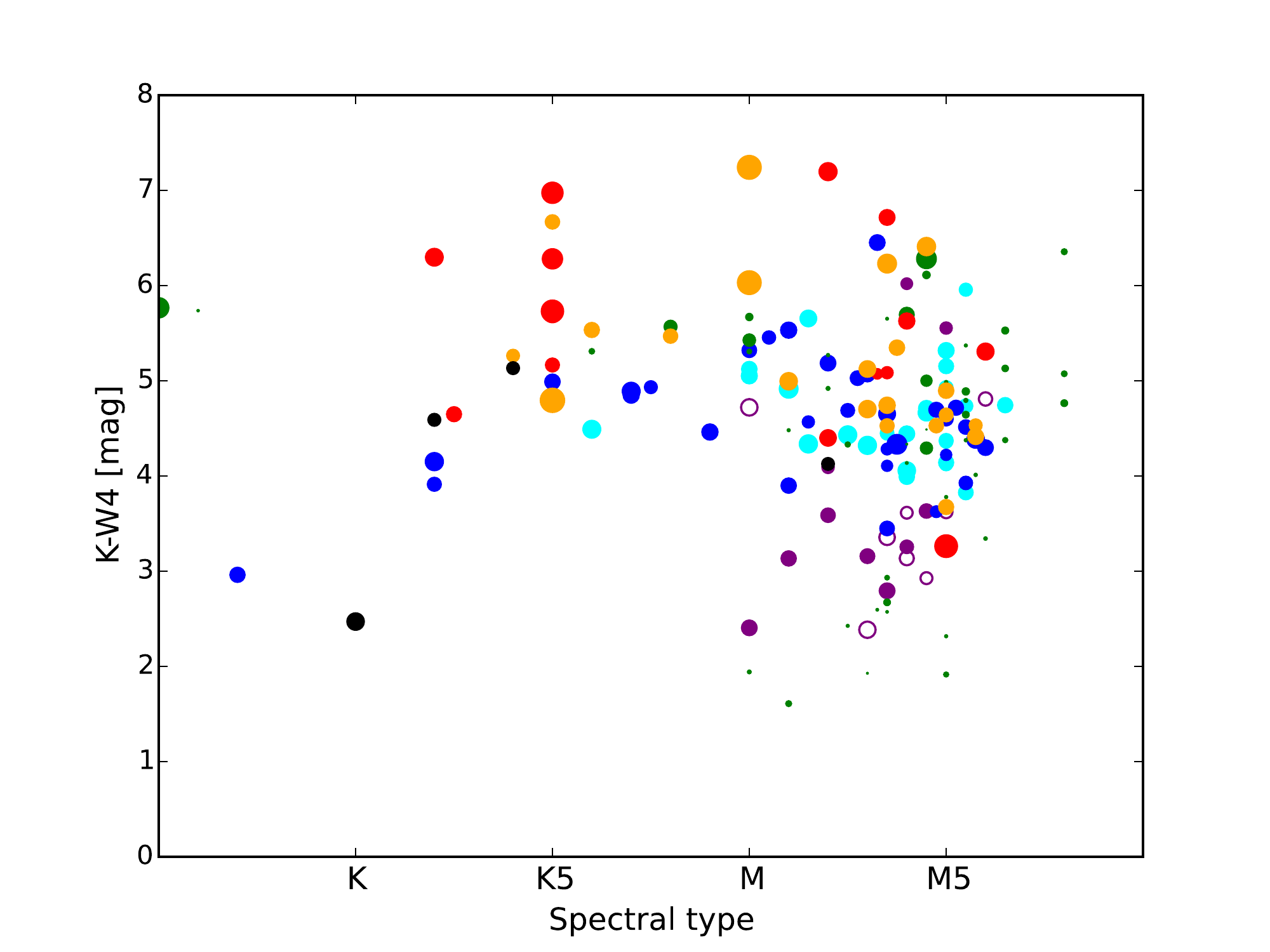}
\caption{$K$-W4 disk excess versus spectral type ; colors are as in Figs.\ 8--10. We show another version (bottom) zoomed in on types K through M. We note that the points for the two A8 stars EPIC~204399980 and EPIC~204514546 (both aperiodic dippers) fall exactly on top of each other.}
\label{spt_kw4_var}
\end{figure}

\subsection{Variability and Stellar Properties}

We can also compare the variability properties in $\rho$~Oph and Upper Sco, across stellar masses, using spectral type as a proxy. Figure \ref{spt_kw4_var} shows variability with spectroscopically determined spectral type.  Variability amplitudes decrease significantly at spectral types earlier than mid-G, with a range of variability types seen. Several early type stars display very clear but low amplitude ($<$1\%) quasi-periodic variability. 
We also encounter the dipper phenomenon with clear aperiodic fading events in the light curves of the late A stars EPIC~204399980 and EPIC~204514546. 

While the sample of early-type stars is too small to conduct a full statistical evaluation of the prevalence of each variability type, we can speculate on the reasons for the drop-off in amplitude. If the quasi-periodic behavior is due to the rotational modulation of accretion hotspots, then one would expect the smaller spot-photosphere contrast to produce lower amplitude light curves. If it is instead due to obscuration effects by the inner disk, then the lower amplitudes could point to a decrease in size and/or larger radial distance of occulting structures.

In terms of stellar rotation, while we found that 96\% of our disk-selected sample of stars is variable, only 12\% among these are periodic (P), with 40\% of these multi-periodic (MP), in a manner consistent with rotating starspot behavior. However, another 25\% of the sample is quasi-periodic with symmetric light curves (QPS), and an additional 17\% is quasi-periodic with dipper type light curves (QPD). The quasi-periodic and disk-related variability could also be related to a rotation-regulated mechanism -- either accretion at the stellar surface, or dipping associated with material entrained in magnetic fields that are co-rotating with the star, or with dust located in asymmetric structures at or near the inner disk edge.

\section{Comparisons}

\subsection{Comparison with $CoRoT$ study of young disked star light curves in NGC~2264}

The variability demographics established here can be compared with results from 
\cite{cody2014} on the $\sim$3~Myr NGC~2264 cluster, which are also given in Table~3.

We find that towards the $\rho$ Oph core, which is perhaps 1-3~Myr old, that the fractions in each of the variability categories defined from $Q$ and $M$ statistics are approximately the same.  We do note that a somewhat lower mass population is probed in $\rho$ Oph, due to its closer distance (though higher extinction), as discussed in \S2.4.

Towards the somewhat older 5-10 Myr, and more populous Upper Sco region, however, 
there are notable differences in the population of the light curve categories.  
The burster fraction is in fact identical across the groups, at $\approx$ 13\%.  However, 
relative to the younger NGC~2264 and $\rho$ Oph, there is a higher fraction among the
Upper Sco light curves that are symmetric (approximately equal upward and downward flux excursions)
for periodic and multi-periodic sources, and fractionally fewer aperiodic 
(also known as stochastic) sources, perhaps because the disk activity is lower.
There are also more dippers overall among Upper Sco stars, including each of the
periodic, quasi-periodic, and aperiodic sub-categories.  This could be evidence of the
dipper behavior being related to evolving disks, perhaps as accretion activity declines
and inner disk cleared regions grow and present more surface area for photons originating
at the stellar surface and accretion zone to be absorbed.

\subsection{Comparison with $K2$ study of periodic sources in Upper Sco}

\cite{rebull2018} undertook an investigation of periodic behavior among a much
larger sample of Upper Sco and $\rho$ Oph members than considered here.
Their sample was selected based on proper motions and color-magnitude diagrams
and includes both non-disk and disked (likely) members.  They concluded that
more than 85\% of cluster members overall are periodic, with 20\% of these multi-periodic.

We found a much lower fraction of periodic stars among the disk-bearing sample, only 7\%.
Even attributing all of the quasi-periodic variability to some
rotation-regulated mechanism (e.g. accretion or dipping associated with structures
that are co-rotating with the star), the fraction of periodic sources
in the disk-selected sample is only about half of that among the broader 
Upper Sco and $\rho$ Oph sample of \cite{rebull2018}.

\subsection{Comparison with other analyses of $K2$ Campaign 2 young star lightcurves}

Independent analyses of the variability types exhibited by members of
the young Upper Sco and $\rho$ Oph regions in $K2$ data have been conducted by 
\cite{ansdell2016} and \cite{hedges2018}; the latter study appeared just as we were finalizing the present work for submission.

Comparing our dipper and burster classifications with these groups, we find that
our sample includes 26 dippers and 14 bursters that do not appear in previous lists (excluding Cody et al.\ 2017). 
We also find that we do not agree with some of the dipper or burster classifications by these groups.
Although guided by our $M$ and $Q$ statistics, our light curve categorization is
ultimately determined by-eye.

\cite{ansdell2016} applied a high-pass filter and used 
measured depth and significance of repeated dips to identify  25 dipper sources.
We agree with all but EPIC 203895983 and EPIC 204630363, which we have classified as quasi-periodic-symmetric (QPS) rather than as dippers (APD or QPD).  However, we do have disagreements for about half of the remaining cases between quasi-periodic ($Q<0.8$ in our scheme) vs aperiodic ($Q<0.8$ in our scheme) dipper categorization.

\cite{hedges2018} use a machine-learning method to identify 95 dippers, based on distinguishing them from other non-young-star periodic variable classes consisting of 
various pulsators and binaries. Given the richness of young star variability behavior, especially the quasi-periodicity and aperiodicity, it is not surprising that our 
results differ substantially from those of \cite{hedges2018}. We agree with 69 of their 95 dippers being in either the APD or QPD categories.  Of the remaining 26, we find 
that 5 are not in our final sample due to lack of membership, disk, or good-quality lightcurve.  The rest we categorize as symmetric, either QPS (quasi-periodic-symmetric) 
or S (stochastic, also called aperiodic-symmetric). \cite{hedges2018} also identify 30 bursters, seemingly 
because they stood out as dippers in their random forest classifier, but which upon examination brightened rather than faded. We agree that 25 are bursters, while four 
(EPIC 203833873, 203870058, 203889938, and 203912674) are labeled by us as stochastic, and the remaining one (EPIC 203668987; Haro1-5) was not in our disk sample.

Among our {\it disk-selected sample}, we find that
32$\pm$3\% of the sources exhibit ``dipper'' behavior, including both 
our quasi-periodic dipper (QPD) and aperiodic dipper (APD) categories.
We found a higher fraction of dippers in the older Upper Sco compared to the young $\rho$ Oph.
\cite{hedges2018}, by contrast, reports 6$\pm$1\% and 11$\pm$2.5\% as the dipper fractions
in Upper Sco and $\rho$ Oph, respectively,
but among a sample that includes both disked and non-disk stars, given that the sample is simply a
compilation of candidate members from the literature.  They claim that among
stars with disks, the fractions are higher, 21$\pm$5.5\% for the combined regions 
-- but this is still lower than our derived 32$\pm$3\% (Table~3). 

Also among our {\it disk-selected sample}, we find 13$\pm$2\% of the sources to exhibit
``burster'' behavior, with no difference between Upper Sco and $\rho$ Oph.  
As for the dippers, \cite{hedges2018} report lower numbers
of 1$\pm$0.5\% and 7.5$\pm$2\% for the burster fractions
in Upper Sco and $\rho$ Oph, respectively, again considering a disk-plus-non-disk sample.
They claim that among stars with disks, the fractions are indeed higher, 
4.5$\pm$1.5\% and 14.5$\pm$4\% -- but these are still lower than our combined 13$\pm$2\% (Table~3).

In addition to the $K2$ studies mentioned above, \cite{ansdell2018} classified light curves for
Upper Sco and $\rho$ Oph members based on ground-based data
at lower precision and cadence than the $K2$ data, though over longer duration.
Among their four identified dippers, we concur with the classification
based on our $K2$ data analysis of shorter time series for three of the cases, but a fourth
(J15554883-2512240 = EPIC 203710077) we have placed in our
multi-periodic category.  None of their three identified ``long period"
systems are in our disk-selected sample.  And among their six
``long-timescale variations", four are not in our disk-selected sample,
one is classified by us as a dipper, and the last is a quasi-periodic symmetric (QPS) system.
We would not be sensitive to the noted long timescale behavior.

\section{Summary and Discussion}

In this paper, we extend our previous work on time series photometric analysis of NGC 2264 to young members of the $\rho$ Oph and Upper Sco regions. Analysis of the dataset 
from $K2$'s Campaign 2 reveals that nearly 100\% of disk-bearing members of Upper Sco and $\rho$ Oph are photometrically variable, with amplitudes ranging from $\sim$0.1\% 
to a factor of three. We have provided a framework for categorizing young star light curves that is most effective for high precision and high cadence data from space-based 
photometric monitoring missions. The statistical metrics $Q$ and $M$ originally developed for the NGC~2264 dataset from {\em CoRoT} allow us to place individual sources into 
variability categories. A range in light curve repeatability ($Q$) from periodic, to quasi-periodic, to stochastic, and a range in light curve flux symmetry ($M$) from 
primarily dipping, to symmetric, to primarily bursting, is seen.  These metrics may form the basis for feature definition in future machine-learning approaches to 
lightcurve classification for young stars.  Thus far, young stars have stymied unsupervised machine-learning approaches that use general feature sets designed around 
periodic variable classification \citep[e.g.][]{richards2011} due to the admixture of periodic and aperiodic phenomena contributing to the lightcurves.

Upon assigning categories to all young stars in our $K2$ dataset, we have explored correlations between variability type and circumstellar properties. We find that bursters, 
stochastic sources, and quasi-periodic sources tend to have larger infrared colors, which can indicate more rapidly accreting circumstellar disks once geometric 
effects are accounted for. However, these groups appear to separate in their inclination distributions, with bursters and stochastic sources found exclusively at $i < 
50-60^\circ$, while quasi-periodic sources tend to have $i > 50^\circ$. On the other hand, dippers, both quasi-periodic and aperiodic, also tend to have $i > 50^\circ$. 

While the burster fraction is the same, we find a higher fraction of dippers in the older Upper Sco region than towards sources in the $\rho$ Oph molecular cloud region and towards NGC~2264, consistent with a scenario in which somewhat evolved inner disks give rise to a dipper lightcurve morphology. Furthermore, the dippers appear to be distinguished from the quasi-periodic sources by having lower H$\alpha$ equivalent widths (indicating lower accretion rates on average). 

The clustering of particular light curve morphologies in the parameter space of inclination, disk size, and accretion rate lends support to previous suggestions that variability mechanisms include obscuration by inner disk dust (dippers), rapid accretion rate changes (bursters; stochastic stars), and flux modulation by accretion hot spots at the stellar surface (large amplitude quasi-periodic class). The periodic, multiperiodic, and low-amplitude quasi-periodic sources are consistent with cool spots on the stellar surface, seen due to a relative lack of obscuring material and/or accretion. The remaining unclassifiable and long-timescale variables remain to be elucidated and may be contaminants such as asymptotic giant stars. 

Dippers have frequently been explained as the manifestation of a warped or otherwise clumpy inner disk occulting parts of the central star \citep[e.g.,][]{bouvier1999}, as seen from a nearly edge-on viewpoint. \cite{bodman2017} argued that either a disk warp or dust in the accretion stream could be responsible for dipping behavior at moderately high view angle. 
The appearance of most dipping behavior in systems with disk inclinations between 50$^\circ$ and 90$^\circ$ strongly supports these ideas. There are two notable exceptions with $i < 10^\circ$, EPIC 204245509 and EPIC 204638512, the latter of which was highlighted by \cite{2016MNRAS.462L.101A}. These may represent rare cases in which the inner disk is not aligned with the resolved outer disk. 

With approximately equal proportions of periodic and aperiodic dippers, it is as yet unclear as to what determines the repeatability (or lack thereof) of fading events. \cite{2015A&A...577A..11M} suggested that the behavior is determined according to whether accretion is Rayleigh-Taylor unstable, as predicted by \cite{2016MNRAS.459.2354B} based on the accretion rate and magnetic dipole tilt relative to the stellar rotation axis.  In contrast, we do not detect any correlation between periodicity status and H$\alpha$ equivalent width in this dataset. However, non-simultaneity of the spectroscopic and photometric observations may obscure such an effect. We do find that most dippers have relatively low H$\alpha$ values, consistent with the requirement that the magnetospheric radius (determined in part by accretion rate) must lie far enough out from the star such that dust does not sublimate at that location.

The highest amplitude quasi-periodic stars are interesting in that they have very large reported H$\alpha$ equivalent widths. The semi-regular behavior of their light curves implies a stable accretion hot spot configuration. Stability is expected for fairly {\em low} accretion rates \citep{2003ApJ...595.1009R,2004ApJ...610..920R}. However, \cite{2016MNRAS.459.2354B} have recently shown that there is an additional ``unstable ordered'' regime in which one or two tongues of accreting gas penetrate the magnetosphere to reach the stellar photosphere. We believe that the quasi-periodic stars with amplitudes from 0.3 to 1.3 (in normalized flux units) may be examples of this effect. They may be seen only at inclinations greater than 50$^\circ$ if the resulting hot spots are situated away from the stellar pole. 

Turning finally to the bursters and stochastic stars, we have seen that these have moderately high H$\alpha$ equivalent widths as well, but inclinations typically between 15$^\circ$ and 50$^\circ$. The chaotic nature of the light curves is suggestive of an unstable accretion regime, while the strong near-infrared excesses imply dust fairly close to the central star (particularly for the bursters). What distinguishes the bursters and stochastic stars? \cite{2016AJ....151...60S} hypothesized that they are both due to variable mass accretion, with flow for the latter including many more low-amplitude events. The similar ranges of H$\alpha$ and [near-]infrared colors for these two classes lends support to the idea that they are both manifestations of unstable accretion flow, seen at low inclination. 

{\em K2}'s Campaign 2 young star dataset has offered a unique opportunity to correlate detailed variability properties with a large collection of stellar and circumstellar parameters, some of which include resolved measurements. Further {\em K2} observations, e.g., of the young cluster NGC~6530 and the Taurus association will provide further tests for our framework of variability classification and corresponding suggestions for physical mechanisms. 

\acknowledgements
We extend our appreciation to the referee for helpful comments. We thank Trevor David for help collecting and categorizing {\em K2} light curves. We acknowledge John Carpenter for early conversations about $ALMA$ samples in Upper Sco, as well as Luisa Rebull and John Stauffer for general discussions about young star variability, membership, and disk properties. This paper includes data collected by the $K2$ mission. Funding for the $K2$ mission is provided by the NASA Science Mission directorate.

\bibliography{K2morph.bib}

\begin{thebibliography}{68}
\expandafter\ifx\csname natexlab\endcsname\relax\def\natexlab#1{#1}\fi

\bibitem[{{Aigrain} {et~al.}(2016){Aigrain}, {Parviainen}, \&
  {Pope}}]{aigrain2016}
{Aigrain}, S., {Parviainen}, H., \& {Pope}, B.~J.~S. 2016, \mnras, 459, 2408

\bibitem[{{Andrews} \& {Williams}(2007)}]{andrews2007}
{Andrews}, S.~M., \& {Williams}, J.~P. 2007, \apj, 671, 1800

\bibitem[{{Ansdell} {et~al.}(2016{\natexlab{a}}){Ansdell}, {Gaidos},
  {Rappaport}, {Jacobs}, {LaCourse}, {Jek}, {Mann}, {Wyatt}, {Kennedy},
  {Williams}, \& {Boyajian}}]{ansdell2016}
{Ansdell}, M., {Gaidos}, E., {Rappaport}, S.~A., {Jacobs}, T.~L., {LaCourse},
  D.~M., {Jek}, K.~J., {Mann}, A.~W., {Wyatt}, M.~C., {Kennedy}, G.,
  {Williams}, J.~P., \& {Boyajian}, T.~S. 2016{\natexlab{a}}, \apj, 816, 69

\bibitem[{{Ansdell} {et~al.}(2016{\natexlab{b}}){Ansdell}, {Gaidos},
  {Williams}, {Kennedy}, {Wyatt}, {LaCourse}, {Jacobs}, \&
  {Mann}}]{2016MNRAS.462L.101A}
{Ansdell}, M., {Gaidos}, E., {Williams}, J.~P., {Kennedy}, G., {Wyatt}, M.~C.,
  {LaCourse}, D.~M., {Jacobs}, T.~L., \& {Mann}, A.~W. 2016{\natexlab{b}},
  \mnras, 462, L101

\bibitem[{{Ansdell} {et~al.}(2018){Ansdell}, {Oelkers}, {Rodriguez}, {Gaidos},
  {Somers}, {Mamajek}, {Cargile}, {Stassun}, {Pepper}, {Stevens}, {Beatty},
  {Siverd}, {Lund}, {Kuhn}, {James}, \& {Gaudi}}]{ansdell2018}
{Ansdell}, M., {Oelkers}, R.~J., {Rodriguez}, J.~E., {Gaidos}, E., {Somers},
  G., {Mamajek}, E., {Cargile}, P.~A., {Stassun}, K.~G., {Pepper}, J.,
  {Stevens}, D.~J., {Beatty}, T.~G., {Siverd}, R.~J., {Lund}, M.~B., {Kuhn},
  R.~B., {James}, D., \& {Gaudi}, B.~S. 2018, \mnras, 473, 1231

\bibitem[{{Ardila} {et~al.}(2000){Ardila}, {Mart{\'{\i}}n}, \&
  {Basri}}]{ardila2000}
{Ardila}, D., {Mart{\'{\i}}n}, E., \& {Basri}, G. 2000, \aj, 120, 479

\bibitem[{{Baglin} {et~al.}(2006){Baglin}, {Auvergne}, {Barge}, {Deleuil},
  {Catala}, {Michel}, {Weiss}, \& {COROT Team}}]{baglin2006}
{Baglin}, A., {Auvergne}, M., {Barge}, P., {Deleuil}, M., {Catala}, C.,
  {Michel}, E., {Weiss}, W., \& {COROT Team}. 2006, in ESA Special Publication,
  Vol. 1306, The CoRoT Mission Pre-Launch Status - Stellar Seismology and
  Planet Finding, ed. M.~{Fridlund}, A.~{Baglin}, J.~{Lochard}, \& L.~{Conroy},
  33

\bibitem[{{Barenfeld} {et~al.}(2017){Barenfeld}, {Carpenter}, {Sargent},
  {Isella}, \& {Ricci}}]{barenfeld2017}
{Barenfeld}, S.~A., {Carpenter}, J.~M., {Sargent}, A.~I., {Isella}, A., \&
  {Ricci}, L. 2017, \apj, 851, 85

\bibitem[{{Blinova} {et~al.}(2016){Blinova}, {Romanova}, \&
  {Lovelace}}]{2016MNRAS.459.2354B}
{Blinova}, A.~A., {Romanova}, M.~M., \& {Lovelace}, R.~V.~E. 2016, \mnras, 459,
  2354

\bibitem[{{Bodman} {et~al.}(2017){Bodman}, {Quillen}, {Ansdell}, {Hippke},
  {Boyajian}, {Mamajek}, {Blackman}, {Rizzuto}, \& {Kastner}}]{bodman2017}
{Bodman}, E.~H.~L., {Quillen}, A.~C., {Ansdell}, M., {Hippke}, M., {Boyajian},
  T.~S., {Mamajek}, E.~E., {Blackman}, E.~G., {Rizzuto}, A., \& {Kastner},
  J.~H. 2017, \mnras, 470, 202

\bibitem[{{Bouvier} {et~al.}(1999){Bouvier}, {Chelli}, {Allain}, {Carrasco},
  {Costero}, {Cruz-Gonzalez}, {Dougados}, {Fern{\'a}ndez}, {Mart{\'{\i}}n},
  {M{\'e}nard}, {Mennessier}, {Mujica}, {Recillas}, {Salas}, {Schmidt}, \&
  {Wichmann}}]{bouvier1999}
{Bouvier}, J., {Chelli}, A., {Allain}, S., {Carrasco}, L., {Costero}, R.,
  {Cruz-Gonzalez}, I., {Dougados}, C., {Fern{\'a}ndez}, M., {Mart{\'{\i}}n},
  E.~L., {M{\'e}nard}, F., {Mennessier}, C., {Mujica}, R., {Recillas}, E.,
  {Salas}, L., {Schmidt}, G., \& {Wichmann}, R. 1999, \aap, 349, 619

\bibitem[{{Carpenter} {et~al.}(2014){Carpenter}, {Ricci}, \&
  {Isella}}]{carpenter2014}
{Carpenter}, J.~M., {Ricci}, L., \& {Isella}, A. 2014, \apj, 787, 42

\bibitem[{{Cody} \& {Hillenbrand}(2010)}]{cody2010}
{Cody}, A.~M., \& {Hillenbrand}, L.~A. 2010, \apjs, 191, 389

\bibitem[{{Cody} \& {Hillenbrand}(2011)}]{cody2011}
---. 2011, \apj, 741, 9

\bibitem[{{Cody} {et~al.}(2017){Cody}, {Hillenbrand}, {David}, {Carpenter},
  {Everett}, \& {Howell}}]{cody2017}
{Cody}, A.~M., {Hillenbrand}, L.~A., {David}, T.~J., {Carpenter}, J.~M.,
  {Everett}, M.~E., \& {Howell}, S.~B. 2017, \apj, 836, 41

\bibitem[{{Cody} {et~al.}(2014){Cody}, {Stauffer}, {Baglin}, {Micela},
  {Rebull}, {Flaccomio}, {Morales-Calder{\'o}n}, {Aigrain}, {Bouvier},
  {Hillenbrand}, {Gutermuth}, {Song}, {Turner}, {Alencar}, {Zwintz},
  {Plavchan}, {Carpenter}, {Findeisen}, {Carey}, {Terebey}, {Hartmann},
  {Calvet}, {Teixeira}, {Vrba}, {Wolk}, {Covey}, {Poppenhaeger}, {G{\"u}nther},
  {Forbrich}, {Whitney}, {Affer}, {Herbst}, {Hora}, {Barrado}, {Holtzman},
  {Marchis}, {Wood}, {Medeiros Guimar{\~a}es}, {Lillo Box}, {Gillen},
  {McQuillan}, {Espaillat}, {Allen}, {D'Alessio}, \& {Favata}}]{cody2014}
{Cody}, A.~M., {Stauffer}, J., {Baglin}, A., {Micela}, G., {Rebull}, L.~M.,
  {Flaccomio}, E., {Morales-Calder{\'o}n}, M., {Aigrain}, S., {Bouvier}, J.,
  {Hillenbrand}, L.~A., {Gutermuth}, R., {Song}, I., {Turner}, N., {Alencar},
  S.~H.~P., {Zwintz}, K., {Plavchan}, P., {Carpenter}, J., {Findeisen}, K.,
  {Carey}, S., {Terebey}, S., {Hartmann}, L., {Calvet}, N., {Teixeira}, P.,
  {Vrba}, F.~J., {Wolk}, S., {Covey}, K., {Poppenhaeger}, K., {G{\"u}nther},
  H.~M., {Forbrich}, J., {Whitney}, B., {Affer}, L., {Herbst}, W., {Hora}, J.,
  {Barrado}, D., {Holtzman}, J., {Marchis}, F., {Wood}, K., {Medeiros
  Guimar{\~a}es}, M., {Lillo Box}, J., {Gillen}, E., {McQuillan}, A.,
  {Espaillat}, C., {Allen}, L., {D'Alessio}, P., \& {Favata}, F. 2014, \aj,
  147, 82

\bibitem[{{David} {et~al.}(2017){David}, {Petigura}, {Hillenbrand}, {Cody},
  {Collier Cameron}, {Stauffer}, {Fulton}, {Isaacson}, {Howard}, {Howell},
  {Everett}, {Wang}, {Benneke}, {Hellier}, {West}, {Pollacco}, \&
  {Anderson}}]{david2017}
{David}, T.~J., {Petigura}, E.~A., {Hillenbrand}, L.~A., {Cody}, A.~M.,
  {Collier Cameron}, A., {Stauffer}, J.~R., {Fulton}, B.~J., {Isaacson}, H.~T.,
  {Howard}, A.~W., {Howell}, S.~B., {Everett}, M.~E., {Wang}, J., {Benneke},
  B., {Hellier}, C., {West}, R.~G., {Pollacco}, D., \& {Anderson}, D.~R. 2017,
  \apj, 835, 168

\bibitem[{{Dawson} {et~al.}(2013){Dawson}, {Scholz}, {Ray}, {Marsh}, {Wood},
  {Natta}, {Padgett}, \& {Ressler}}]{dawson2013}
{Dawson}, P., {Scholz}, A., {Ray}, T.~P., {Marsh}, K.~A., {Wood}, K., {Natta},
  A., {Padgett}, D., \& {Ressler}, M.~E. 2013, \mnras, 429, 903

\bibitem[{{de Bruijne}(1999)}]{debruijne1999}
{de Bruijne}, J.~H.~J. 1999, \mnras, 310, 585

\bibitem[{{Erickson} {et~al.}(2011){Erickson}, {Wilking}, {Meyer}, {Robinson},
  \& {Stephenson}}]{erickson2011}
{Erickson}, K.~L., {Wilking}, B.~A., {Meyer}, M.~R., {Robinson}, J.~G., \&
  {Stephenson}, L.~N. 2011, \aj, 142, 140

\bibitem[{{Fang} {et~al.}(2017){Fang}, {Herczeg}, \& {Rizzuto}}]{fang2017}
{Fang}, Q., {Herczeg}, G.~J., \& {Rizzuto}, A. 2017, \apj, 842, 123

\bibitem[{{Findeisen} {et~al.}(2015){Findeisen}, {Cody}, \&
  {Hillenbrand}}]{findeisen2015}
{Findeisen}, K., {Cody}, A.~M., \& {Hillenbrand}, L. 2015, \apj, 798, 89

\bibitem[{{Guenther} {et~al.}(2007){Guenther}, {Esposito}, {Mundt}, {Covino},
  {Alcal{\'a}}, {Cusano}, \& {Stecklum}}]{guenther2007}
{Guenther}, E.~W., {Esposito}, M., {Mundt}, R., {Covino}, E., {Alcal{\'a}},
  J.~M., {Cusano}, F., \& {Stecklum}, B. 2007, \aap, 467, 1147

\bibitem[{{Gutermuth} {et~al.}(2009){Gutermuth}, {Megeath}, {Myers}, {Allen},
  {Pipher}, \& {Fazio}}]{gutermuth2009}
{Gutermuth}, R.~A., {Megeath}, S.~T., {Myers}, P.~C., {Allen}, L.~E., {Pipher},
  J.~L., \& {Fazio}, G.~G. 2009, \apjs, 184, 18

\bibitem[{{Hedges} {et~al.}(2018){Hedges}, {Hodgkin}, \&
  {Kennedy}}]{hedges2018}
{Hedges}, C., {Hodgkin}, S., \& {Kennedy}, G. 2018, ArXiv e-prints

\bibitem[{{Herbig}(1954)}]{herbig1954}
{Herbig}, G.~H. 1954, \apj, 119, 483

\bibitem[{{Honda} {et~al.}(2015){Honda}, {Maaskant}, {Okamoto}, {Kataza},
  {Yamashita}, {Miyata}, {Sako}, {Fujiyoshi}, {Sakon}, {Fujiwara}, {Kamizuka},
  {Mulders}, {Lopez-Rodriguez}, {Packham}, \& {Onaka}}]{honda2015}
{Honda}, M., {Maaskant}, K., {Okamoto}, Y.~K., {Kataza}, H., {Yamashita}, T.,
  {Miyata}, T., {Sako}, S., {Fujiyoshi}, T., {Sakon}, I., {Fujiwara}, H.,
  {Kamizuka}, T., {Mulders}, G.~D., {Lopez-Rodriguez}, E., {Packham}, C., \&
  {Onaka}, T. 2015, \apj, 804, 143

\bibitem[{{Houk} \& {Smith-Moore}(1988)}]{1988mcts.book.....H}
{Houk}, N., \& {Smith-Moore}, M. 1988, {Michigan Catalogue of Two-dimensional
  Spectral Types for the HD Stars. Volume 4, Declinations -26deg.0 to
  -12deg.0.}

\bibitem[{{Howell} {et~al.}(2014){Howell}, {Sobeck}, {Haas}, {Still},
  {Barclay}, {Mullally}, {Troeltzsch}, {Aigrain}, {Bryson}, {Caldwell},
  {Chaplin}, {Cochran}, {Huber}, {Marcy}, {Miglio}, {Najita}, {Smith},
  {Twicken}, \& {Fortney}}]{howell2014}
{Howell}, S.~B., {Sobeck}, C., {Haas}, M., {Still}, M., {Barclay}, T.,
  {Mullally}, F., {Troeltzsch}, J., {Aigrain}, S., {Bryson}, S.~T., {Caldwell},
  D., {Chaplin}, W.~J., {Cochran}, W.~D., {Huber}, D., {Marcy}, G.~W.,
  {Miglio}, A., {Najita}, J.~R., {Smith}, M., {Twicken}, J.~D., \& {Fortney},
  J.~J.~. 2014, \pasp, 126, 398

\bibitem[{{Isella} {et~al.}(2009){Isella}, {Carpenter}, \&
  {Sargent}}]{isella2009}
{Isella}, A., {Carpenter}, J.~M., \& {Sargent}, A.~I. 2009, \apj, 701, 260

\bibitem[{{Joy}(1945)}]{joy1945}
{Joy}, A.~H. 1945, \apj, 102, 168

\bibitem[{{K{\"o}hler} {et~al.}(2000){K{\"o}hler}, {Kunkel}, {Leinert}, \&
  {Zinnecker}}]{kohler2000}
{K{\"o}hler}, R., {Kunkel}, M., {Leinert}, C., \& {Zinnecker}, H. 2000, \aap,
  356, 541

\bibitem[{{Lodieu}(2013)}]{lodieu2013}
{Lodieu}, N. 2013, \mnras, 431, 3222

\bibitem[{{Lodieu} {et~al.}(2011){Lodieu}, {Dobbie}, \& {Hambly}}]{lodieu2011}
{Lodieu}, N., {Dobbie}, P.~D., \& {Hambly}, N.~C. 2011, \aap, 527, A24

\bibitem[{{Lodieu} {et~al.}(2006){Lodieu}, {Hambly}, \& {Jameson}}]{lodieu2006}
{Lodieu}, N., {Hambly}, N.~C., \& {Jameson}, R.~F. 2006, \mnras, 373, 95

\bibitem[{{Luhman} \& {Mamajek}(2012)}]{luhman2012}
{Luhman}, K.~L., \& {Mamajek}, E.~E. 2012, \apj, 758, 31

\bibitem[{{Luhman} \& {Rieke}(1999)}]{luhman1999}
{Luhman}, K.~L., \& {Rieke}, G.~H. 1999, \apj, 525, 440

\bibitem[{{Manara} {et~al.}(2015){Manara}, {Testi}, {Natta}, \&
  {Alcal{\'a}}}]{manara2015}
{Manara}, C.~F., {Testi}, L., {Natta}, A., \& {Alcal{\'a}}, J.~M. 2015, \aap,
  579, A66

\bibitem[{{McGinnis} {et~al.}(2015){McGinnis}, {Alencar}, {Guimar{\~a}es},
  {Sousa}, {Stauffer}, {Bouvier}, {Rebull}, {Fonseca}, {Venuti}, {Hillenbrand},
  {Cody}, {Teixeira}, {Aigrain}, {Favata}, {F{\H u}r{\'e}sz}, {Vrba},
  {Flaccomio}, {Turner}, {Gameiro}, {Dougados}, {Herbst},
  {Morales-Calder{\'o}n}, \& {Micela}}]{2015A&A...577A..11M}
{McGinnis}, P.~T., {Alencar}, S.~H.~P., {Guimar{\~a}es}, M.~M., {Sousa}, A.~P.,
  {Stauffer}, J., {Bouvier}, J., {Rebull}, L., {Fonseca}, N.~N.~J., {Venuti},
  L., {Hillenbrand}, L., {Cody}, A.~M., {Teixeira}, P.~S., {Aigrain}, S.,
  {Favata}, F., {F{\H u}r{\'e}sz}, G., {Vrba}, F.~J., {Flaccomio}, E.,
  {Turner}, N.~J., {Gameiro}, J.~F., {Dougados}, C., {Herbst}, W.,
  {Morales-Calder{\'o}n}, M., \& {Micela}, G. 2015, \aap, 577, A11

\bibitem[{{Morales-Calder{\'o}n} {et~al.}(2011){Morales-Calder{\'o}n},
  {Stauffer}, {Hillenbrand}, {Gutermuth}, {Song}, {Rebull}, {Plavchan},
  {Carpenter}, {Whitney}, {Covey}, {Alves de Oliveira}, {Winston},
  {McCaughrean}, {Bouvier}, {Guieu}, {Vrba}, {Holtzman}, {Marchis}, {Hora},
  {Wasserman}, {Terebey}, {Megeath}, {Guinan}, {Forbrich}, {Hu{\'e}lamo},
  {Riviere-Marichalar}, {Barrado}, {Stapelfeldt}, {Hern{\'a}ndez}, {Allen},
  {Ardila}, {Bayo}, {Favata}, {James}, {Werner}, \& {Wood}}]{morales2011}
{Morales-Calder{\'o}n}, M., {Stauffer}, J.~R., {Hillenbrand}, L.~A.,
  {Gutermuth}, R., {Song}, I., {Rebull}, L.~M., {Plavchan}, P., {Carpenter},
  J.~M., {Whitney}, B.~A., {Covey}, K., {Alves de Oliveira}, C., {Winston}, E.,
  {McCaughrean}, M.~J., {Bouvier}, J., {Guieu}, S., {Vrba}, F.~J., {Holtzman},
  J., {Marchis}, F., {Hora}, J.~L., {Wasserman}, L.~H., {Terebey}, S.,
  {Megeath}, T., {Guinan}, E., {Forbrich}, J., {Hu{\'e}lamo}, N.,
  {Riviere-Marichalar}, P., {Barrado}, D., {Stapelfeldt}, K., {Hern{\'a}ndez},
  J., {Allen}, L.~E., {Ardila}, D.~R., {Bayo}, A., {Favata}, F., {James}, D.,
  {Werner}, M., \& {Wood}, K. 2011, \apj, 733, 50

\bibitem[{{Najita} {et~al.}(2015){Najita}, {Andrews}, \&
  {Muzerolle}}]{najita2015}
{Najita}, J.~R., {Andrews}, S.~M., \& {Muzerolle}, J. 2015, \mnras, 450, 3559

\bibitem[{{Parks} {et~al.}(2014){Parks}, {Plavchan}, {White}, \&
  {Gee}}]{parks2014}
{Parks}, J.~R., {Plavchan}, P., {White}, R.~J., \& {Gee}, A.~H. 2014, \apjs,
  211, 3

\bibitem[{{Pecaut} \& {Mamajek}(2016)}]{pecaut2016}
{Pecaut}, M.~J., \& {Mamajek}, E.~E. 2016, \mnras, 461, 794

\bibitem[{{Pecaut} {et~al.}(2012){Pecaut}, {Mamajek}, \& {Bubar}}]{pecaut2012}
{Pecaut}, M.~J., {Mamajek}, E.~E., \& {Bubar}, E.~J. 2012, \apj, 746, 154

\bibitem[{{Preibisch} {et~al.}(2002){Preibisch}, {Brown}, {Bridges},
  {Guenther}, \& {Zinnecker}}]{preibisch2002}
{Preibisch}, T., {Brown}, A.~G.~A., {Bridges}, T., {Guenther}, E., \&
  {Zinnecker}, H. 2002, \aj, 124, 404

\bibitem[{{Preibisch} {et~al.}(1998){Preibisch}, {Guenther}, {Zinnecker},
  {Sterzik}, {Frink}, \& {Roeser}}]{preibisch1998}
{Preibisch}, T., {Guenther}, E., {Zinnecker}, H., {Sterzik}, M., {Frink}, S.,
  \& {Roeser}, S. 1998, \aap, 333, 619

\bibitem[{{Preibisch} \& {Mamajek}(2008)}]{pm2008}
{Preibisch}, T., \& {Mamajek}, E. 2008, {The Nearest OB Association:
  Scorpius-Centaurus (Sco OB2)}, ed. B.~{Reipurth}, 235

\bibitem[{{Rebollido} {et~al.}(2015){Rebollido}, {Mer{\'{\i}}n}, {Ribas},
  {Bustamante}, {Bouy}, {Riviere-Marichalar}, {Prusti}, {Pilbratt},
  {Andr{\'e}}, \& {{\'A}brah{\'a}m}}]{rebollido2015}
{Rebollido}, I., {Mer{\'{\i}}n}, B., {Ribas}, {\'A}., {Bustamante}, I., {Bouy},
  H., {Riviere-Marichalar}, P., {Prusti}, T., {Pilbratt}, G.~L., {Andr{\'e}},
  P., \& {{\'A}brah{\'a}m}, P. 2015, \aap, 581, A30

\bibitem[{{Reboussin} {et~al.}(2015){Reboussin}, {Guilloteau}, {Simon},
  {Grosso}, {Wakelam}, {Di Folco}, {Dutrey}, \& {Pi{\'e}tu}}]{reboussin2015}
{Reboussin}, L., {Guilloteau}, S., {Simon}, M., {Grosso}, N., {Wakelam}, V.,
  {Di Folco}, E., {Dutrey}, A., \& {Pi{\'e}tu}, V. 2015, \aap, 578, A31

\bibitem[{{Rebull} {et~al.}(2018){Rebull}, {Stauffer}, {Cody}, {Hillenbrand},
  {David}, \& {Pinsonneault}}]{rebull2018}
{Rebull}, L.~M., {Stauffer}, J.~R., {Cody}, A.~M., {Hillenbrand}, L.~A.,
  {David}, T.~J., \& {Pinsonneault}, M. 2018, \apj

\bibitem[{{Rice} {et~al.}(2012){Rice}, {Wolk}, \& {Aspin}}]{rice2012}
{Rice}, T.~S., {Wolk}, S.~J., \& {Aspin}, C. 2012, \apj, 755, 65

\bibitem[{{Richards} {et~al.}(2011){Richards}, {Starr}, {Butler}, {Bloom},
  {Brewer}, {Crellin-Quick}, {Higgins}, {Kennedy}, \&
  {Rischard}}]{richards2011}
{Richards}, J.~W., {Starr}, D.~L., {Butler}, N.~R., {Bloom}, J.~S., {Brewer},
  J.~M., {Crellin-Quick}, A., {Higgins}, J., {Kennedy}, R., \& {Rischard}, M.
  2011, \apj, 733, 10

\bibitem[{{Rizzuto} {et~al.}(2015){Rizzuto}, {Ireland}, \&
  {Kraus}}]{rizzuto2015}
{Rizzuto}, A.~C., {Ireland}, M.~J., \& {Kraus}, A.~L. 2015, \mnras, 448, 2737

\bibitem[{{Rizzuto} {et~al.}(2011){Rizzuto}, {Ireland}, \&
  {Robertson}}]{rizzuto2011}
{Rizzuto}, A.~C., {Ireland}, M.~J., \& {Robertson}, J.~G. 2011, \mnras, 416,
  3108

\bibitem[{{Romanova} {et~al.}(2004){Romanova}, {Ustyugova}, {Koldoba}, \&
  {Lovelace}}]{2004ApJ...610..920R}
{Romanova}, M.~M., {Ustyugova}, G.~V., {Koldoba}, A.~V., \& {Lovelace},
  R.~V.~E. 2004, \apj, 610, 920

\bibitem[{{Romanova} {et~al.}(2003){Romanova}, {Ustyugova}, {Koldoba}, {Wick},
  \& {Lovelace}}]{2003ApJ...595.1009R}
{Romanova}, M.~M., {Ustyugova}, G.~V., {Koldoba}, A.~V., {Wick}, J.~V., \&
  {Lovelace}, R.~V.~E. 2003, \apj, 595, 1009

\bibitem[{{Scaringi} {et~al.}(2016){Scaringi}, {Manara}, {Barenfeld}, {Groot},
  {Isella}, {Kenworthy}, {Knigge}, {Maccarone}, {Ricci}, \&
  {Ansdell}}]{scaringi2016}
{Scaringi}, S., {Manara}, C.~F., {Barenfeld}, S.~A., {Groot}, P.~J., {Isella},
  A., {Kenworthy}, M.~A., {Knigge}, C., {Maccarone}, T.~J., {Ricci}, L., \&
  {Ansdell}, M. 2016, \mnras, 463, 2265

\bibitem[{{Skiff}(2014)}]{skiffBMK}
{Skiff}, B.~A. 2014, VizieR Online Data Catalog, 1

\bibitem[{{Slesnick} {et~al.}(2006){Slesnick}, {Carpenter}, \&
  {Hillenbrand}}]{slesnick2006}
{Slesnick}, C.~L., {Carpenter}, J.~M., \& {Hillenbrand}, L.~A. a nd~{Mamajek},
  E.~E. 2006, \aj, 132, 2665

\bibitem[{{Slesnick} {et~al.}(2008){Slesnick}, {Hillenbrand}, \&
  {Carpenter}}]{slesnick2008}
{Slesnick}, C.~L., {Hillenbrand}, L.~A., \& {Carpenter}, J.~M. 2008, \apj, 688,
  377

\bibitem[{{Stauffer} {et~al.}(2015){Stauffer}, {Cody}, {McGinnis}, {Rebull},
  {Hillenbrand}, {Turner}, {Carpenter}, {Plavchan}, {Carey}, {Terebey},
  {Morales-Calder{\'o}n}, {Alencar}, {Bouvier}, {Venuti}, {Hartmann}, {Calvet},
  {Micela}, {Flaccomio}, {Song}, {Gutermuth}, {Barrado}, {Vrba}, {Covey},
  {Padgett}, {Herbst}, {Gillen}, {Lyra}, {Medeiros Guimaraes}, {Bouy}, \&
  {Favata}}]{stauffer2015}
{Stauffer}, J., {Cody}, A.~M., {McGinnis}, P., {Rebull}, L., {Hillenbrand},
  L.~A., {Turner}, N.~J., {Carpenter}, J., {Plavchan}, P., {Carey}, S.,
  {Terebey}, S., {Morales-Calder{\'o}n}, M., {Alencar}, S.~H.~P., {Bouvier},
  J., {Venuti}, L., {Hartmann}, L., {Calvet}, N., {Micela}, G., {Flaccomio},
  E., {Song}, I., {Gutermuth}, R., {Barrado}, D., {Vrba}, F.~J., {Covey}, K.,
  {Padgett}, D., {Herbst}, W., {Gillen}, E., {Lyra}, W., {Medeiros Guimaraes},
  M., {Bouy}, H., \& {Favata}, F. 2015, \aj, 149, 130

\bibitem[{{Stauffer} {et~al.}(2016){Stauffer}, {Cody}, {Rebull}, {Hillenbrand},
  {Turner}, {Carpenter}, {Carey}, {Terebey}, {Morales-Calder{\'o}n}, {Alencar},
  {McGinnis}, {Sousa}, {Bouvier}, {Venuti}, {Hartmann}, {Calvet}, {Micela},
  {Flaccomio}, {Song}, {Gutermuth}, {Barrado}, {Vrba}, {Covey}, {Herbst},
  {Gillen}, {Medeiros Guimar{\~a}es}, {Bouy}, \&
  {Favata}}]{2016AJ....151...60S}
{Stauffer}, J., {Cody}, A.~M., {Rebull}, L., {Hillenbrand}, L.~A., {Turner},
  N.~J., {Carpenter}, J., {Carey}, S., {Terebey}, S., {Morales-Calder{\'o}n},
  M., {Alencar}, S.~H.~P., {McGinnis}, P., {Sousa}, A., {Bouvier}, J.,
  {Venuti}, L., {Hartmann}, L., {Calvet}, N., {Micela}, G., {Flaccomio}, E.,
  {Song}, I., {Gutermuth}, R., {Barrado}, D., {Vrba}, F.~J., {Covey}, K.,
  {Herbst}, W., {Gillen}, E., {Medeiros Guimar{\~a}es}, M., {Bouy}, H., \&
  {Favata}, F. 2016, \aj, 151, 60

\bibitem[{{Stauffer} {et~al.}(2017){Stauffer}, {Collier Cameron}, {Jardine},
  {David}, {Rebull}, {Cody}, {Hillenbrand}, {Barrado}, {Wolk}, {Davenport}, \&
  {Pinsonneault}}]{stauffer2017}
{Stauffer}, J., {Collier Cameron}, A., {Jardine}, M., {David}, T.~J., {Rebull},
  L., {Cody}, A.~M., {Hillenbrand}, L.~A., {Barrado}, D., {Wolk}, S.,
  {Davenport}, J., \& {Pinsonneault}, M. 2017, \aj, 153, 152

\bibitem[{{Torres} {et~al.}(2006){Torres}, {Quast}, {da Silva}, {de La Reza},
  {Melo}, \& {Sterzik}}]{torres2006}
{Torres}, C.~A.~O., {Quast}, G.~R., {da Silva}, L., {de La Reza}, R., {Melo},
  C.~H.~F., \& {Sterzik}, M. 2006, \aap, 460, 695

\bibitem[{{Wahhaj} {et~al.}(2010){Wahhaj}, {Cieza}, {Koerner}, {Stapelfeldt},
  {Padgett}, {Case}, {Keller}, {Mer{\'{\i}}n}, {Evans}, {Harvey}, {Sargent},
  {van Dishoeck}, {Allen}, {Blake}, {Brooke}, {Chapman}, {Mundy}, \&
  {Myers}}]{wahhaj2010}
{Wahhaj}, Z., {Cieza}, L., {Koerner}, D.~W., {Stapelfeldt}, K.~R., {Padgett},
  D.~L., {Case}, A., {Keller}, J.~R., {Mer{\'{\i}}n}, B., {Evans}, II, N.~J.,
  {Harvey}, P., {Sargent}, A., {van Dishoeck}, E.~F., {Allen}, L., {Blake}, G.,
  {Brooke}, T., {Chapman}, N., {Mundy}, L., \& {Myers}, P.~C. 2010, \apj, 724,
  835

\bibitem[{{Walker}(1954)}]{walker1954}
{Walker}, M.~F. 1954, \aj, 59, 333

\bibitem[{{Walter} {et~al.}(1994){Walter}, {Vrba}, {Mathieu}, {Brown}, \&
  {Myers}}]{walter1994}
{Walter}, F.~M., {Vrba}, F.~J., {Mathieu}, R.~D., {Brown}, A., \& {Myers},
  P.~C. 1994, \aj, 107, 692

\bibitem[{{Wilking} {et~al.}(2008){Wilking}, {Gagn{\'e}}, \&
  {Allen}}]{wilking2008}
{Wilking}, B.~A., {Gagn{\'e}}, M., \& {Allen}, L.~E. 2008, {Star Formation in
  the {$\rho$} Ophiuchi Molecular Cloud}, ed. B.~{Reipurth}, 351

\end{thebibliography}

\appendix

\section{Hybrid variability classes}

While we have assigned a single morphology to each light curve in Table~2 and Figure~\ref{alllcs}, in some cases the behavior appears to fall into more than one category.
We describe the various combinations of variability
types and hybrid behaviors, below.

{$\bullet$}
Around 10 stars in our disk sample display both dips and spot signatures.  \cite{rebull2018} identified these from the perspective of examining periodic stars in the Upper Sco and $\rho$ Oph $K2$ data, and presented evidence that the dip events have nearly the same period as the spot pattern.  They argued this was evidence of disk locking, with the stellar photosphere and material at the inner disk radius co-rotating due to linkage by the magnetosphere. Specific objects include EPIC 203542463, EPIC 203770366 (periodic behavior for many weeks, after which narrow dips suddenly appear during the last 25 days), EPIC 204274536, EPIC 204489514, EPIC 204344180 (narrow dips in the first third of the light curve, at the spot period), EPIC 204449389 (evolving spot plus occasional narrow dips), and EPIC 205051240 (occasional narrow dips on top of spot pattern). EPIC 203927902 is a special case in which periodic behavior dominates for most of the light curve, apart from a large (20\% depth) double dip that lasts approximately 10 days.
EPIC 204278916 similarly exhibits gradually diminishing aperiodic dips over the first 30 days, yielding to a spot pattern; this source is also discussed by \cite{scaringi2016}.

{$\bullet$}
In some objects, there is stochastic or otherwise noisy behavior on timescales of several weeks, after which dipping suddenly appears to turn on. An example is EPIC~204538777.

{$\bullet$}
A small set of stars shows both bursts and spots; these include EPIC~204397408, EPIC~203856109 and EPIC~205156547. One unique star, EPIC 204449274, displays all three behaviors. 

{$\bullet$}
There are stars with a long-term trend on top of the dominant variability type that we list in Table~2. It is difficult to discern whether these trends are real or simply uncorrected systematics.

{$\bullet$}
Finally, a significant fraction ($\sim$5\%) of periodic stars show more than one period, as indicated by the variability type ``MP" in Table 2.

 \begin{figure*}
 \epsscale{0.90}
 \plotone{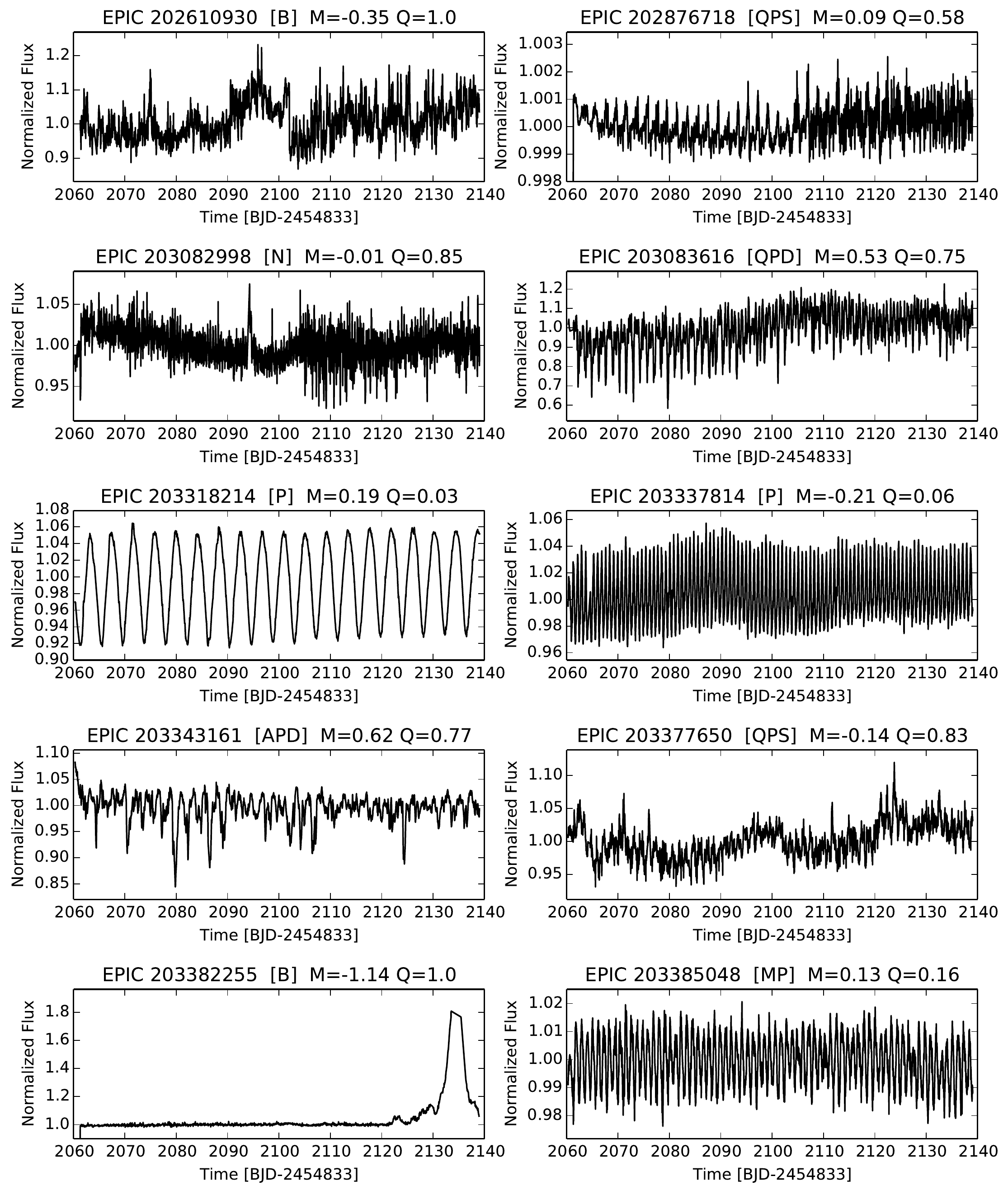}
 \caption{Light curves of disk-bearing stars over the 80-day duration of $K2$ Campaign 2, in order of EPIC identifier.  Figure labels include the variability type from Table~2, namely
 "P" = strictly periodic behavior, "MP" = multiple distinct periods, "QPD" = quasi-periodic dippers, "QPS" = quasi-periodic symmetric, "APD" = aperiodic dippers, "B" = bursters, "S" = stochastic stars, "L" = long-timescale behavior that doesn't fall into the other categories.
 Values of flux symmetry metric $M$ and the quasi-periodicity metric $Q$ from Tabler~2 are also provided.}
 \label{alllcs}
 \end{figure*}

\clearpage

 \addtocounter{figure}{-1} 
 \begin{figure*}
 \epsscale{0.90}
 \plotone{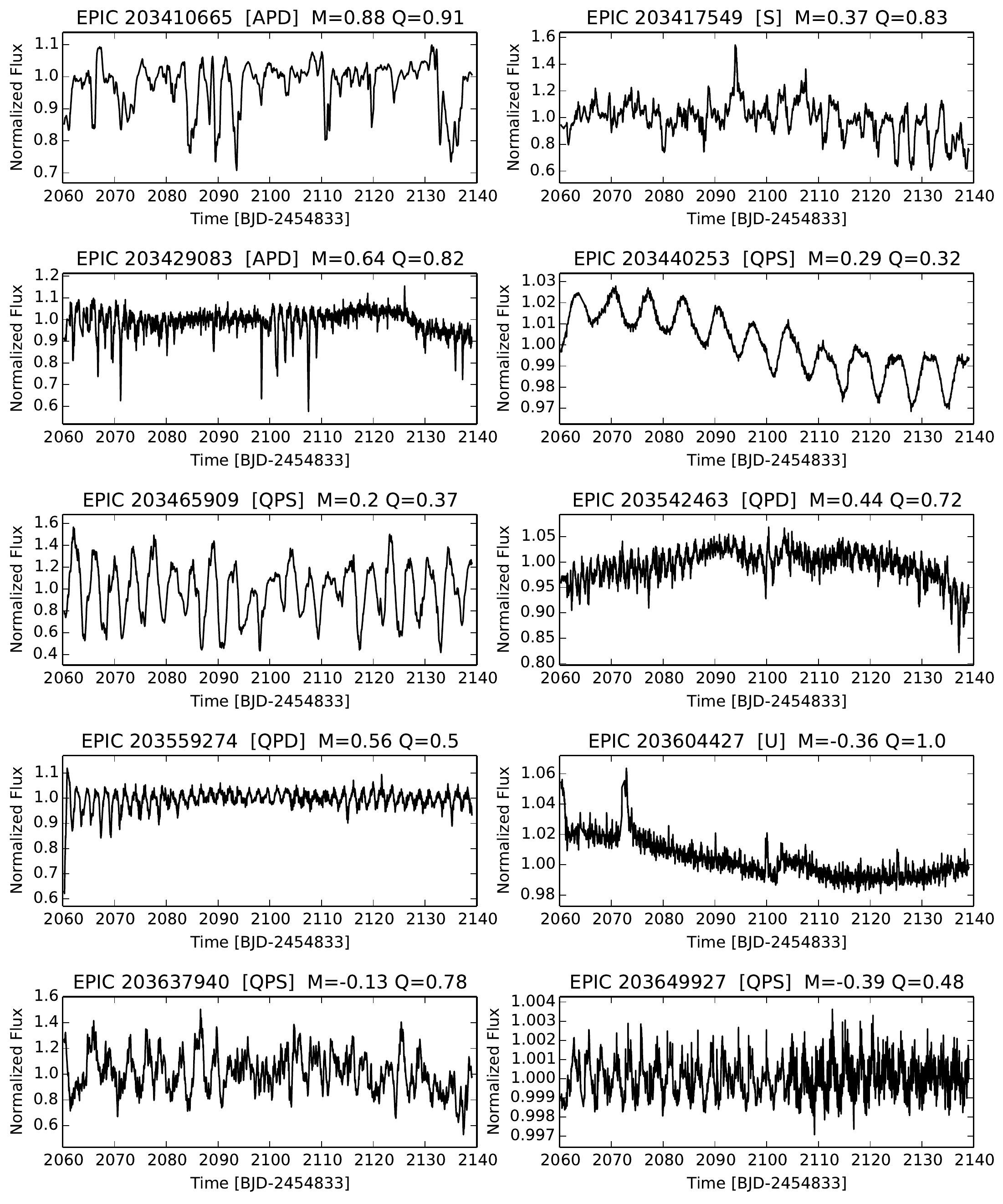}
 \caption{Cont.}
  \end{figure*}

\clearpage
  
 \addtocounter{figure}{-1} 
 \begin{figure*}
 \epsscale{0.90}
 \plotone{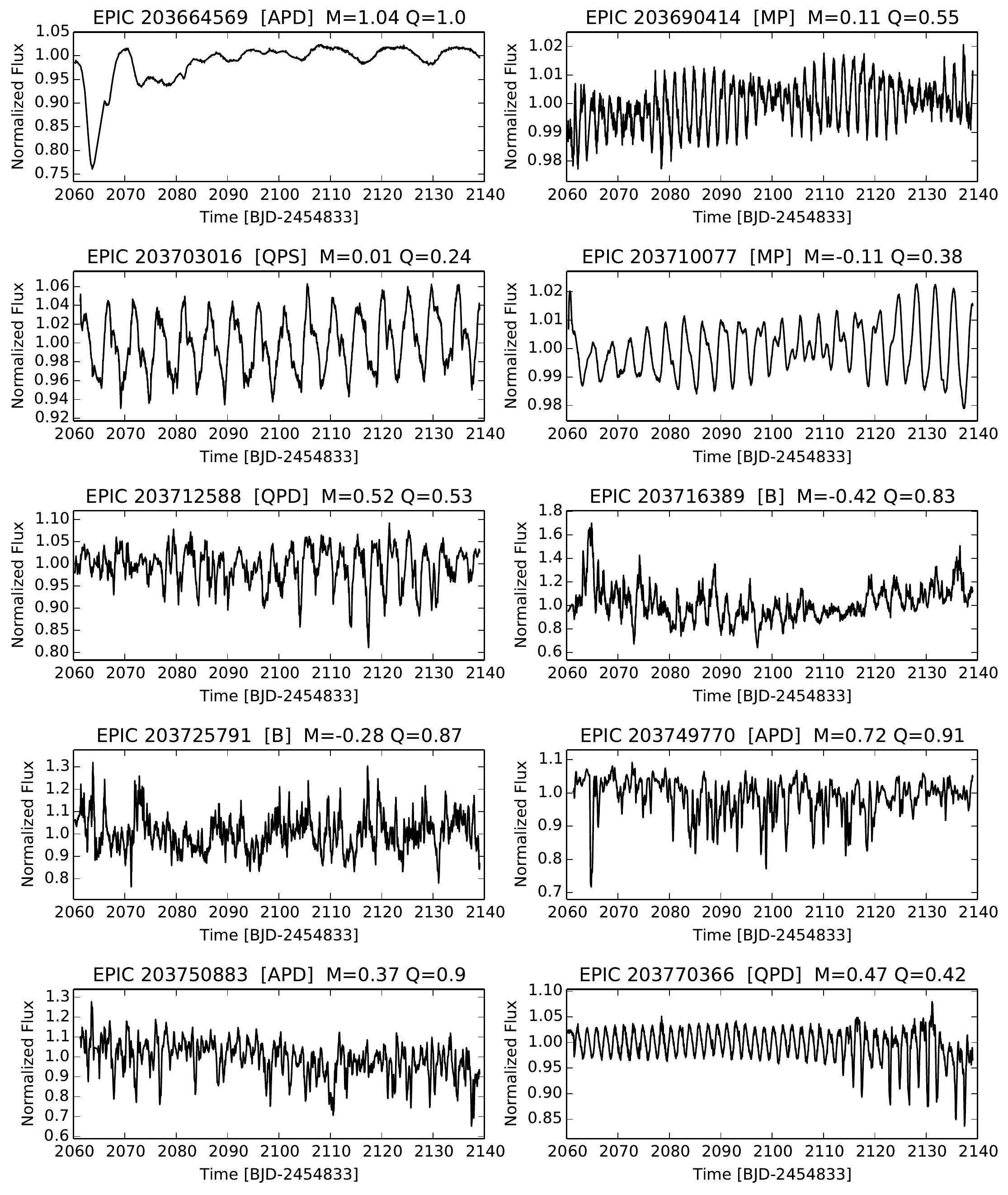}
 \caption{Cont.}
  \end{figure*}

\clearpage

 \addtocounter{figure}{-1} 
 \begin{figure*}
 \epsscale{0.90}
 \plotone{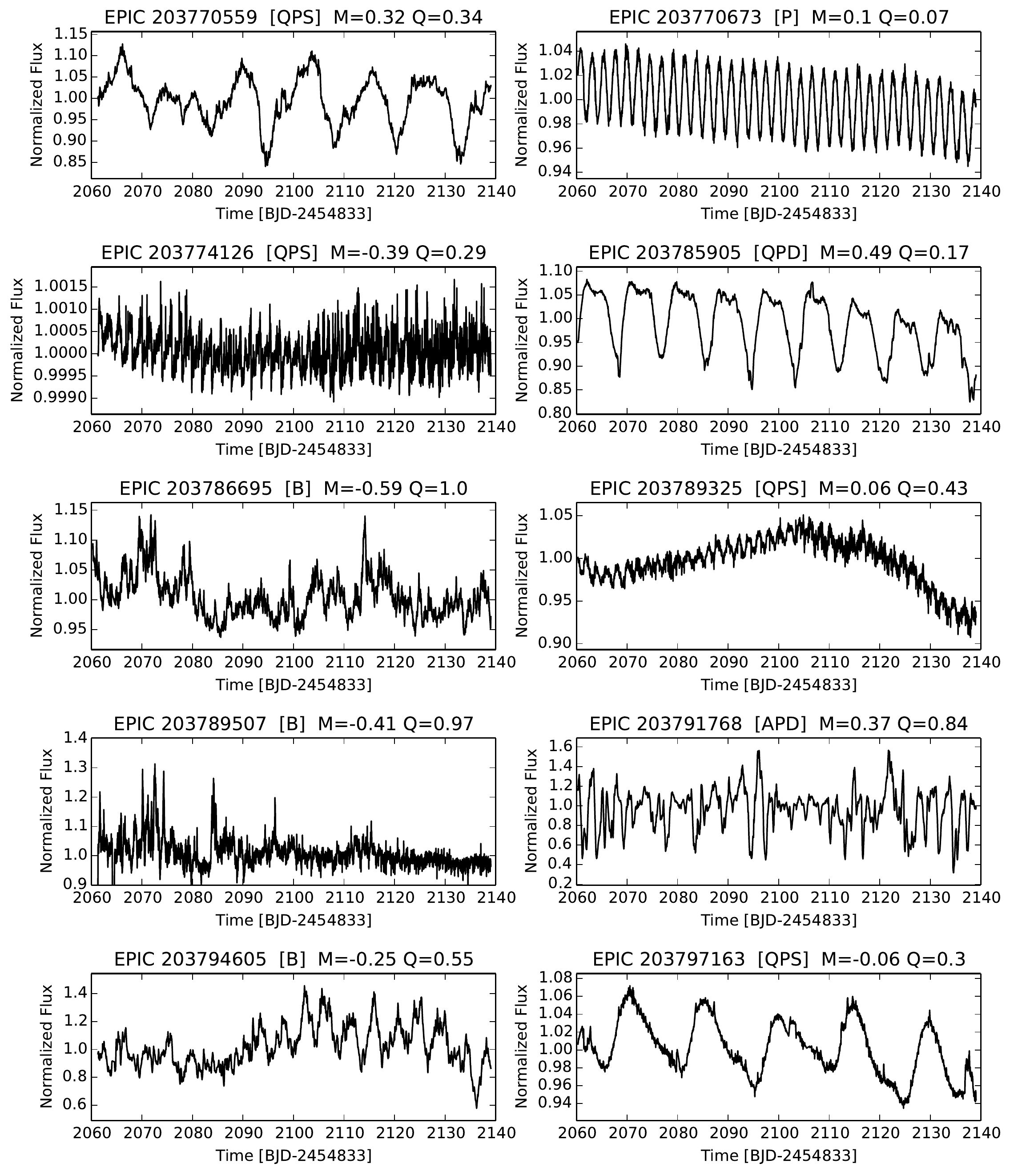}
 \caption{Cont.}
  \end{figure*}

\clearpage
  
 \addtocounter{figure}{-1} 
 \begin{figure*}
 \epsscale{0.90}
 \plotone{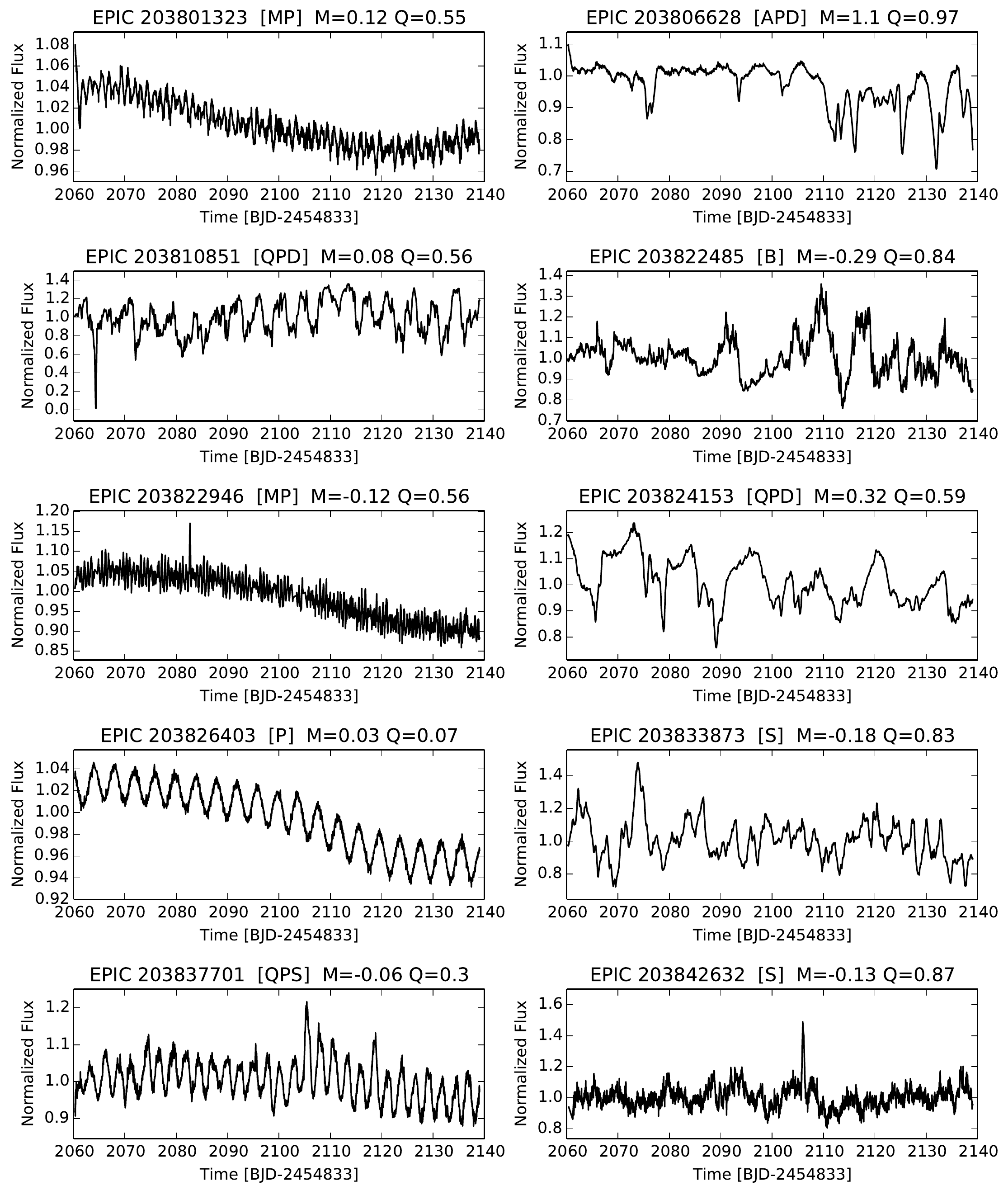}
 \caption{Cont.}
  \end{figure*}

\clearpage
  
 \addtocounter{figure}{-1} 
 \begin{figure*}
 \epsscale{0.90}
 \plotone{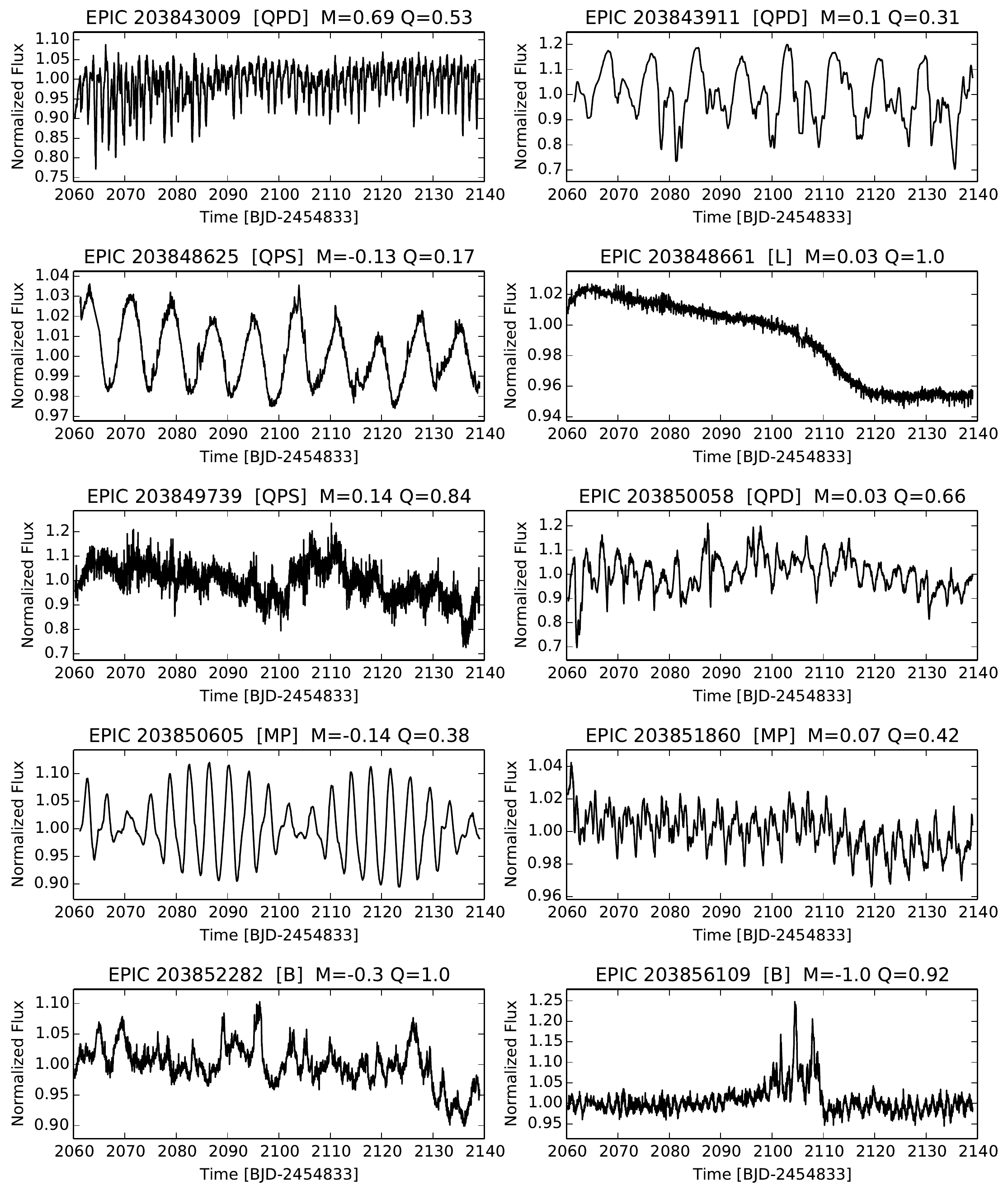}
 \caption{Cont.}
  \end{figure*}

\clearpage
  
 \addtocounter{figure}{-1} 
 \begin{figure*}
 \epsscale{0.90}
 \plotone{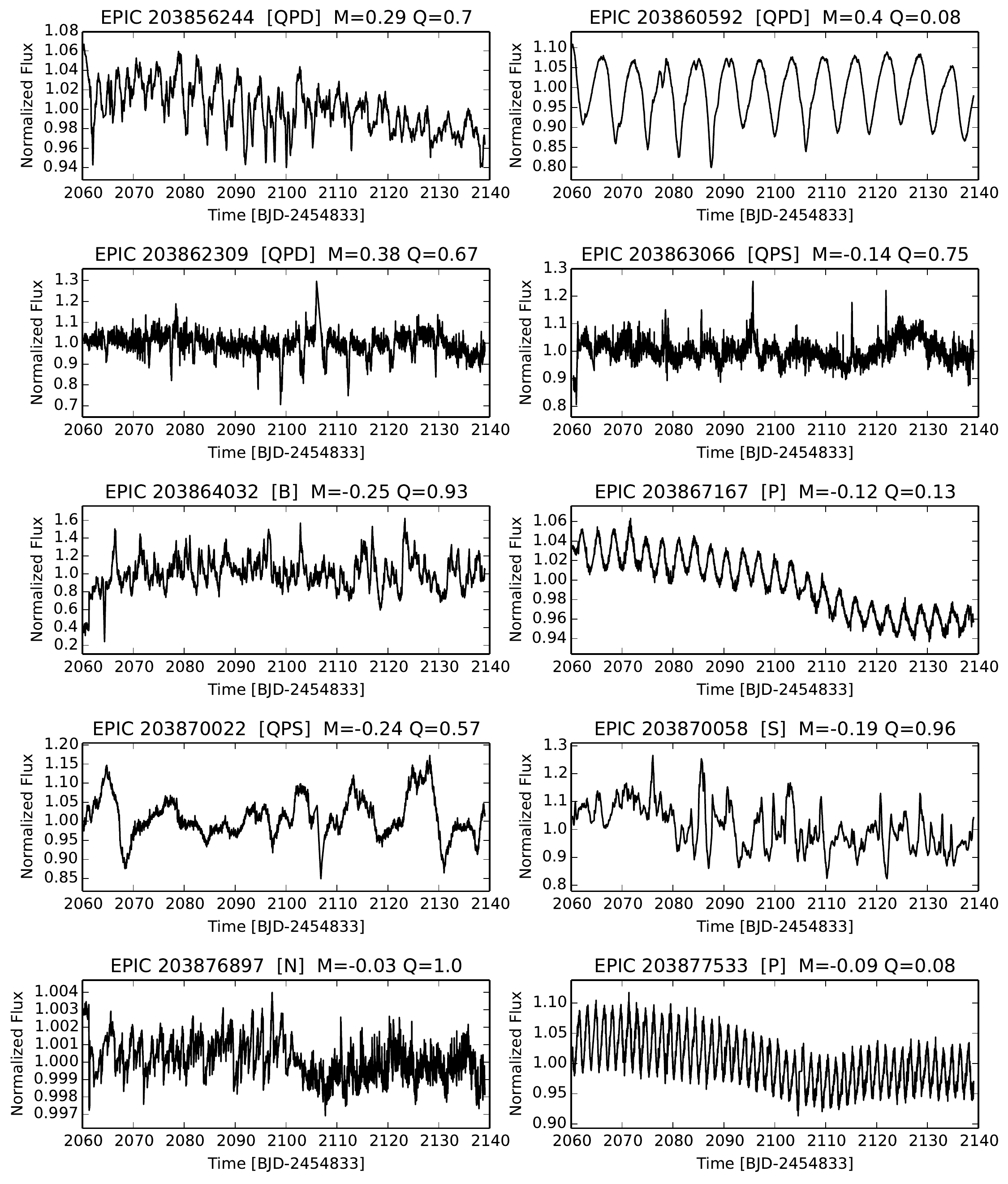}
 \caption{Cont.}
  \end{figure*}

\clearpage
  
 \addtocounter{figure}{-1} 
 \begin{figure*}
 \epsscale{0.90}
 \plotone{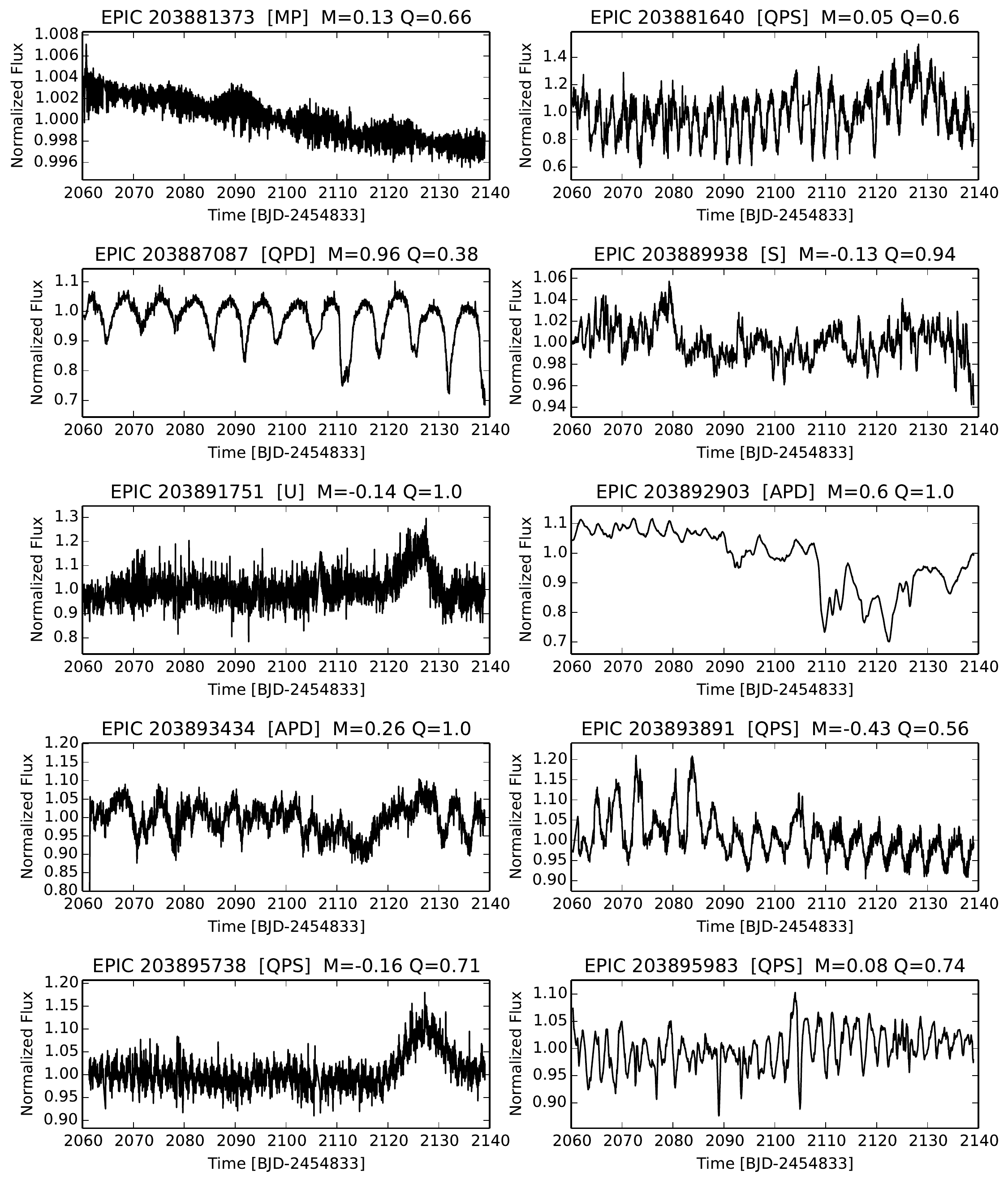}
 \caption{Cont.}
  \end{figure*}

\clearpage
  
 \addtocounter{figure}{-1} 
 \begin{figure*}
 \epsscale{0.90}
 \plotone{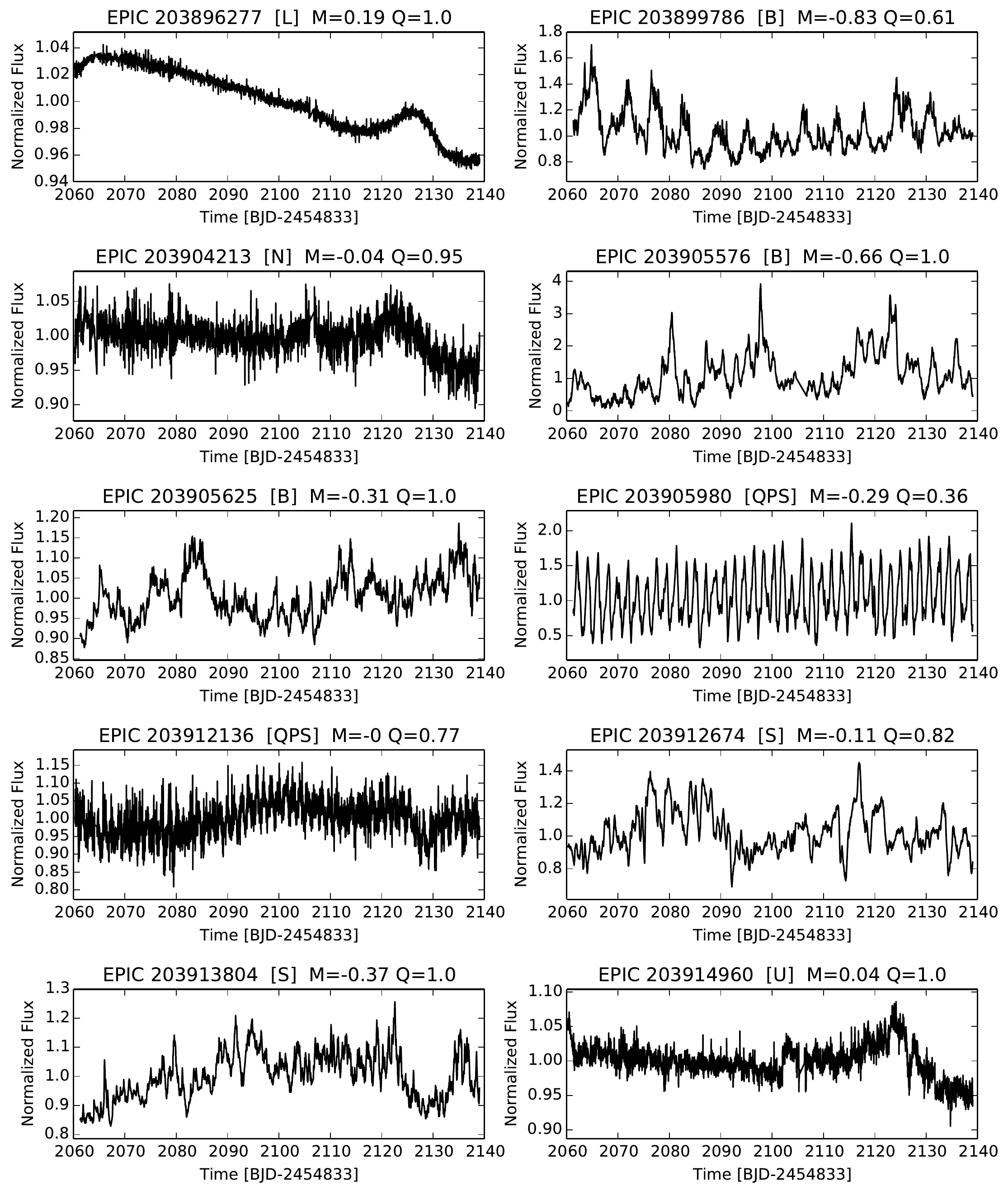}
 \caption{Cont.}
  \end{figure*}

\clearpage
  
 \addtocounter{figure}{-1} 
 \begin{figure*}
 \epsscale{0.90}
 \plotone{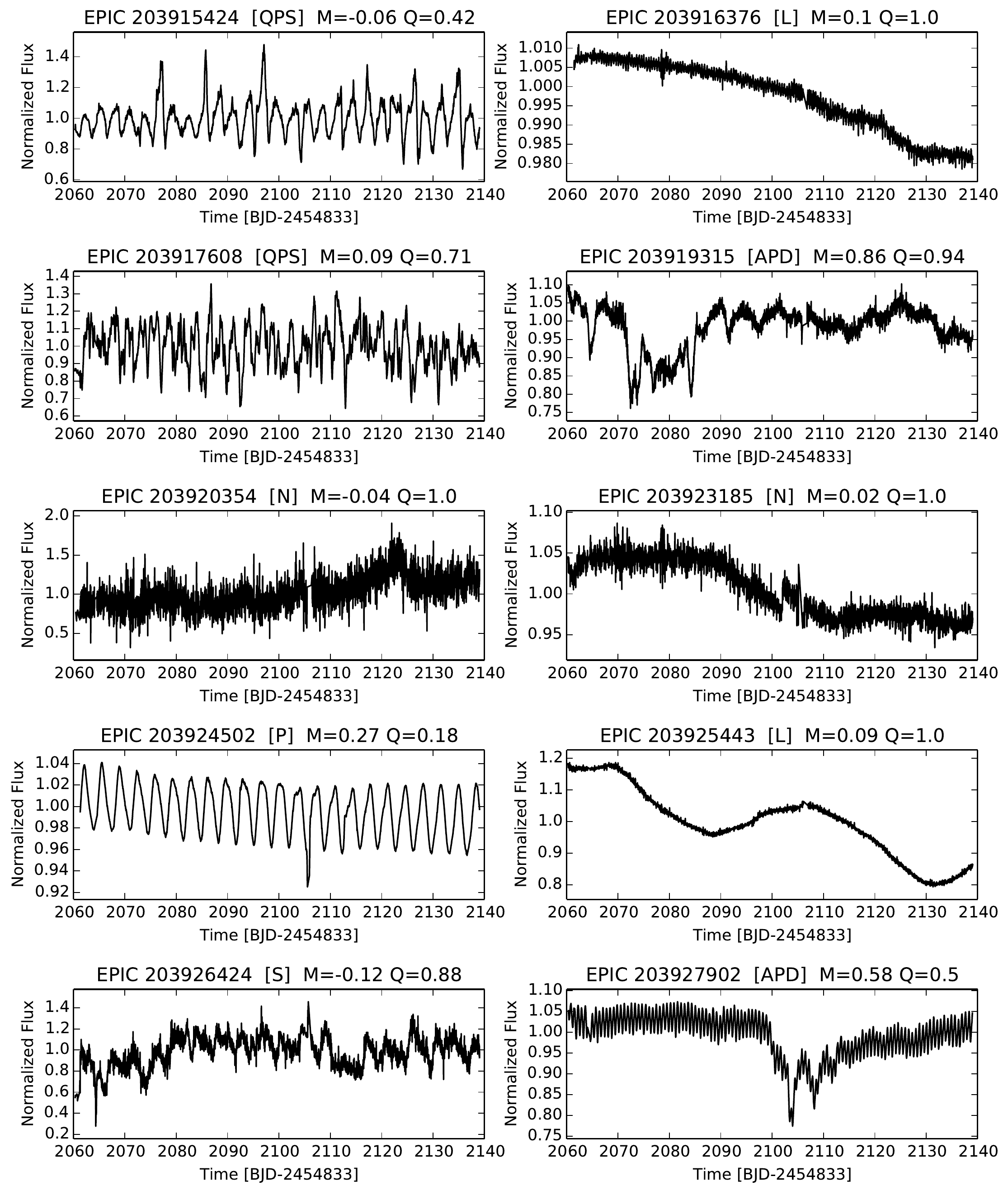}
 \caption{Cont.}
  \end{figure*}
  
\clearpage

 \addtocounter{figure}{-1} 
 \begin{figure*}
 \epsscale{0.90}
 \plotone{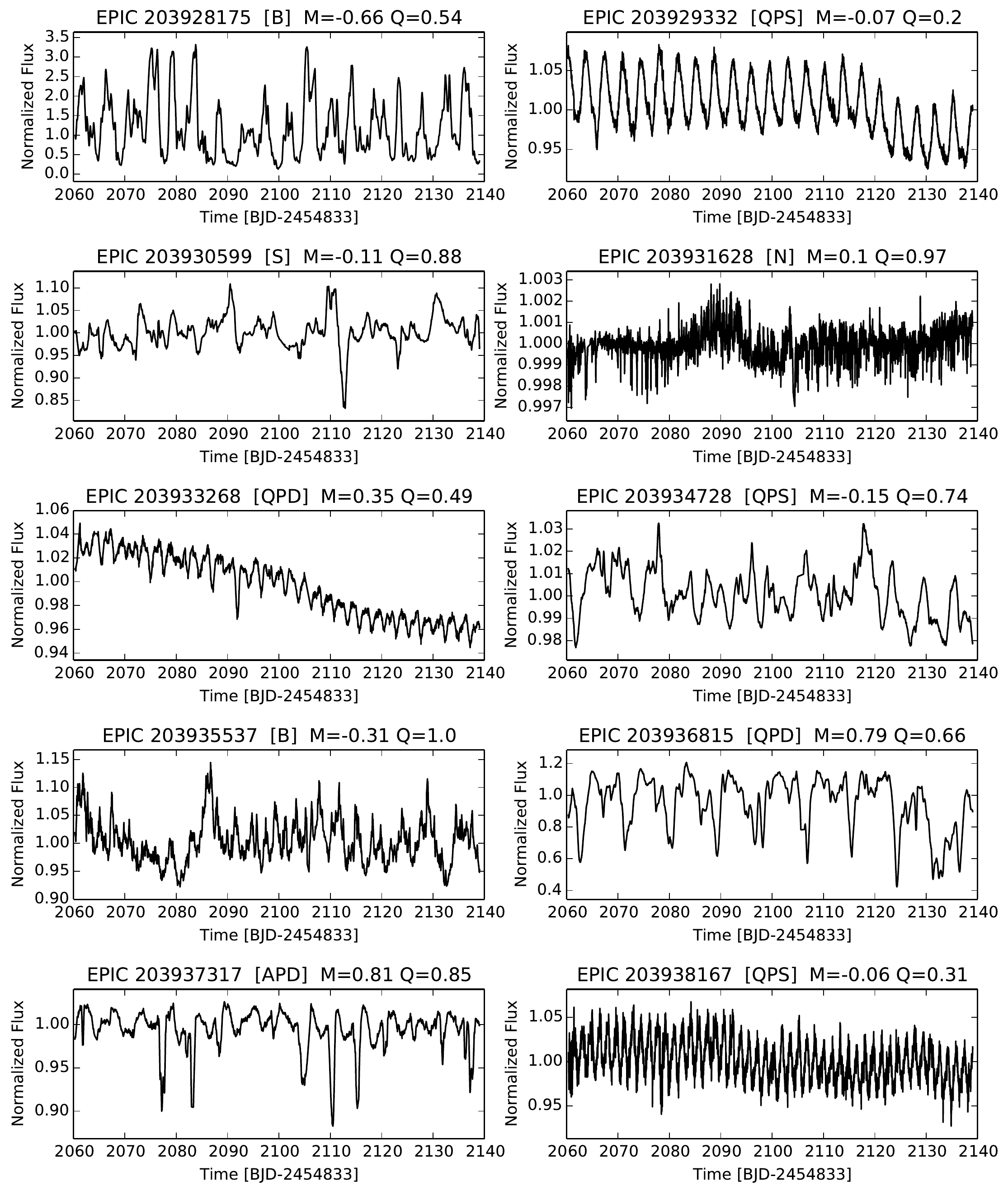}
 \caption{Cont.}
  \end{figure*}

\clearpage
  
 \addtocounter{figure}{-1} 
 \begin{figure*}
 \epsscale{0.90}
 \plotone{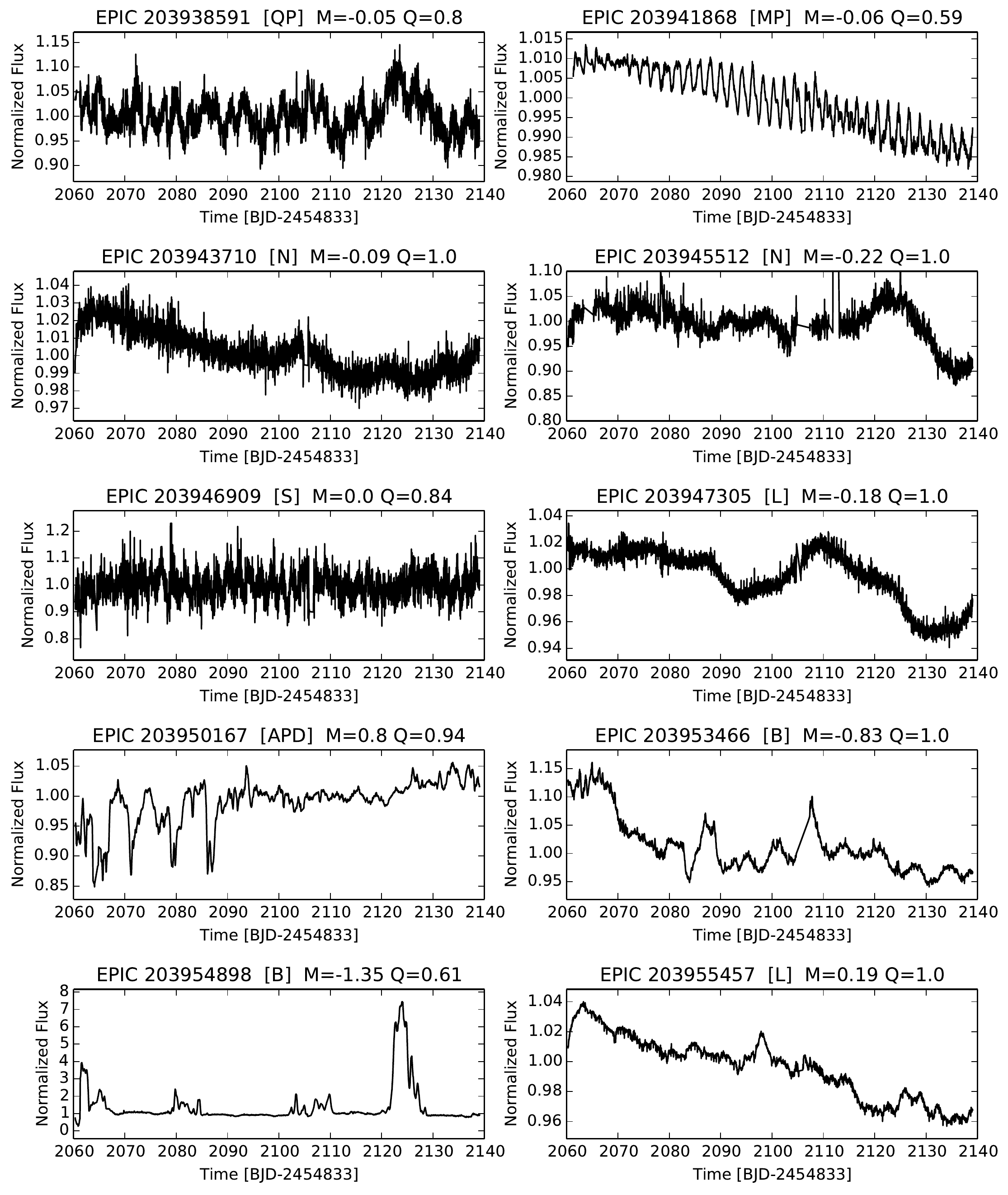}
 \caption{Cont.}
  \end{figure*}

\clearpage
  
 \addtocounter{figure}{-1} 
 \begin{figure*}
 \epsscale{0.90}
 \plotone{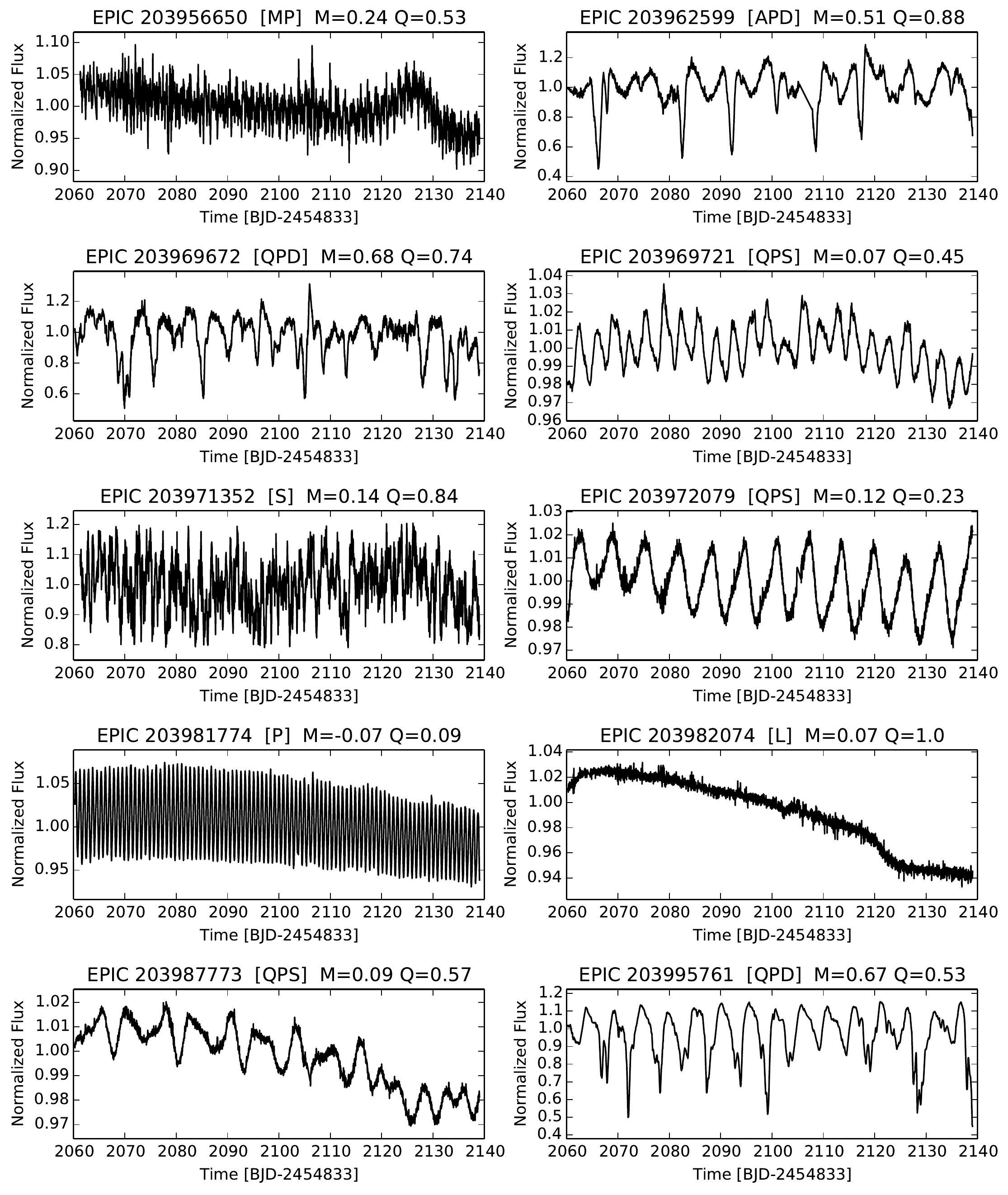}
 \caption{Cont.}
  \end{figure*}

\clearpage
  
 \addtocounter{figure}{-1} 
 \begin{figure*}
 \epsscale{0.90}
 \plotone{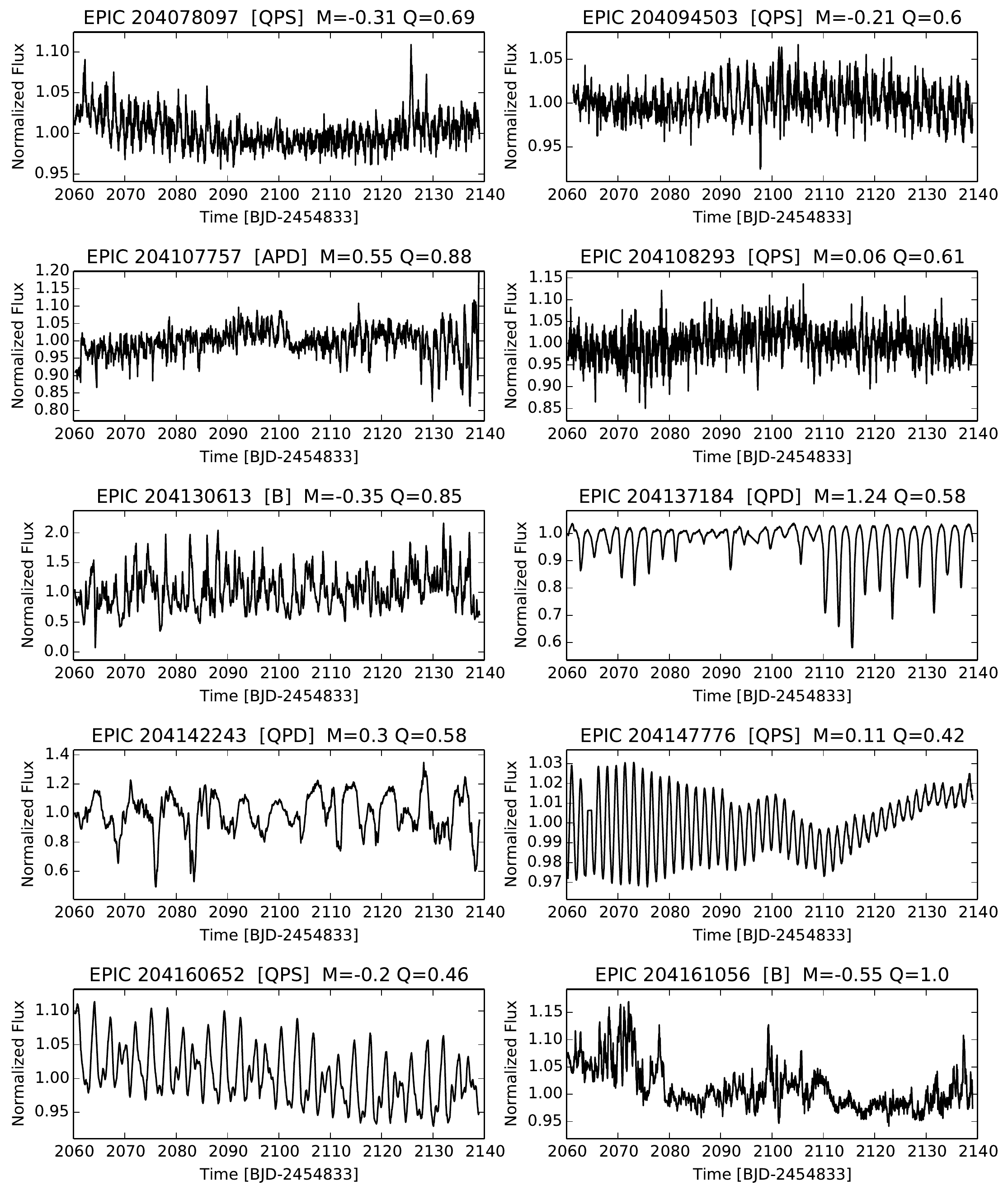}
 \caption{Cont.}
  \end{figure*}
  
\clearpage

 \addtocounter{figure}{-1} 
 \begin{figure*}
 \epsscale{0.90}
 \plotone{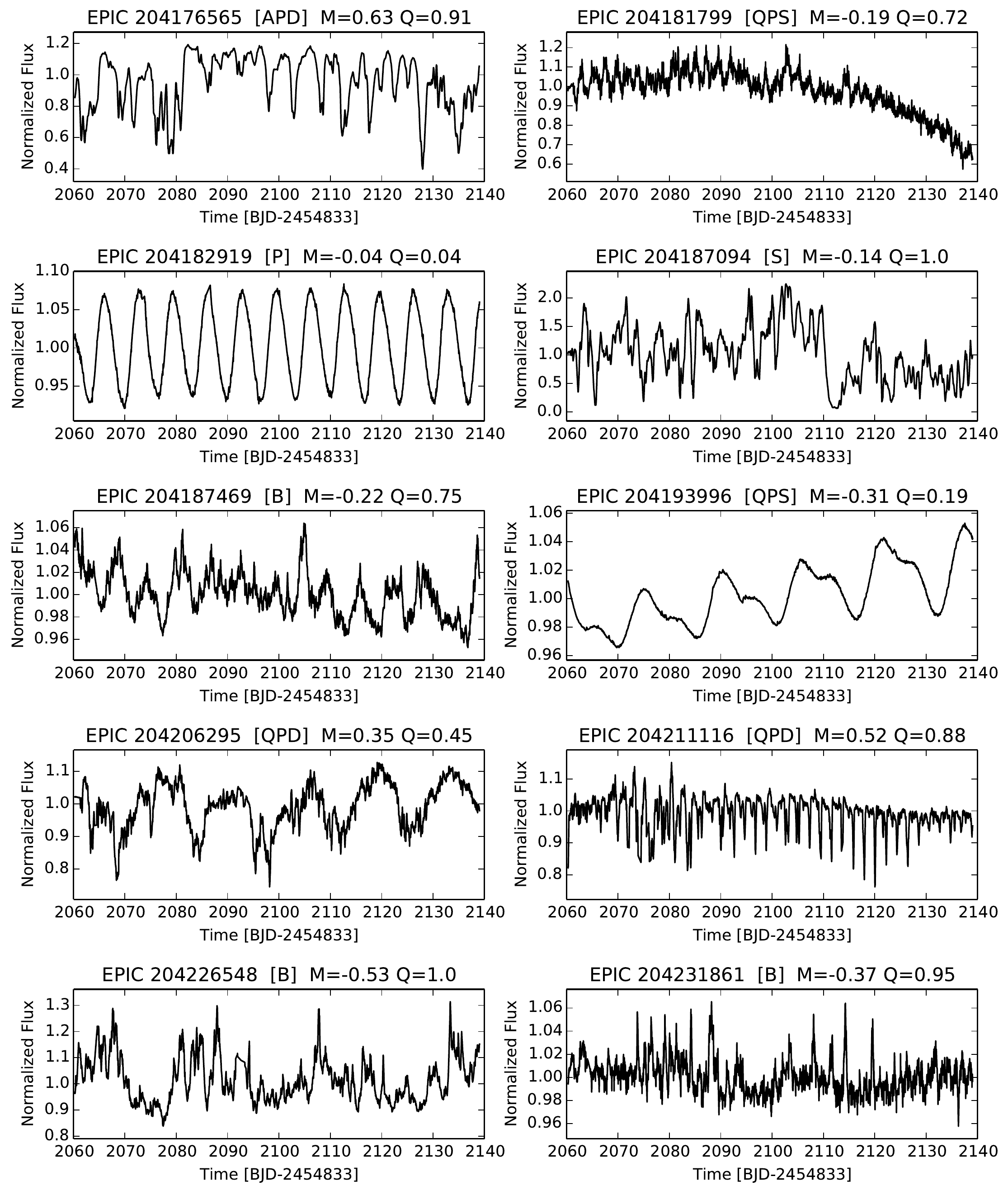}
 \caption{Cont.}
  \end{figure*}
  
\clearpage

 \addtocounter{figure}{-1} 
 \begin{figure*}
 \epsscale{0.90}
 \plotone{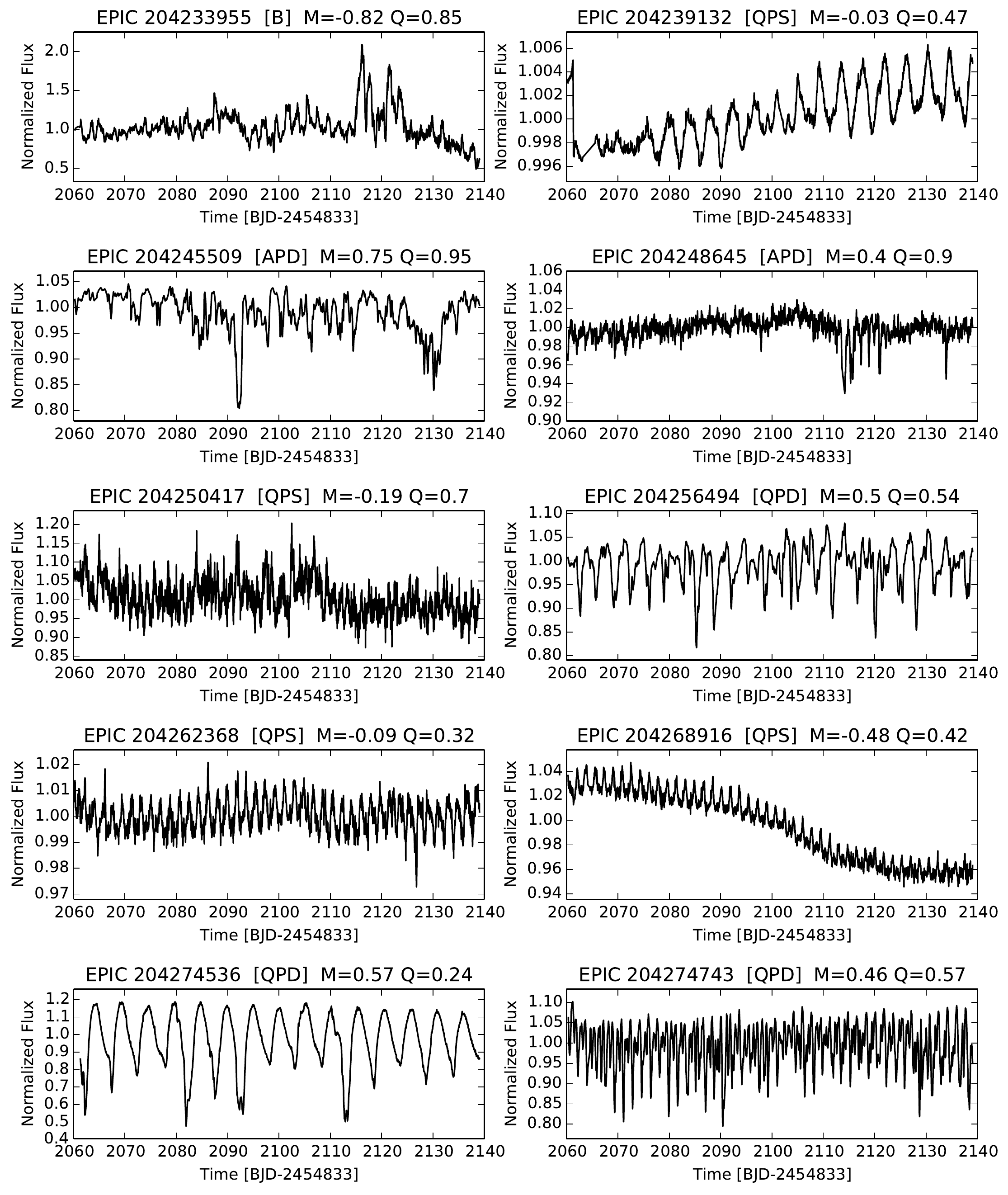}
 \caption{Cont.}
  \end{figure*}

\clearpage
  
 \addtocounter{figure}{-1} 
 \begin{figure*}
 \epsscale{0.90}
 \plotone{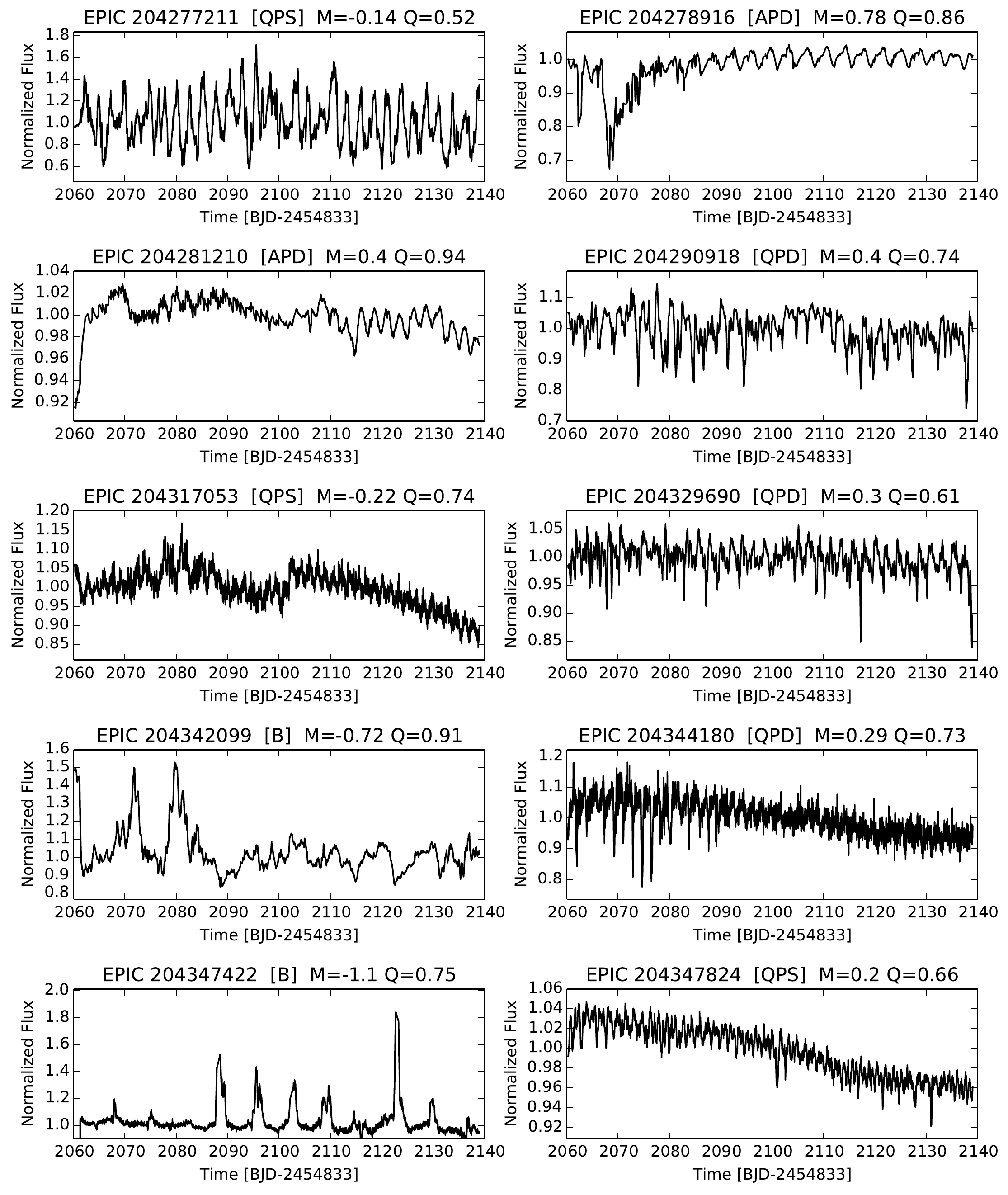}
 \caption{Cont.}
  \end{figure*}

\clearpage
  
 \addtocounter{figure}{-1} 
 \begin{figure*}
 \epsscale{0.90}
 \plotone{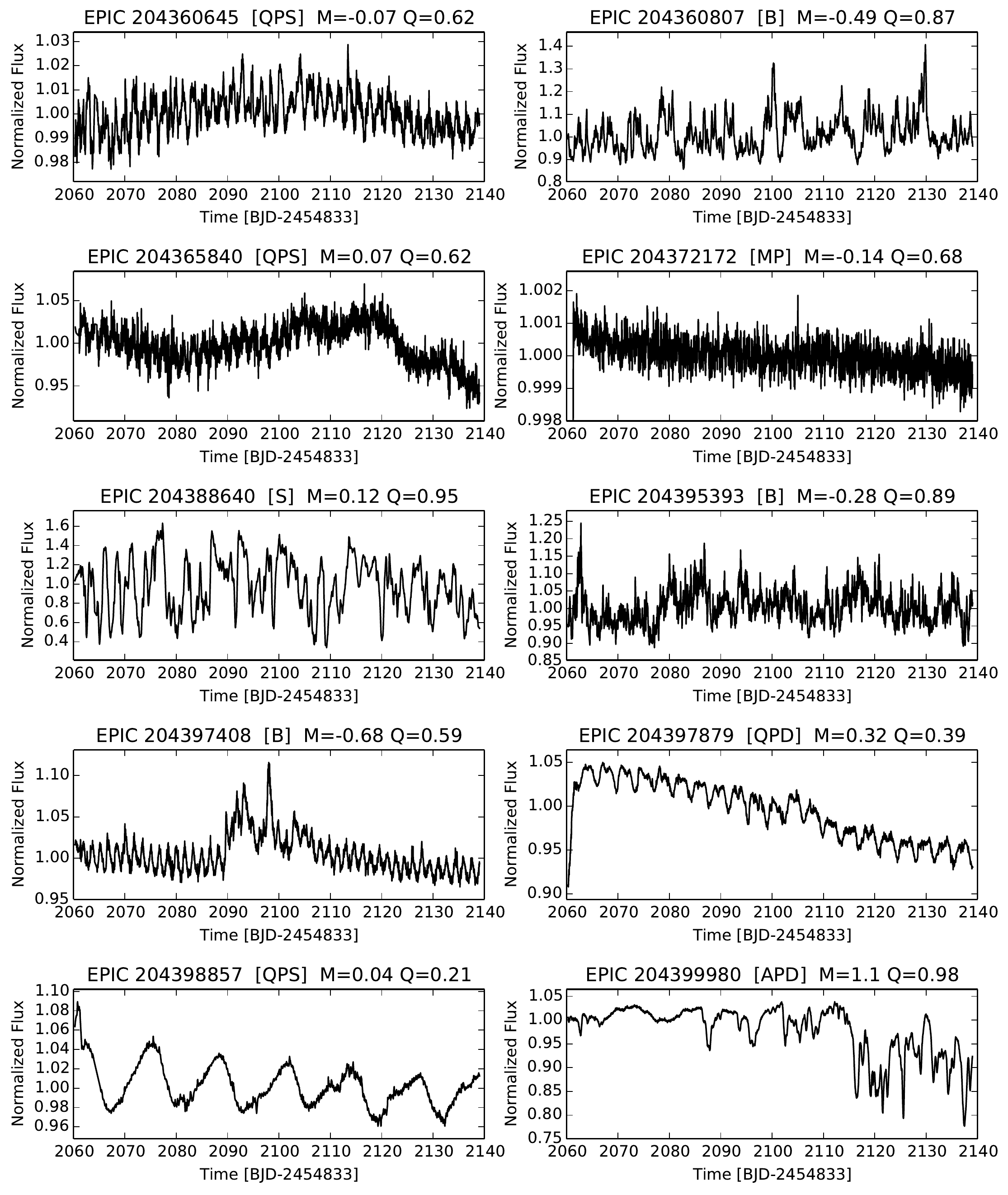}
 \caption{Cont.}
  \end{figure*}

\clearpage
  
 \addtocounter{figure}{-1} 
 \begin{figure*}
 \epsscale{0.90}
 \plotone{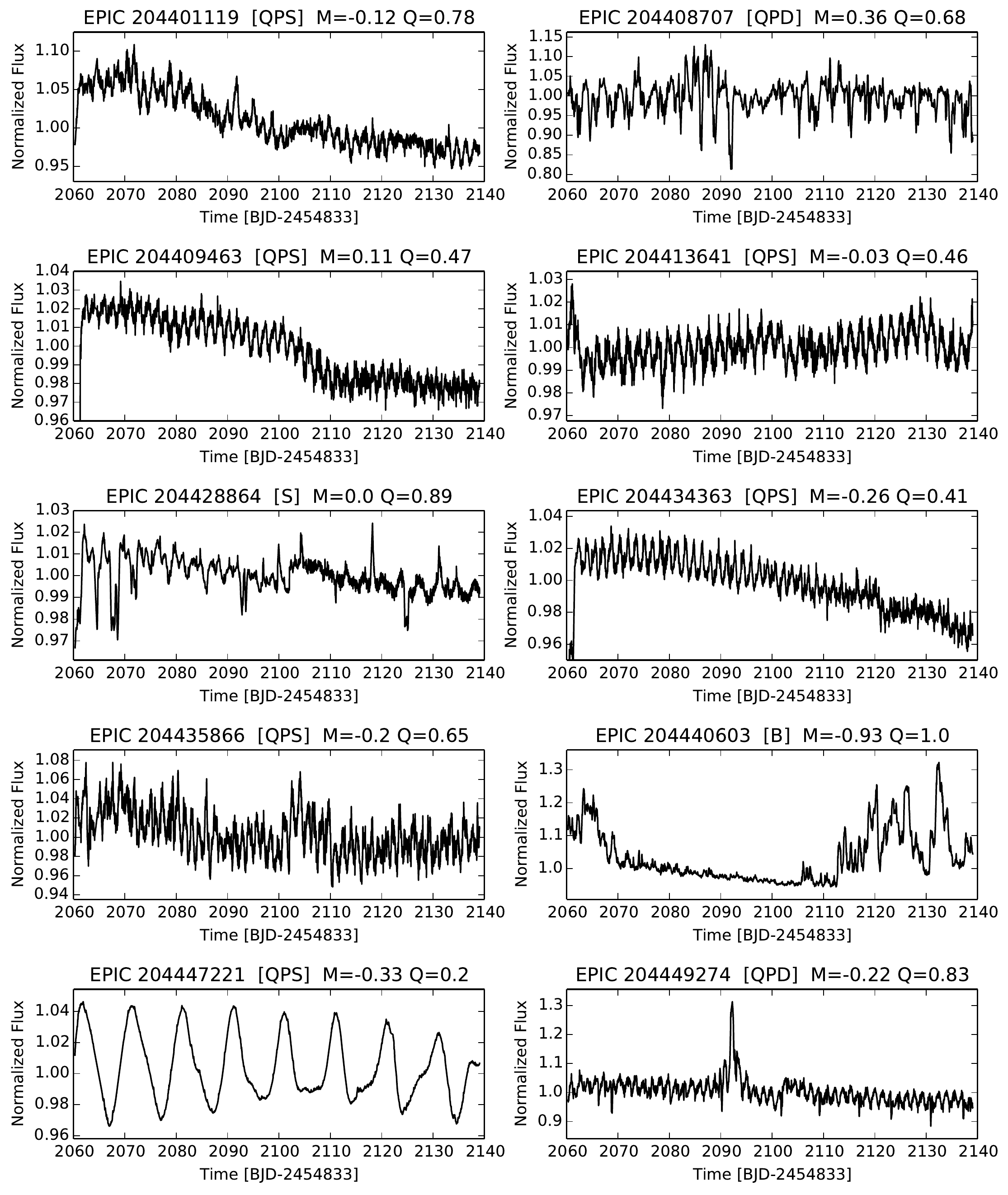}
 \caption{Cont.}
  \end{figure*}

\clearpage
  
 \addtocounter{figure}{-1} 
 \begin{figure*}
 \epsscale{0.90}
 \plotone{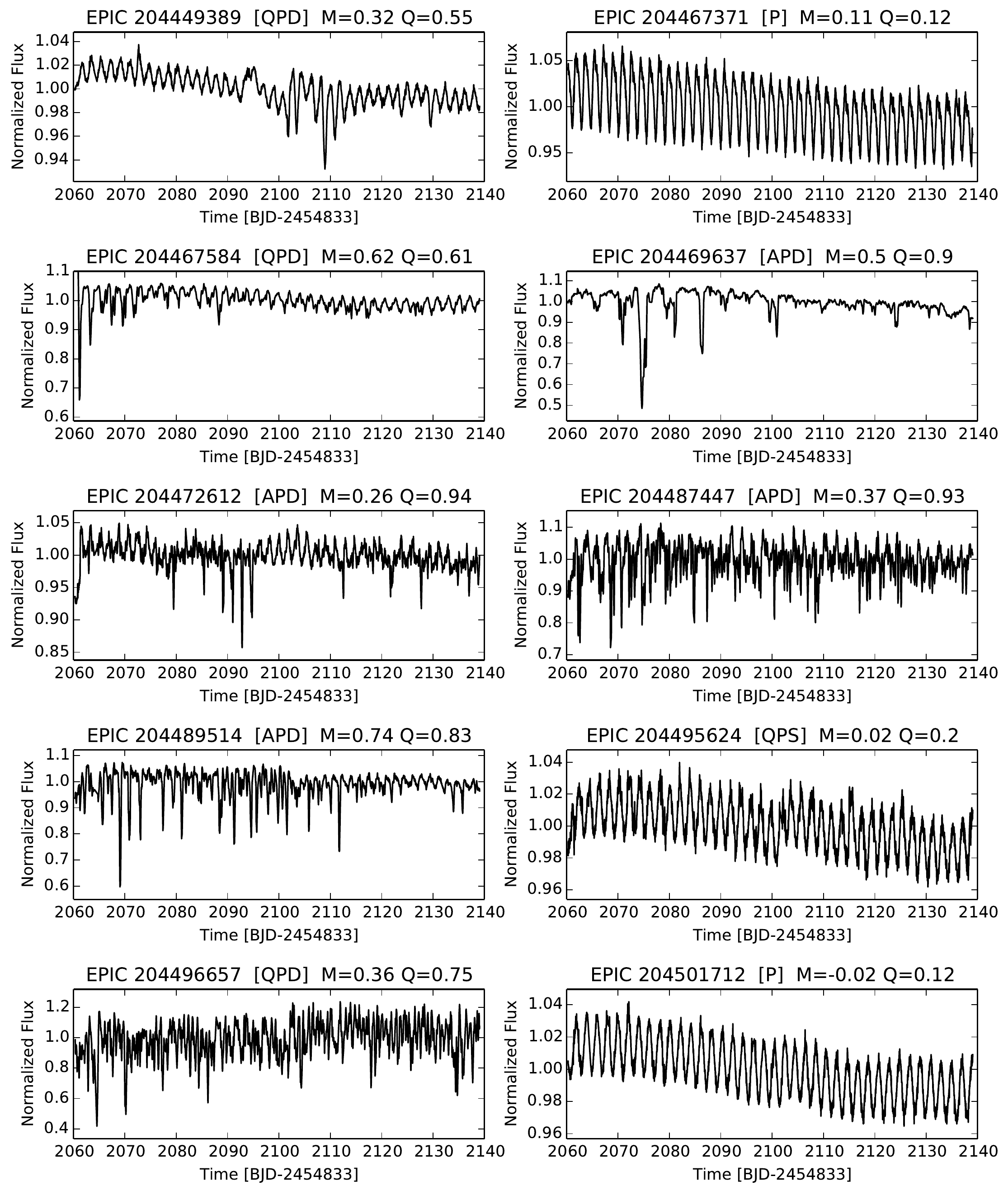}
 \caption{Cont.}
  \end{figure*}

\clearpage
  
 \addtocounter{figure}{-1} 
 \begin{figure*}
 \epsscale{0.90}
 \plotone{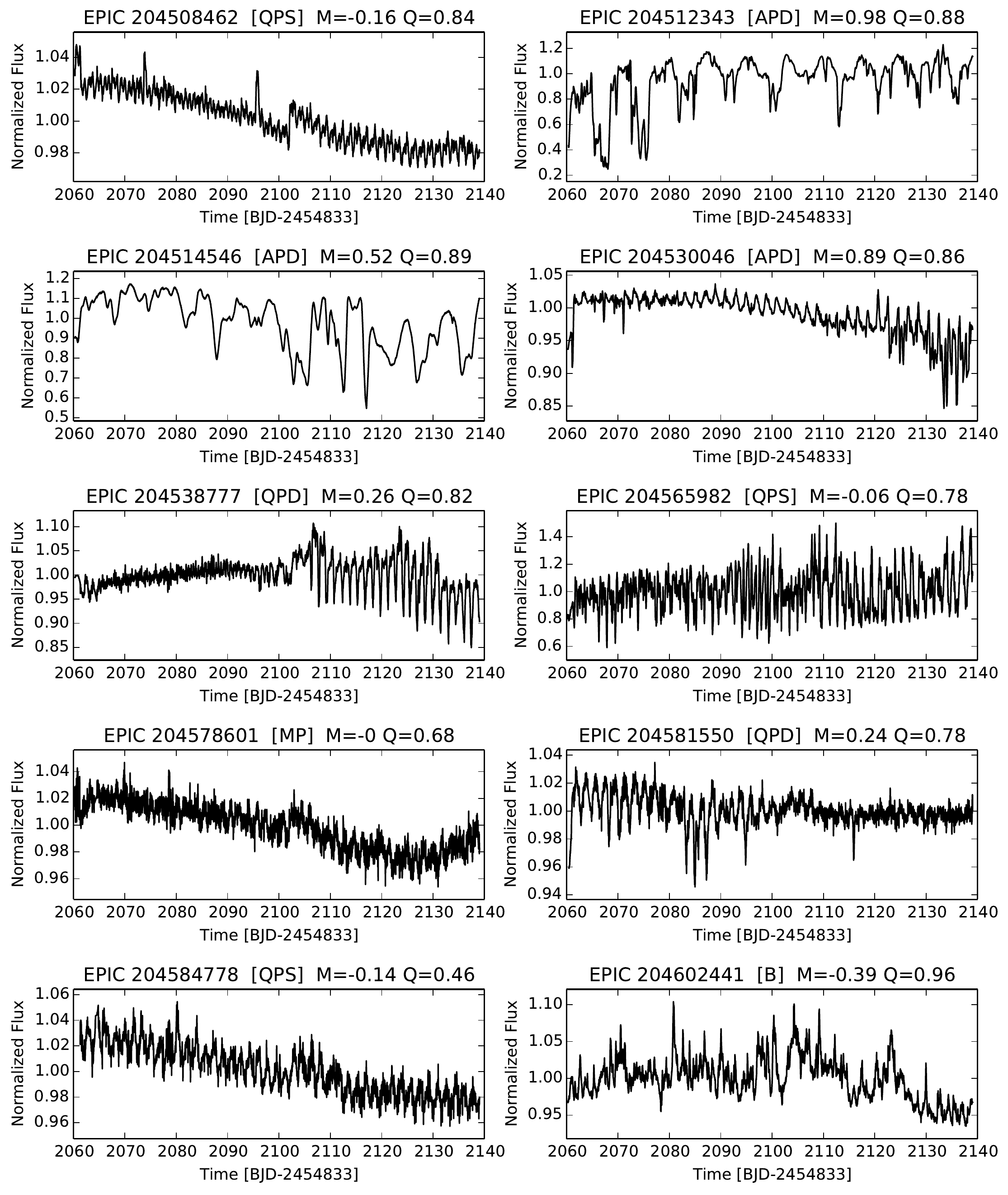}
 \caption{Cont.}
  \end{figure*}

\clearpage
  
 \addtocounter{figure}{-1} 
 \begin{figure*}
 \epsscale{0.90}
 \plotone{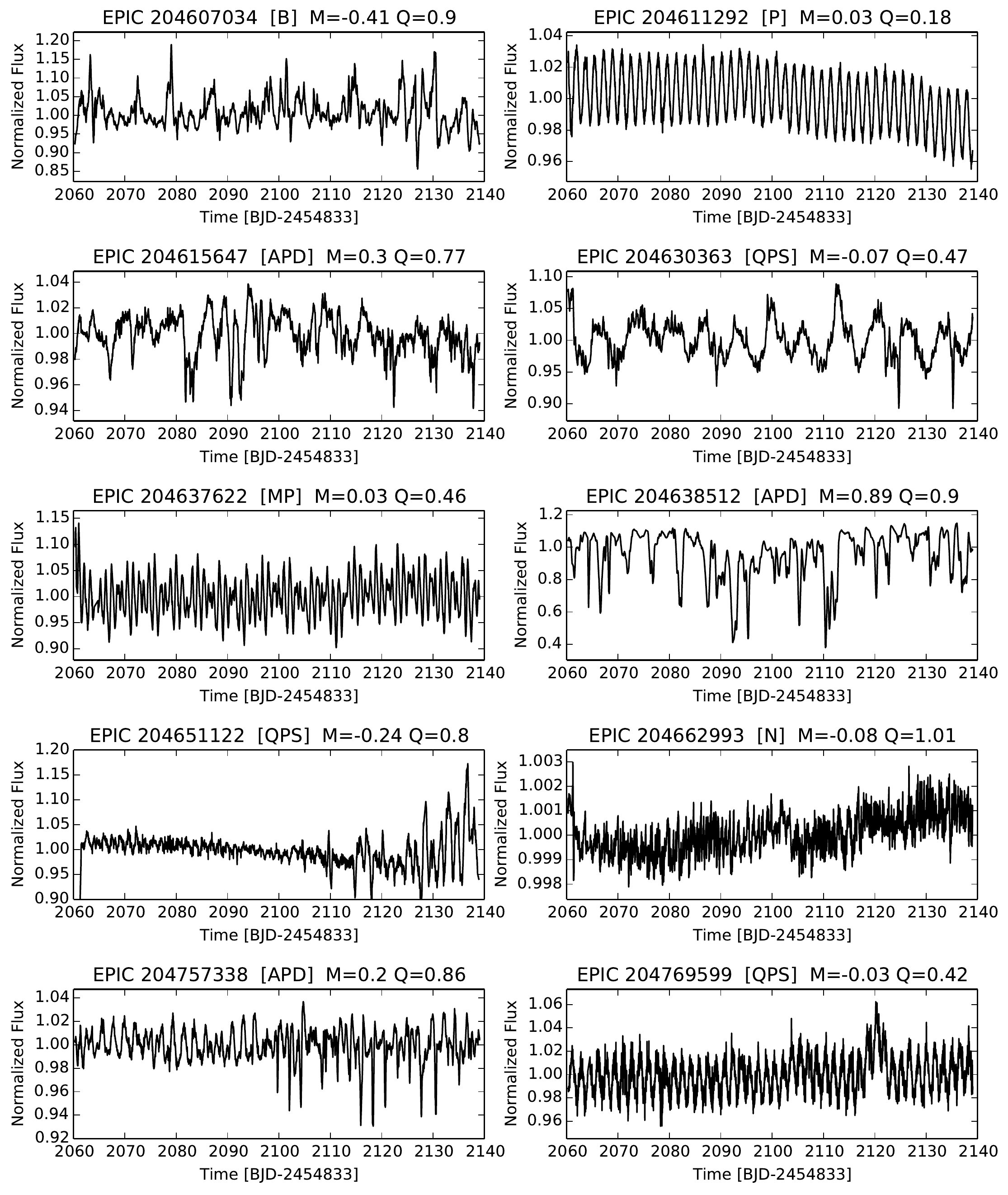}
 \caption{Cont.}
  \end{figure*}

\clearpage
  
 \addtocounter{figure}{-1} 
 \begin{figure*}
 \epsscale{0.90}
 \plotone{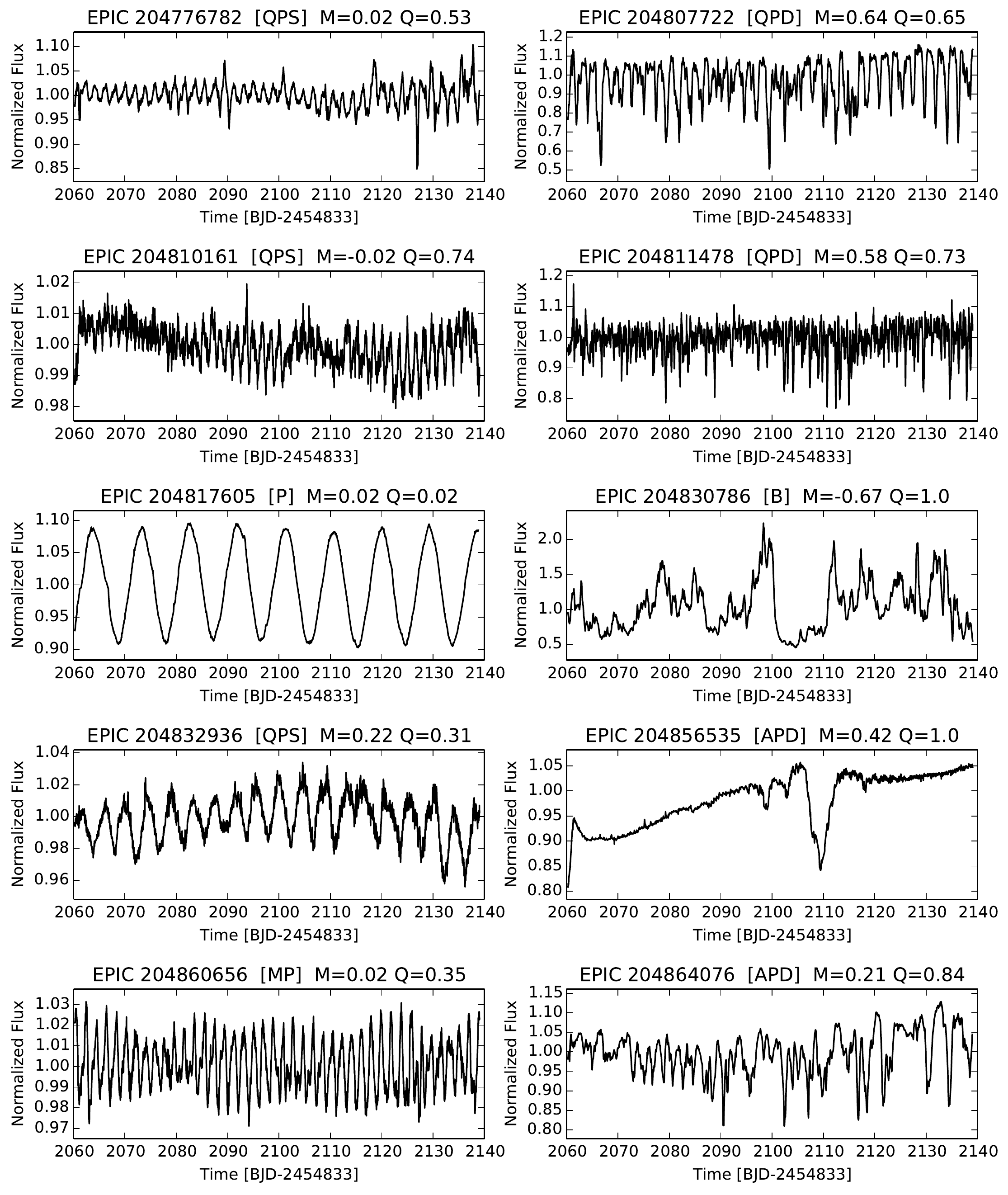}
 \caption{Cont.}
  \end{figure*}

\clearpage
  
 \addtocounter{figure}{-1} 
 \begin{figure*}
 \epsscale{0.90}
 \plotone{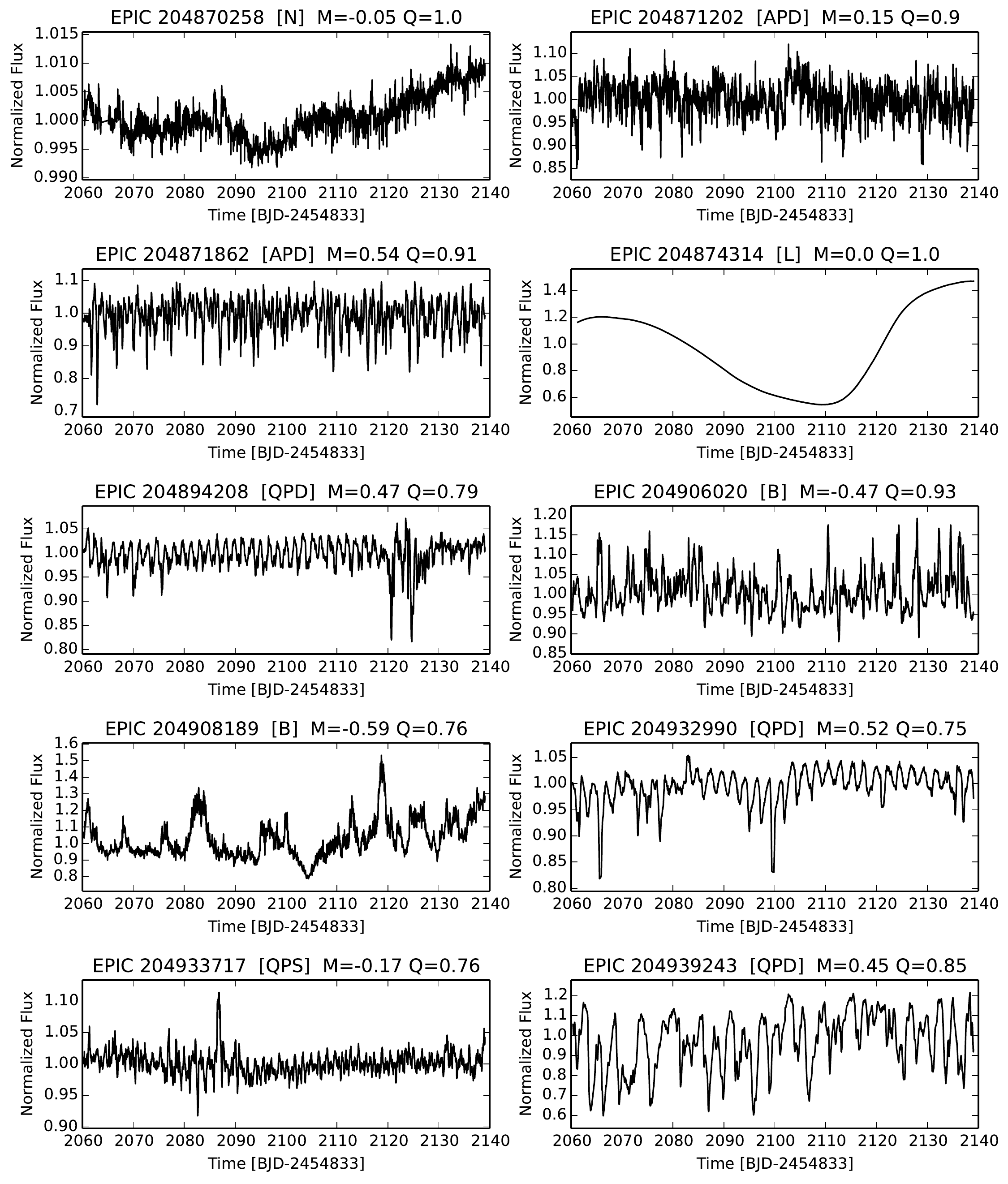}
 \caption{Cont.}
  \end{figure*}

\clearpage
  
 \addtocounter{figure}{-1} 
 \begin{figure*}
 \epsscale{0.90}
 \plotone{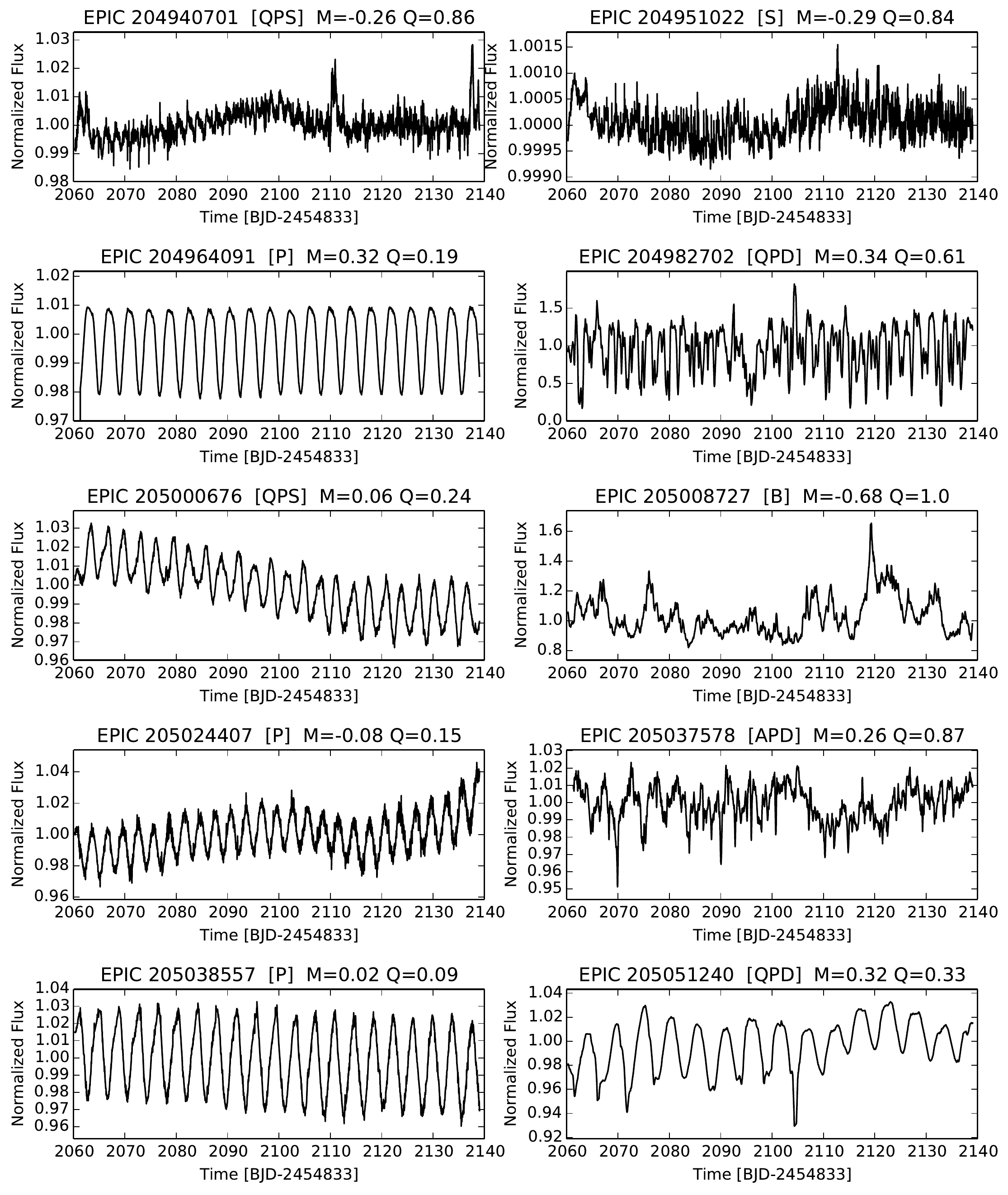}
 \caption{Cont.}
  \end{figure*}

\clearpage
  
 \addtocounter{figure}{-1} 
 \begin{figure*}
 \epsscale{0.90}
 \plotone{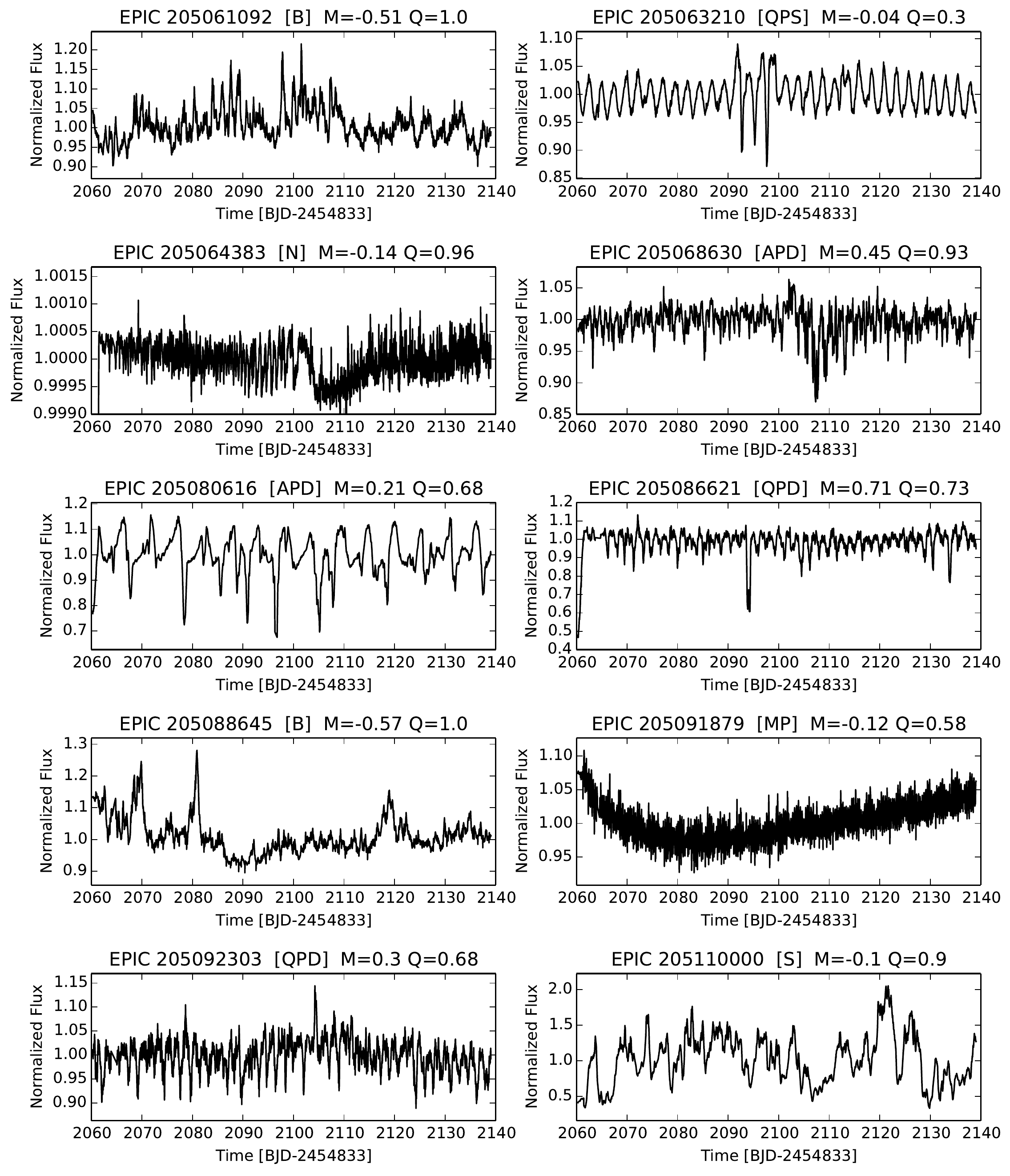}
 \caption{Cont.}  \end{figure*}

\clearpage
  
 \addtocounter{figure}{-1} 
 \begin{figure*}
 \epsscale{0.90}
 \plotone{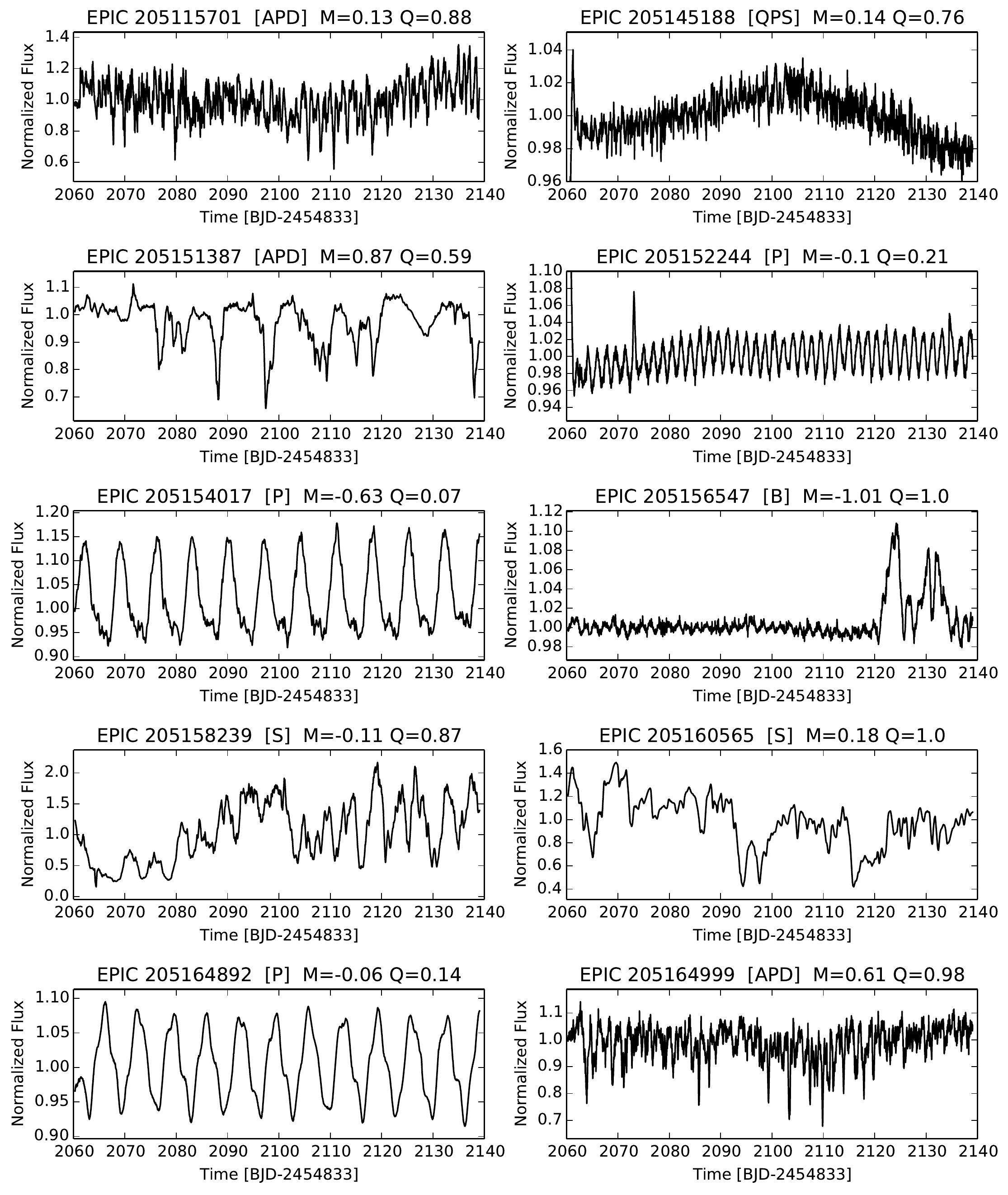}
 \caption{Cont.}
  \end{figure*}

\clearpage
  
 \addtocounter{figure}{-1} 
 \begin{figure*}
 \epsscale{0.90}
 \plotone{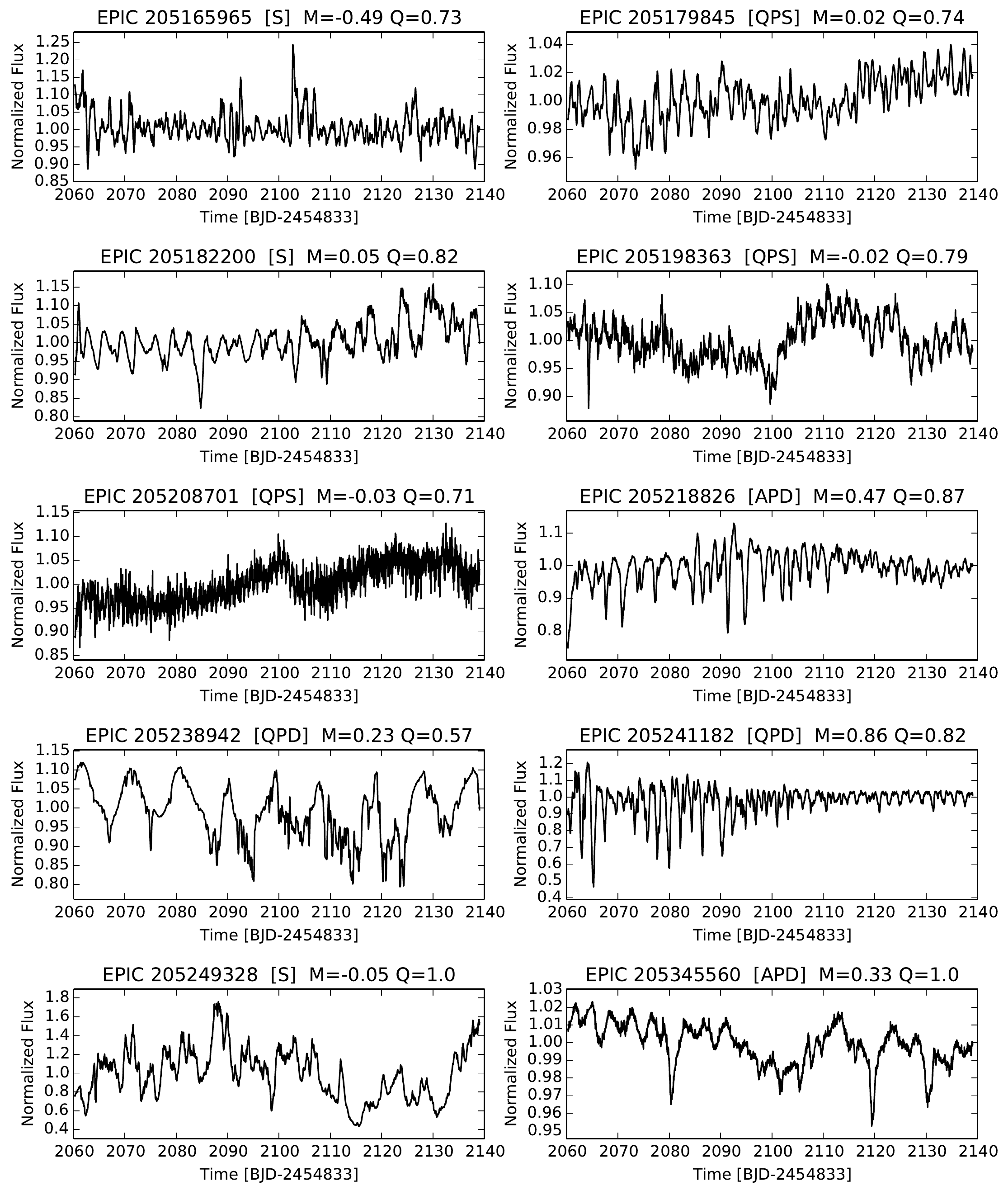}
 \caption{Cont.}
  \end{figure*}

\clearpage 
 
 \addtocounter{figure}{-1} 
 \begin{figure*}
 \epsscale{0.90}
 \plotone{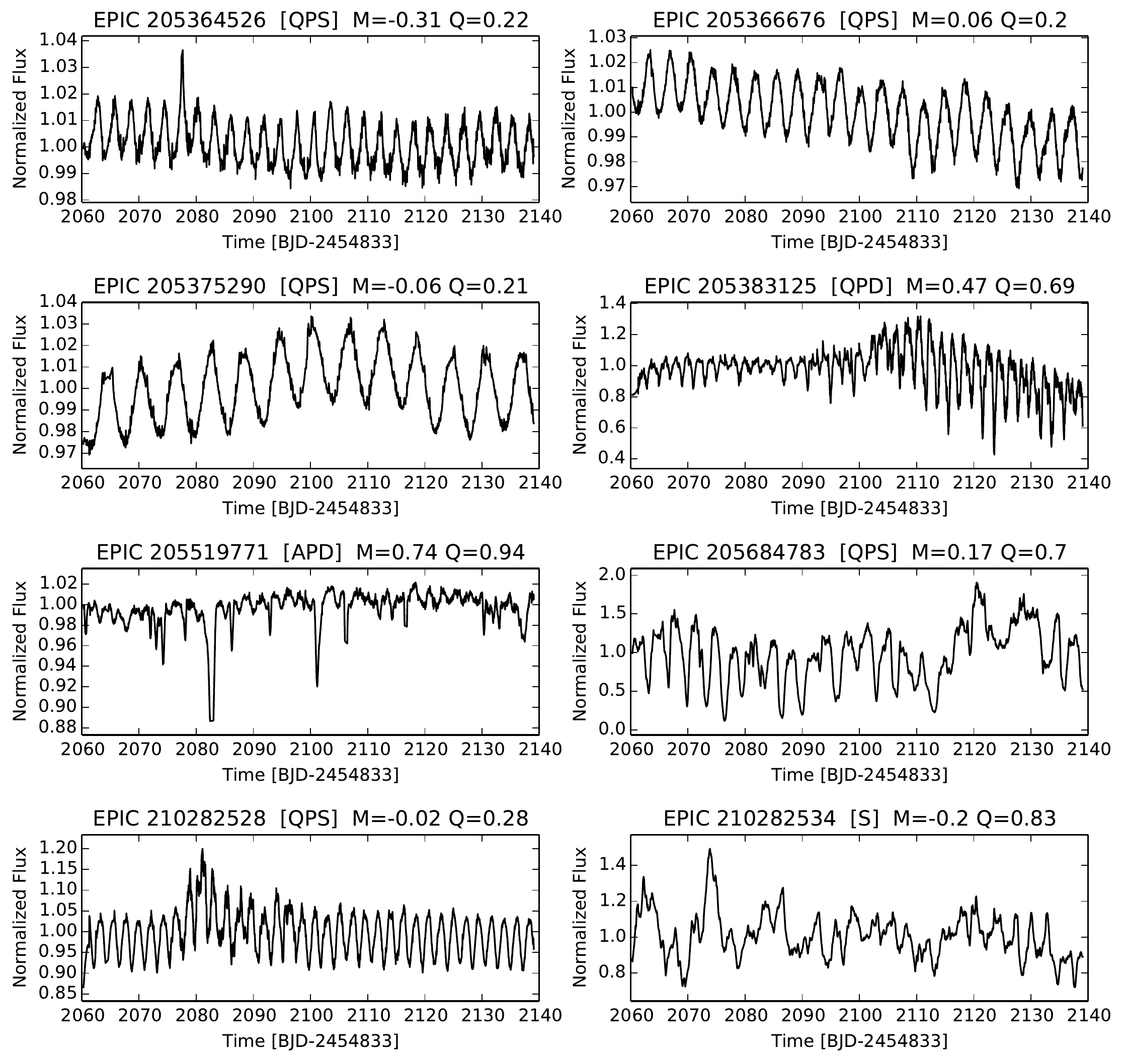}
 \caption{Cont.}
  \end{figure*}

\end{document}